\documentclass[twocolumn, twocolappendix]{aastex63}
\received{\today}
\shorttitle{Detecting DCOs with LISA}
\graphicspath{{./figures/}}

\usepackage{lipsum}
\usepackage{physics}
\usepackage{multirow}
\usepackage{xspace}
\usepackage{natbib}
\usepackage{fontawesome5}
\usepackage{xcolor}
\usepackage{wrapfig}
\usepackage[figuresright]{rotating}

\usepackage[hang,flushmargin]{footmisc}

\makeatletter
\newcommand{\unit}[1]{%
    \,\mathrm{#1}\checknextarg}
\newcommand{\checknextarg}{\@ifnextchar\bgroup{\gobblenextarg}{}}
\newcommand{\gobblenextarg}[1]{\,\mathrm{#1}\@ifnextchar\bgroup{\gobblenextarg}{}}
\makeatother

\newcommand{\avg}[1]{\left\langle#1\right\rangle}

\newcommand{\modFid}{A}
\newcommand{\modBetaLow}{B}
\newcommand{\modBetaMed}{C}
\newcommand{\modBetaHigh}{D}
\newcommand{\modCaseBB}{E$^{\prime}$}
\newcommand{\modCaseBBOpt}{F}
\newcommand{\modAlphaLowest}{G}
\newcommand{\modAlphaLow}{H}
\newcommand{\modAlphaHigh}{I}
\newcommand{\modAlphaHighest}{J}
\newcommand{\modOpt}{K}
\newcommand{\modRapid}{L}
\newcommand{\modNSLow}{M}
\newcommand{\modNSHigh}{N}
\newcommand{\modNoPISN}{O}
\newcommand{\modSigLow}{P}
\newcommand{\modSigLower}{Q}
\newcommand{\modNoBH}{R}
\newcommand{\modWRLow}{S}
\newcommand{\modWRHigh}{T}

\newcommand{\modRangeMT}{B-D}
\newcommand{\modRangeCE}{E$^{\prime}$-K}
\newcommand{\modRangeSN}{L-R}
\newcommand{\modRangeML}{S-T}

\newcommand{\nModels}{20}
\newcommand{\nMinusOneModels}{19}

\newcommand{\rangeFourYear}{30-370}
\newcommand{\rangeTenYear}{50-550}

\newcommand{\BHBHFourYear}{74}
\newcommand{\BHNSFourYear}{42}
\newcommand{\NSNSFourYear}{8}

\newcommand{\BHBHqAbovePointEight}{8\%}

\newcommand{\NSNSqAbovePointEight}{90\%}

\newcommand{\BHBHNotCirc}{87\%}

\newcommand{\BHBHHighlyEccentric}{21\%}

\newcommand{\BHBHAboveMaxWDWDFourPerc}{24}
\newcommand{\BHNSAboveMaxWDWDFourPerc}{28}
\newcommand{\NSNSAboveMaxWDWDFourPerc}{4}
\newcommand{\BHBHAboveMaxWDWDTenPerc}{38}
\newcommand{\BHNSAboveMaxWDWDTenPerc}{41}
\newcommand{\NSNSAboveMaxWDWDTenPerc}{5}

\newcommand{\BHBHMultipleHarmonicsFourPerc}{55}
\newcommand{\BHNSMultipleHarmonicsFourPerc}{27}
\newcommand{\NSNSMultipleHarmonicsFourPerc}{66}
\newcommand{\BHBHMultipleHarmonicsTenPerc}{61}
\newcommand{\BHNSMultipleHarmonicsTenPerc}{29}
\newcommand{\NSNSMultipleHarmonicsTenPerc}{68}

\newcommand{\BHBHEccInDiscFourPerc}{40}
\newcommand{\BHNSEccInDiscFourPerc}{23}
\newcommand{\NSNSEccInDiscFourPerc}{59}
\newcommand{\BHBHEccInDiscTenPerc}{40}
\newcommand{\BHNSEccInDiscTenPerc}{23}
\newcommand{\NSNSEccInDiscTenPerc}{59}

\newcommand{\BHBHNotWDWDFour}{37}
\newcommand{\BHNSNotWDWDFour}{18}
\newcommand{\NSNSNotWDWDFour}{5}
\newcommand{\BHBHNotWDWDTen}{70}
\newcommand{\BHNSNotWDWDTen}{38}
\newcommand{\NSNSNotWDWDTen}{8}

\newcommand{\BHBHEitherBHOrNSFour}{16}
\newcommand{\BHNSEitherBHOrNSFour}{7}
\newcommand{\NSNSEitherBHOrNSFour}{4}
\newcommand{\BHBHEitherBHOrNSTen}{39}
\newcommand{\BHNSEitherBHOrNSTen}{17}
\newcommand{\NSNSEitherBHOrNSTen}{8}
\newcommand{\BHBHEitherBHOrNSFourPerc}{21}
\newcommand{\BHNSEitherBHOrNSFourPerc}{18}
\newcommand{\NSNSEitherBHOrNSFourPerc}{47}
\newcommand{\BHBHEitherBHOrNSTenPerc}{33}
\newcommand{\BHNSEitherBHOrNSTenPerc}{24}
\newcommand{\NSNSEitherBHOrNSTenPerc}{62}

\newcommand{\BHBHatLeastOneLowerMassGapPerc}{69}
\newcommand{\BHNSatLeastOneLowerMassGapPerc}{39}
\newcommand{\NSNSatLeastOneLowerMassGapPerc}{0}

\newcommand{\citefloorp}{(\citealt{Broekgaarden+2021}, Broekgaarden et al.\ (in prep.))}

\newcommand{\confinv}[3]{$#1${\raisebox{0.5ex}{\tiny$_{-#2}^{+#3}$}}}
\newcommand{\boldconfinv}[3]{$\mathbf{#1}${\raisebox{0.5ex}{\tiny$\mathbf{_{-#2}^{+#3}}$}}}

\definecolor{Blush}{rgb}{0.87, 0.36, 0.51}

\newcommand{\achem}{\ensuremath{[\alpha/\mathrm{Fe}]}}

\begin{document}

\title{{Gravitational wave sources in our Galactic backyard:}\\{Predictions for BHBH, BHNS and NSNS binaries detectable with LISA}}

% affiliations
\newcommand{\cfa}{Center for Astrophysics | Harvard \& Smithsonian, 60 Garden Street, Cambridge, MA 02138, USA}
\newcommand{\mpa}{Max-Planck-Institut für Astrophysik, Karl-Schwarzschild-Straße 1, 85741 Garching, Germany}
\newcommand{\cca}{Center for Computational Astrophysics, Flatiron Institute, 162 Fifth Ave, New York, NY, 10010, USA}
\newcommand{\UvA}{Anton Pannekoek Institute for Astronomy \& Grappa, University of Amsterdam, Postbus 94249, 1090 GE Amsterdam, The Netherlands}
\newcommand{\UW}{Department of Astronomy, University of Washington, Seattle, WA, 98195}

\author[0000-0001-6147-5761]{T. Wagg}
\affiliation{\UW}
\affiliation{\cfa}
\affiliation{\mpa}

\author[0000-0002-4421-4962]{F.S. Broekgaarden}
\affiliation{\cfa}

\author[0000-0001-9336-2825]{S.E. de Mink}
\affiliation{\mpa}
\affiliation{\UvA}
\affiliation{\cfa}

\author[0000-0002-6411-8695]{N. Frankel}
\affiliation{Canadian Institute for Theoretical Astrophysics, University of Toronto, 60 St. George Street, Toronto, ON M5S 3H8, Canada}
\affiliation{Max Planck Institute for Astronomy, K\"onigstuhl 17, D-69117 Heidelberg, Germany}
\affiliation{David A. Dunlap Department for Astronomy and Astrophysics, University
of Toronto, 50 St. George Street, Toronto, ON M5S 3H4, Canada}

\author[0000-0001-5484-4987]{L.A.C. van Son}
\affiliation{\cfa}
\affiliation{\UvA}
\affiliation{\mpa}

\author[0000-0001-7969-1569]{S. Justham}
\affiliation{School of Astronomy \& Space Science, University of the Chinese Academy of Sciences, Beijing 100012, China}
\affiliation{\UvA}
\affiliation{\mpa}

\correspondingauthor{Tom Wagg}
\email{tomjwagg@gmail.com}

\begin{abstract}
Future searches for gravitational waves from space will be sensitive to double compact objects (DCOs) in our Milky-Way. We present new simulations of the populations of double black holes (BHBHs), black hole neutron stars (BHNSs) and double neutron stars (NSNSs) that will be detectable by the planned space-based gravitational wave detector LISA. 
For our estimates, we use an empirically-informed model of the metallicity dependent star formation history of the Milky Way. We populate using an extensive suite of binary population-synthesis predictions for varying assumptions relating to mass transfer, common-envelope, supernova kicks, remnant masses and wind mass loss physics.

For a 4(10)-year LISA mission, we predict between \rangeFourYear{}(\rangeTenYear{}) detections over these variations, out of which 6-154(9-238) are BHBHs, 2-198(3-289) are BHNSs and 3-35(4-57) are NSNSs.
We discuss how the variations in the physics assumptions alter the distribution of properties of the detectable systems, even when the detection rates are unchanged. In particular we discuss the observable characteristics such as the chirp mass, eccentricity and sky localisation and how the BHBH, BHNS and NSNS populations can be distinguished, both from each other and from the more numerous double white dwarf population. 
We further discuss the possibility of multi-messenger observations of pulsar populations with the Square Kilometre Array (SKA) and assess the benefits of extending the LISA mission. 

\end{abstract}


\section{Introduction} \label{sec:intro}
Since the first direct observation of gravitational waves \citep{Abbott+2016_first_detection}, the number of black hole (BH) and neutron star (NS) binaries observed by ground-based gravitational-wave detectors has rapidly grown \citep{Abbott+2019_GWTC1,Abbott+2020_GWTC2,Abbott+2021_GWTC3}, offering exciting insights into the formation, lives and deaths of massive (binary) stars \citep[e.g.][]{Abbott+2021_GWTC2_inference}.

The Laser Interferometer Space Antenna (LISA, \citealp{Amaro-Seoane+2017, Colpi+2019}) will provide observations in an entirely new regime of gravitational waves. LISA will observe at lower frequencies ($10^{-5} \lesssim f / \unit{Hz} \lesssim 10^{-1}$) than ground-based detectors and so will enable the study of sources that are imperceptible by ground-based detectors, such as the mergers of supermassive black holes and extreme mass-ratio inspirals \citep[e.g.][]{Begelman+1980, Klein+2016}. Moreover, this frequency regime is also of interest for the detection of \textit{local} stellar-mass double compact objects (DCOs) millions of years before their merger. This presents an opportunity for both multi-messenger detections to search for electromagnetic counterparts, as well as multiband gravitational-wave detections that can help to constrain binary characteristics \citep[e.g.][]{Sesana+2016, Gerosa+2019}. In addition, LISA will be able to measure the eccentricities of DCOs, which may yield further constraints on binary evolution, differentiate between formation channels and distinguish between DCO types \citep[e.g.][]{Nelemans+2001, Breivik+2016, Antonini+2017, Rodriguez+2018}. Unlike ground-based detectors, LISA only detects stellar-mass sources in local galaxies, with the majority residing in the Milky Way. These sources could be used as a probe for the structure of our galaxy \citep[e.g.][]{Korol+2019}.

Traditionally, predictions about the detection of stellar-mass sources with LISA focus on double white dwarf (WDWD) binaries, as they are abundantly present in our galaxy and are expected to be the dominant source of stellar-mass binaries that are detectable by LISA \citep{Nelemans+2001,Ruiter+2010,Yu+2010,Nissanke+2012,Korol+2017,Lamberts+2018}. More recently, interest has grown in the detection of NS and BH binaries. Although they are more rare, LISA detections of these sources are potentially valuable for learning more about the evolution and endpoints of massive stars. In this paper we focus on making LISA predictions for double black hole binaries (BHBHs), black hole neutron star binaries (BHNSs) and double neutron star binaries (NSNSs).

The detection of NSNSs in LISA could improve our understanding of many phenomena. Galactic NSNSs have been observed with electromagnetic signals for several decades (e.g.\ \citealp{Hulse+1975, Antoniadis+2016}, see also refs.\ in \citealp{Tauris+2017}) and more recently the mergers of NSNSs have been detected with ground-based gravitational-wave detectors \citep[e.g.][]{Abbott+2017_NSNS}. A LISA detectable NSNS with a pulsar component close to merger would be ideal for connecting these populations, as the binary could be observed from inspiral to merger. NSNSs (and possibly BHNSs) are useful sources for understanding the origin of r-process elements \citep[e.g.][]{Eichler+1989} as well as the electromagnetic counterparts to gravitational-wave signals, such as kilonovae \citep[e.g.][]{Li+1998, Metzger+2017}, short gamma-ray bursts \citep[e.g.][]{Berger+2014}, radio emission \citep[e.g.][]{Hotokezaka+2016} and neutrinos \citep[e.g.][]{Kyutoku+2018}.

BHBHs in the Milky Way present a greater observational challenge. To date, no BH has been observed in a BHBH binary in the Milky Way, and so LISA could provide the first detection of a Galactic BHBH. The only confirmed BHs in our galaxy have been discovered as components of X-ray binaries with companion stars \citep[e.g.][]{Bolton+1972,Webster+1972}. These detections have observed BHs with masses mainly constrained between $5$ and $10 \unit{M_\odot}$ \citep{Corral-Santana+2016}, a stark contrast to the more massive BHs observed with LIGO/Virgo that tend to contain at least one BH with a mass greater than $10 \unit{M_{\odot}}$ \citep{Abbott+2020_GWTC2}. These observations of X-ray binaries suggest the presence of a lower mass gap (from $2$-$5 \unit{M_{\odot}}$) in which there are no strong candidates for either black holes or neutron stars \citep{Ozel+2010,Farr+2011} but the gap's existence remains an open question \citep[e.g.][]{Kreidberg+2012, Mandel+2020}. Recently there has also been increased discussion over the maximum BH mass in our galaxy, with the claims of a $70 \unit{M_{\odot}}$ BH \citep{Liu+2019} which has subsequently been challenged (\citealp{El-Badry+2020, Abdul-Masih+2020, Shenar+2020,Eldridge+2020}, see also \citealp{Liu+2020}) and revised measurements of the mass of Cygnus X-1 \citep{Miller-Jones+2021}. A sample of BHBHs detected with LISA could possibly help to constrain the BH mass distribution.

One particularly interesting class of potential LISA sources is BHNSs. With the recent detection of two BHNSs by the LIGO scientific collaboration, the existence of these DCOs has been confirmed \citep{TheLIGOScientificCollaboration+2021}. However, with only two detections (not including the low-confidence candidate GW190426, \citealt{Abbott+2020_GWTC2}, or GW190425, GW190814 and GW190917 which have not been ruled out as BHNSs, \citealt{Abbott+2020_GW190425,Abbott+2020_GW190814, GWTC_2_1}) and no electromagnetic counterparts, the formation rate and properties of BHNSs are still uncertain. Current predictions for the merger rate of BHNSs range across three orders of magnitude \citep[e.g.][]{Abadie+2010, Mandel+2021} so the number of detections in LISA will be important in reducing this uncertainty, thereby refining our understanding of the remnants and evolution of massive stars. Similar to NSNSs, these binaries are also expected to have electromagnetic counterparts. A distinctly exciting possibility is the detection of a pulsar--BH system or millisecond pulsar--BH system \citep{Narayan+1991, Pol+2021}. These systems could be observed not only by LISA, but also radio telescopes such as MeerKAT and the Square Kilometre Array (SKA, \citealt{Dewdney+2009}), which would help to improve the measurement of individual system parameters and to constrain uncertain binary evolution processes \citep[e.g.][]{Pfahl+2005,Chattopadhyay+2020}.

For the purposes of this investigation, we consider the `classical' isolated binary evolution channel \citep[e.g.][]{Tutukov+1973,Tutukov+1993,Smarr+1976,Srinivasan+1989,Kalogera+2007,Belczynski+2016} in which double compact objects are formed following common-envelope ejection or a phase of highly non-conservative mass transfer \citep{Heuvel+2011, vandenHeuvel+2017}. We do not, however, account for several alternative proposed formation channels, which could affect the rate and distribution of detectable NS and BH binaries in LISA. These channels include: dynamical formation in dense star clusters \citep[e.g.][]{Sigurdsson+1993,PortegiesZwart+2000,Miller+2009,Rodriguez+2015}, young/open star clusters \citep[e.g.][]{Ziosi+2014, DiCarlo+2020, Banerjee+2020, Rastello+2020, Rastello+2021} and (active) galactic nuclei discs \citep[e.g.][]{Morris+1993, Antonini+2016, McKernan+2020}, isolated (hierarchical) triple evolution involving Kozai-Lidov oscillations \citep[e.g.][]{Stephan+2016, Silsbee+2017,Antonini+2017, Toonen+2020},  and chemically homogenous evolution through efficient rotational mixing \citep[e.g.][]{deMink+2009,Mandel+2016,Marchant+2016,Marchant+2017,duBuisson+2020}.

In this paper, we present models for the detection rate and distribution of binary properties (masses, frequency, eccentricity, distance, merger time) of BHBHs, BHNSs and NSNSs formed through isolated binary evolution in the Milky Way. We explore the effect of varying physical assumptions in our population synthesis model on our results as well as discuss the effect of extending the LISA mission length and the prospects for distinguishing DCO detections from the WDWD background.

Earlier work on BHBHs, BHNSs and NSNSs in LISA has used a variety of population synthesis codes, Milky Way models and LISA specifications, resulting in a wide range of predictions \citep{Nelemans+2001,Belczynski+2010,Liu+2014,Lamberts+2019,Lau+2020,Breivik+2020,Sesana+2020, Shao+2021}. We build upon previous efforts but with several important improvements. We explore the effects of varying binary physics assumptions by repeating our analysis for \nModels{} different models and comparing the effects on the detection rate and distributions of source parameters \citefloorp{}. We use a model for the Milky Way that accounts for the chemical enrichment history and is calibrated on the APOGEE survey \citep{Majewski+2017,Frankel+2018}, whereas most others did not consider the effect of metallicity in detail (see however \citealp{Lamberts+2019, Sesana+2020}). We provide a detailed treatment of the eccentricity of detectable sources, both for the inspiral evolution as well as gravitational wave signal during the LISA mission. Moreover, the grid of binary population synthesis simulations that we use is the most extensive of its kind to date and makes use of the adaptive sampling algorithm STROOPWAFEL \citep{Broekgaarden+2019, Broekgaarden+2021}. Overall over 2 billion massive binaries were simulated to produce the DCO populations used in this work. We find that this large number of simulations is important to reduce the sampling noise even when using adaptive importance sampling.

All data related to the predictions made in this study are publicly available on Zenodo at \citet{Wagg+2021_zenodo}, as are the populations used in our simulations at \citet[][BHBH]{Broekgaarden:2021-zenodo-BHBH} \citet[][BHNS]{Broekgaarden:2021-zenodo-BHNS} and  \citet[][NSNS]{Broekgaarden:2021-zenodo-NSNS}. We make all code used to produce our results available in a Github repository \href{https://github.com/TomWagg/detecting-DCOs-in-LISA}{\faGithub}\footnote{\url{https://github.com/TomWagg/detecting-DCOs-in-LISA}}. In addition, the repository contains step-by-step Jupyter notebooks that explain how to reproduce and change each figure in the paper. In a companion paper, \citet{Wagg+2021}, we present \href{https://legwork.readthedocs.io}{\texttt{LEGWORK}}\footnote{\url{https://legwork.readthedocs.io}}, the \textbf{L}ISA \textbf{E}volution and \textbf{G}ravitational \textbf{W}ave \textbf{Or}bit \textbf{K}it, a python package designed for making predictions for the detection of sources with LISA, which we use in this work.

Our paper is structured as follows. In Section~\ref{sec:method}, we describe the methods for synthesising a population of binaries, the variations of physical assumptions that we consider, how we simulate the Milky Way distribution of DCOs and our methods for calculating a detection rate for LISA. We present the main results for our fiducial model in Section~\ref{sec:results}, before exploring the variations in the detectable population when changing our physical assumptions in Section~\ref{sec:variations}. In Section~\ref{sec:discussion} we discuss these results. In Section~\ref{sec:compare_studies}, we compare and contrast our methods and findings to previous work and finish with our conclusions in Section~\ref{sec:conclusion}.
\section{Method} \label{sec:method}
To produce predictions for the DCOs that are detectable with LISA, we use a synthesised population of DCOs, simulated using the methods described in Section~\ref{sec:COMPAS_explained}. In Section~\ref{sec:galaxy_synthesis} we describe our model for the Milky Way and how we place DCOs in randomly sampled Milky Way instances. We evolve the orbit of each DCO in a Milky Way instance up to the LISA mission and calculate the detection rate for that instance using the methods presented in Section~\ref{sec:gw_detection}.

\subsection{Binary population synthesis}\label{sec:COMPAS_explained}

We use the grid of \nModels{} binary population synthesis simulations recently presented in \citet{Broekgaarden+2021} and Broekgaarden et al.\ (in prep.). This grid of simulations is synthesised using the rapid population synthesis code \href{https://compas.science}{COMPAS} \citep{Stevenson+2017,Vigna-Gomez+2018,Stevenson+2019,Broekgaarden+2019}. COMPAS follows the approach of the population synthesis code BSE \citep{Hurley+2000,Hurley+2002} and uses fitting formula and rapid algorithms to efficiently predict the final fate of binary systems. The code is open source and documented in the papers listed above, the online documentation\footnote{\url{https://compas.science}} and in the methods paper \citep{COMPAS:2021methodsPaper}. We summarise the main assumptions and settings relevant for this work in Appendix~\ref{app:pop_synth}.

The result of the simulations is a sample of binaries, which, for each metallicity $Z$, have $N_{\rm binaries}$ binaries with parameters
\begin{equation}
    \mathbf{b}_{{Z, i}} = \{m_1, m_2, a_{\rm DCO}, e_{\rm DCO}, t_{\rm evolve}, t_{\rm inspiral}, w\},
\end{equation}
for $i = 1, 2, \dots, N_{\rm binaries}$, where $m_1$ and $m_2$ are the primary and secondary masses, $a_{\rm DCO}$ and $e_{\rm DCO}$ are the semi-major axis and eccentricity at the moment of double compact object (DCO) formation, $t_{\rm evolve}$ is the time between the binary's zero-age main sequence and DCO formation, $t_{\rm inspiral}$ is the time between DCO formation (that is, immediately after the second supernova in the system) and gravitational-wave merger and $w$ is the adaptive importance sampling weight assigned by STROOPWAFEL \cite[][Eq.~7]{Broekgaarden+2019}. We sample from these sets of parameters when creating synthetic galaxies.

\subsection{Galaxy synthesis}\label{sec:galaxy_synthesis}

In order to estimate a detection rate of DCOs with statistical uncertainties, we create a series of random instances of the Milky Way, each populated with a subsample drawn (with replacement) from the synthesised binaries described in Section~\ref{sec:COMPAS_explained}.

Most previous studies that predict a detection rate for LISA place binaries in the Milky Way independently of their age or evolution. We improve upon this as the first study to use an empirically-informed analytical model of the Milky Way that takes into account the galaxy's enrichment history by applying the metallicity-radius-time relation from \citet{Frankel+2018}. Those authors developed this relation in order to measure the global efficiency of radial migration in the Milky Way and calibrated it using a sample of red clump stars measured with APOGEE \citep{Majewski+2017}.

In Section~\ref{sec:mw_model}, we outline our model for the Milky Way and in Section~\ref{sec:combining_pop_gal} we explain how we combine our population of synthesised DCOs with this Milky Way model.

\subsubsection{Milky Way model}\label{sec:mw_model}

\begin{figure*}[t]
    \centering
    \includegraphics[width=\textwidth]{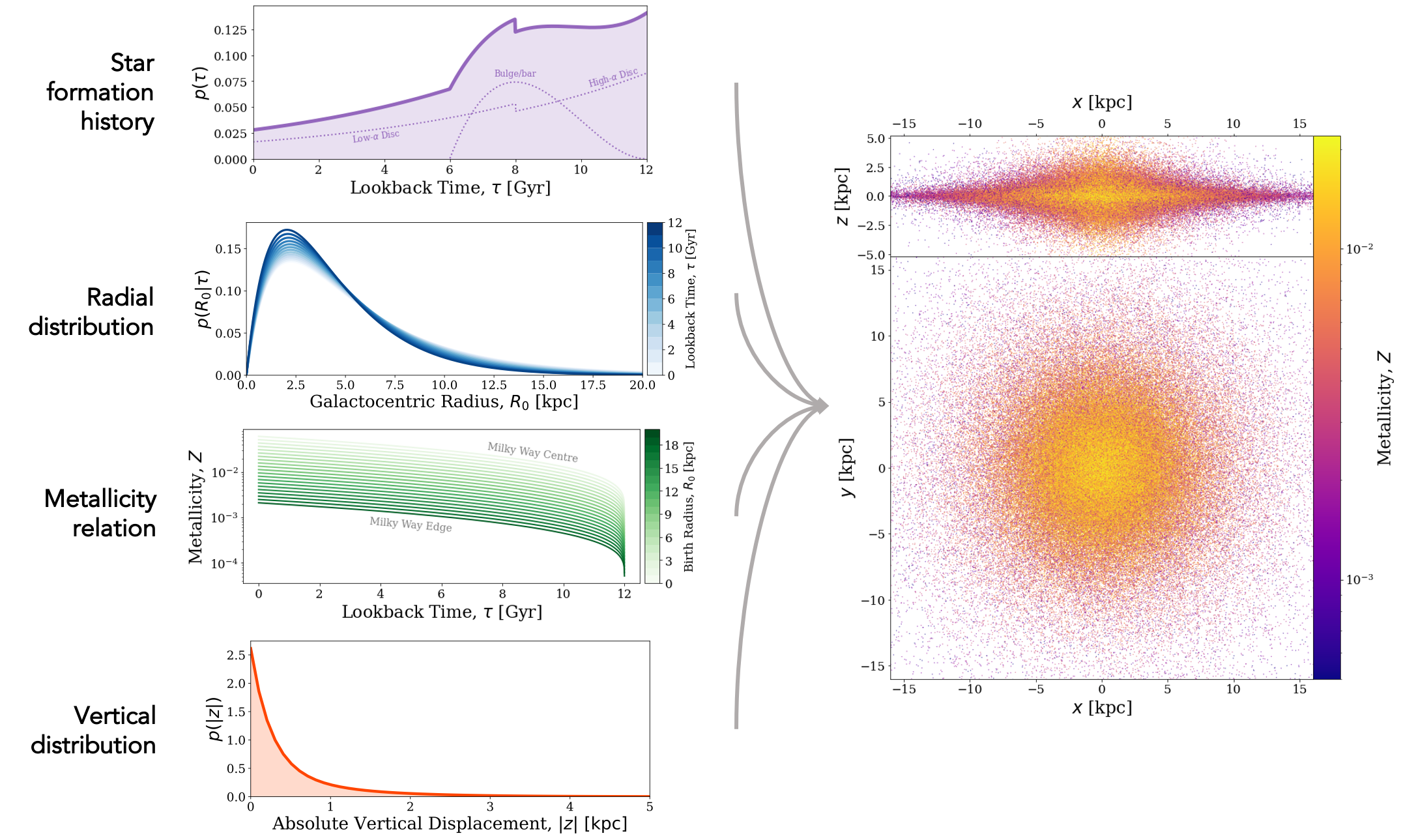}
    \caption{A schematic illustrating how we model the Milky Way. The left panel illustrates the different model aspects: star formation history of three galactic components (individually shown in the dotted lines), radial distribution, metallicity-radius-time relation, and height distribution. The right panel shows an example instance of the Milky Way with $250000$ binaries shown as points coloured by metallicity. The top panel shows a side-on view and the bottom panel a face-on view. \href{https://github.com/TomWagg/detecting-DCOs-in-LISA/blob/main/paper/figures/fig1_galaxy_diagram.png}{\faFileImage} \href{https://github.com/TomWagg/detecting-DCOs-in-LISA/blob/main/paper/figure_notebooks/galaxy_creation_station.ipynb}{\faBook}.}
    \label{fig:galaxy_schematic}
\end{figure*}

Fig.~\ref{fig:galaxy_schematic} shows the distributions and relations outlined in this section and also displays an example random galaxy drawn using this model.

Our model for the Milky Way accounts for the low-\achem\footnote{Nomenclature used to describe the enhancement of $\alpha$ elements compared to iron in stellar atmospheres}~disc, high-\achem~disc and a central component approximating a bar/bulge. The low- and high-\achem~discs are often also referred to as the thin and thick discs because the stellar vertical distribution is better fit by a double exponential rather than a single one. However, this doesn't allow one to assign a star to either the thin or thick disk purely based on its height above the Galactic plane. Therefore, we instead use the chemical definition of the two disks (applying the \achem~nomenclature) as there is a clear bimodal distribution in the chemical plane, allowing stars to be assigned to each of the disc components based on their chemical abundances. For each of the three components, we use a separate star formation history and spatial distribution, which we combine into a single model, weighting each component by its present-day stellar mass. \citet{Licquia+2015} gives that the stellar mass of the bulge is $0.9 \times 10^{10} \unit{M_{\odot}}$ and the stellar mass of the disc is $5.2 \times 10^{10} \unit{M_\odot}$, which we split equally between the low- and high-\achem~discs \citep[e.g.,][]{Snaith+2014}.

\textit{Star formation history:} 
We use an exponentially declining star formation history \citep{Frankel+2018} (inspired by the average cosmic star formation history) for the combined low- and high-\achem~discs,
\begin{equation}\label{eq:thin_disc_tau}
    p(\tau) \propto \exp \qty(-\frac{(\tau_m - \tau)}{\tau_{\rm SFR}}),
\end{equation}
where $\tau$ is the lookback time (the amount of time elapsed between the binary's zero-age main sequence and today), $\tau_m = 12 \unit{Gyr}$ is the assumed age of the Milky Way and $\tau_{\rm SFR} = 6.8 \unit{Gyr}$ is the star formation timescale \citep{Frankel+2018}. The two discs form stars in mutually exclusive time periods, such that the high-\achem~disc forms stars in the early history of the galaxy ($8$--$12 \unit{Gyr}$ ago) and the low-\achem~disc forms stars more recently ($0$--$8 \unit{Gyr}$ ago). Both distributions are normalised so that an equal amount of mass is formed in each of the two components over their respective star forming periods.

The star formation history of the Milky Way bulge (which we assume here to be dominated by the central bar) has many uncertainties due to the (1) sizeable age measurement uncertainties at large ages in observational studies, (2) complex selection processes affecting the observed age distributions, and (3) formation mechanisms that are still under debate. However, the central bar was shown to contain stars with an extended age range, with most observed stars between $6$ and $12 \unit{Gyr}$ with a younger tail of ages that could come from the subsequent secular growth of the Galactic bar \citep[e.g.,][]{Bovy+2019}. To model the bar's age distribution more realistically than in previous studies (which assume an old bulge coming from a single starburst), we choose to adopt a more extended star formation history using a $\beta(2,3)$ distribution, shifted and scaled such that stars are only formed in the range $[6, 12] \unit{Gyr}$. We show these distributions in the top left panel of Fig.~\ref{fig:galaxy_schematic}.

\textit{Radial distribution:} For each of the three components we employ the same single exponential distribution (but with different scale lengths)
\begin{equation}\label{eq:galaxy_R}
    p(R) = \exp(-\frac{R}{R_d}) \frac{R}{R_d^2},
\end{equation}
where $R$ is the Galactocentric radius and $R_d$ is the scale length of the component. For the low-\achem~disc, we set $R_d = R_{\rm exp}(\tau)$, where $R_{\rm exp}(\tau)$ is the scale length presented in \citet[][Eq.~6]{Frankel+2018}
\begin{equation}
    R_{\rm exp}(\tau) = 4 \unit{kpc} \qty(1 - \alpha_{R_{\rm exp}} \qty(\frac{\tau}{8 \unit{Gyr}})),
\end{equation}
where $\alpha_{R_{\rm exp}} = 0.3$ is the inside-out growth parameter\footnote{We find that $R_{\rm exp}(\tau) = 4$ kpc fits the data well and adopt this value rather than the 3 kpc quoted in \cite{Frankel+2018}, which was a fixed parameter (not a fit).}

This scale length accounts for the inside-out growth of the low-\achem~disc and hence is age dependent. We assume $R_d = (1 / 0.43) \unit{kpc}$ for the high-\achem~disc \citep[][Table~1]{Bovy+2016} and $R_d = 1.5 \unit{kpc}$ for the bar component \citep{Bovy+2019}. We show the combination of these distributions in the second panel on the left in Fig.~\ref{fig:galaxy_schematic}.

\textit{Vertical distribution}: Similar to the radial distribution, we use the same single exponential distribution (but with different scale heights) for each component, given by
\begin{equation}\label{eq:galaxy_z}
    p(\abs{z}) = \frac{1}{z_d} \exp\qty(-\frac{z}{z_d}),
\end{equation}
where $z$ is the vertical displacement above the Galactic plane and $z_d$ is the scale height. We set $z_d = 0.3 \unit{kpc}$ for the low-\achem~disc \citep{McMillan+2011} and $z_d = 0.95 \unit{kpc}$ for the high-\achem~disc \citep{Bovy+2016}. For the bar, we set $z_d = 0.2 \unit{kpc}$ \citep{Wegg+15}. We show the combination of these distributions in the bottom left panel of Fig.~\ref{fig:galaxy_schematic}.

\textit{Metallicity-radius-time relation:} To account for the chemical enrichment of star forming gas as the Milky Way evolves, we adopt the relation given by \citep[][Eq. 7]{Frankel+2018}
\begin{equation}\label{eq:galaxy_FeH}
    \begin{split}
        [{\rm Fe} / {\rm H}] (R, \tau) &= F_m + \nabla [{\rm Fe} / {\rm H}] R \\
        &- \qty(F_m + \nabla [{\rm Fe} / {\rm H}] R^{\rm now}_{[{\rm Fe} / {\rm H}] = 0} ) f(\tau),
    \end{split}
\end{equation}
where
\begin{equation}
    f(\tau) = \qty(1 - \frac{\tau}{\tau_m})^{\gamma_{[{\rm Fe} / {\rm H}]}},
\end{equation}
$F_m = -1 \unit{dex}$ is the metallicity of the gas at the center of the disc at $\tau = \tau_m$, $\nabla [{\rm Fe} / {\rm H}] = -0.075 \unit{kpc^{-1}}$ is the metallicity gradient, $R^{\rm now}_{[{\rm Fe} / {\rm H}] = 0} = 8.7 \unit{kpc}$ is the radius at which the present day metallicity is solar and $\gamma_{[{\rm Fe} / {\rm H}]} = 0.3$ sets the time dependence of the chemical enrichment. We convert this to the representation of metallicity that we use in this paper by applying \citep[e.g.][]{Bertelli+1994}
\begin{equation}\label{eq:galaxy_FeH_to_Z}
    \log_{10} (Z) = 0.977 [{\rm Fe} / {\rm H}] + \log_{10}(Z_\odot).
\end{equation}

Although \citet{Frankel+2018} only fit this model for the low-\achem~disc, we also use this metallicity-radius-time relation for the high-$\alpha$ disc and the bar, but focusing on the chemical tracks more representative to the inner disc and large ages. \citet{Sharma+2020} showed that using a simple continuous model for both the low- and high-\achem~discs, the Milky Way abundance distributions could be well reproduced. Empirically, the abundance tracks in the [$\alpha$/Fe]-[Fe/H] plane (and other elements) of the stars in the bulge/bar follow the same track as those of the old stars in the Solar neighbourhood \citep[][Fig.~7,]{Griffith+2021,Bovy+2019}, which motivates our modelling choice to use the same metallicity-radius-time relation.

\subsubsection{Combining population and galaxy synthesis}\label{sec:combining_pop_gal}

For each Milky Way instance, we randomly sample the following set of parameters
\begin{equation}
    \mathbf{g}_{{j}} = \{\tau, R, Z, z, \theta\}
\end{equation}
for $j = 1, 2, \dots, N_{\rm MW}$, where we set $N_{\rm MW} = 2 \times 10^{5}$, $\tau, R, Z$ and $z$ are defined and sampled using the distribution functions specified in Section~\ref{sec:mw_model}, $\theta$ is the azimuthal angle sampled uniformly on $[0, 2\pi)$ and $Z$ is the metallicity. Fig.~\ref{fig:galaxy_schematic} shows an example of a random Milky Way instance created with these distributions. This shows how these distributions translate to positions and illustrates the gradient in metallicity over radius.

We match each set of galaxy parameters $\mathbf{g}_{{j}}$, to a random set of binary parameters $\mathbf{b}_{{Z, i}}$, by drawing a binary from the closest metallicity bin to the metallicity in $\mathbf{g}_{{j}}$. If the metallicity in $\mathbf{g}_{{j}}$ is below the minimum COMPAS metallicity bin ($Z = 10^{-4}$), we use this minimum bin. If the metallicity in $\mathbf{g}_{{j}}$ is above the maximum COMPAS metallicity bin ($Z = 0.03$), we use a randomly selected bin from the five highest metallicity bins.

Each binary is likely to move from its birth orbit. Although all stars in the Galactic disc experience radial migration \citep{Sellwood+2002, Frankel+2018}, DCOs generally experience stronger dynamical evolution as a result of the effects of both Blaauw kicks \citep{Blaauw+1961} and natal kicks \citep[e.g.][]{Hobbs+2005}.

The magnitude of the systemic kicks are typically small compared to the initial circular velocity of a binary at each Galactocentric radius. Therefore, we expect that kicks will not significantly alter the overall distribution of their positions (see however, e.g., \citealt{Brandt+1995, Abbott+2017_GW170817_progenitor}). Given this, and for the sake of computational efficiency, we do not account for the displacement due to systemic kicks in our analysis.

\subsection{Gravitational wave detection}\label{sec:gw_detection}
We use the Python package \href{https://legwork.readthedocs.io/en/latest/}{LEGWORK} \citep{Wagg+2021} to evolve binaries and calculate their LISA detectability. For a full derivation of the equations given below see \citep[][Section~3]{Wagg+2021}, or the LEGWORK documentation \href{https://legwork.readthedocs.io/en/latest/notebooks/Derivations.html}{\faBook}.

\subsubsection{Inspiral evolution}

Each binary loses orbital energy to gravitational waves throughout its lifetime. This causes the binary to shrink and circularise over time. In order to assess the detectability of a binary, we need to know its eccentricity and frequency at the time of the LISA mission. For each binary in our simulated Milky Way, we know that the time from DCO formation to today is $\tau - t_{\rm evolve}$ and that the initial eccentricity and semi-major axis are $e_{\rm DCO}$ and $a_{\rm DCO}$. We find the eccentricity of the binary at the start of the LISA mission, $e_{\rm LISA}$, by numerically integrating its time derivative \citep[][Eq. 5.13]{Peters+1964} given the initial conditions. This can be converted to the semi-major axis at the start of LISA, $a_{\rm LISA} $\citep[][Eq. 5.11]{Peters+1964}, which in turn gives the orbital frequency, $f_{\rm orb, LISA}$, by Kepler's third law since we know the component masses.

\subsubsection{Binary detectability}

We define a binary as detectable if its gravitational wave signal has a signal-to-noise ratio (SNR) of greater than 7 \citep[e.g.][]{Breivik+2020, Korol+2020}. The sky-, polarisation- and orientation-averaged signal-to-noise ratio, $\rho$, of an inspiraling binary can be calculated with the following \citep[e.g.][]{Finn+2000}
\begin{equation}\label{eq:snr}
    \rho^2 = \sum_{n=1}^{\infty} \int_{f_{n, i}}^{f_{n, f}} \frac{h_{c, n}^{2}}{f_{n}^{2} S_{\rm n}\left(f_{n}\right)} \dd{f_n},
\end{equation}
where $n$ is a harmonic of the gravitational wave signal, $f_n = n \cdot f_{\rm orb}$ is the frequency of the $n^{\rm th}$ harmonic of the gravitational wave signal, $f_{\rm orb}$ is the orbital frequency, $S_{\rm n}(f_n)$ is the LISA sensitivity curve at frequency $f_n$ \citep[e.g.][]{Robson+2019} and $h_{c,n}$ is the characteristic strain of the $n^{\rm th}$ harmonic, given by \citep[e.g.][]{Barack+2004}
\begin{equation}\label{eq:charstrain}
    h^2_{c,n} = \frac{2^{5/3}}{3 \pi^{4/3}} \frac{(G \mathcal{M}_c)^{5/3}}{c^3 D_L^2} \frac{1}{f_{\rm orb}^{1/3}} \frac{g(n,e)}{n F(e)},
\end{equation}
where $D_L$ is the luminosity distance to the source, $f_{\rm orb}$ is the orbital frequency, $g(n, e)$ and $F(e)$ are given in \citet{Peters+1963} and $\mathcal{M}_c$ is the chirp mass, defined as
\begin{equation}\label{eq:chirp_mass}
    \mathcal{M}_c = \frac{(m_1 m_2)^{3/5}}{(m_1 + m_2)^{1/5}}.
\end{equation}

Note that increasing the length of the LISA mission allows more time for a DCO to evolve over the mission. Therefore the frequency limits in Eq.~\ref{eq:snr} are dictated by the LISA mission length. The SNR generally scales as $\sqrt{T_{\rm obs}}$ (with exceptions for sources very close to merging) and thus the SNR of a typical source in a 10-year LISA mission is approximately $1.58$ (=$\sqrt{10/4}$) times stronger than in a 4-year mission.

We use \href{https://legwork.readthedocs.io/en/latest/}{LEGWORK} \citep{Wagg+2021} to calculate the signal-to-noise ratio for each binary and the package ensures that enough harmonics are computed for each binary such that the error on the gravitational-wave luminosity remains below 1\%.

\subsubsection{Detection rate calculation}
For each physics variation model and DCO type, we first convert the COMPAS simulation results into a total number of DCOs in the Milky Way, $N_{\rm DCO}$. We do this by integrating the full mass and period distributions and stars and normalising to the total Milky Way mass. For more details see Appendix~\ref{app:rate_normalisation}.

We then determine the fraction of binaries that are detectable in each Milky Way instance by summing the adaptive importance sampling weights of the binaries that have an SNR greater than 7, and dividing by the total weights in the simulation. We multiply this fraction by $N_{\rm DCO}$ to find a detection rate (which we write as a total number of detections per LISA mission)
\begin{equation}
    N_{\rm detect} = \frac{\sum_{i = 0}^{N_{\rm MW}} w_i \phi(i)}{\sum_{i = 0}^{N_{\rm MW}} w_i} N_{\rm DCO},
\end{equation}
where $\phi(i) = 1$ if a binary is detectable and $0$ otherwise. We calculate the detection rate by Monte Carlo sampling 2500 Milky Way instances (each containing 200,000 DCOs) for each DCO type and every physics variation in order to obtain values for the uncertainty on the expected detection rate.
\section{Results I - Predictions for LISA sources} \label{sec:results}
In this section we present our predictions for the population of detectable LISA sources for our fiducial model. In total we expect, on average, $124$ detections in a 4-year LISA mission, of which $\BHBHFourYear{}$, $\BHNSFourYear{}$ and $\NSNSFourYear{}$ are BHBHs, BHNSs and NSNSs respectively, based on our fiducial simulations. In the remainder of this section, we discuss where the sources are expected relative to LISA's sensitivity curve (Sec.~\ref{sec:dcos_on_sc}), their properties (Sec.~\ref{sec:fiducial_distributions}), their locations in the Milky Way (Sec.~\ref{sec:mw_detectable_distribution}), their formation channels (Sec.~\ref{sec:progenitors_and_formation}) and finally we discuss the expected SNR and how accurately we expect that the parameters can be measured (Sec.~\ref{sec:measurement_uncertainties}). Note that all results shown in this section are based on our fiducial simulations. A discussion of the impact of variations in the physics assumptions is provided in Sec.~\ref{sec:variations}.

\subsection{The LISA sensitivity curve and the population of detectable sources}\label{sec:dcos_on_sc}
\begin{figure*}[p]
    \centering
    \includegraphics[width=\textwidth]{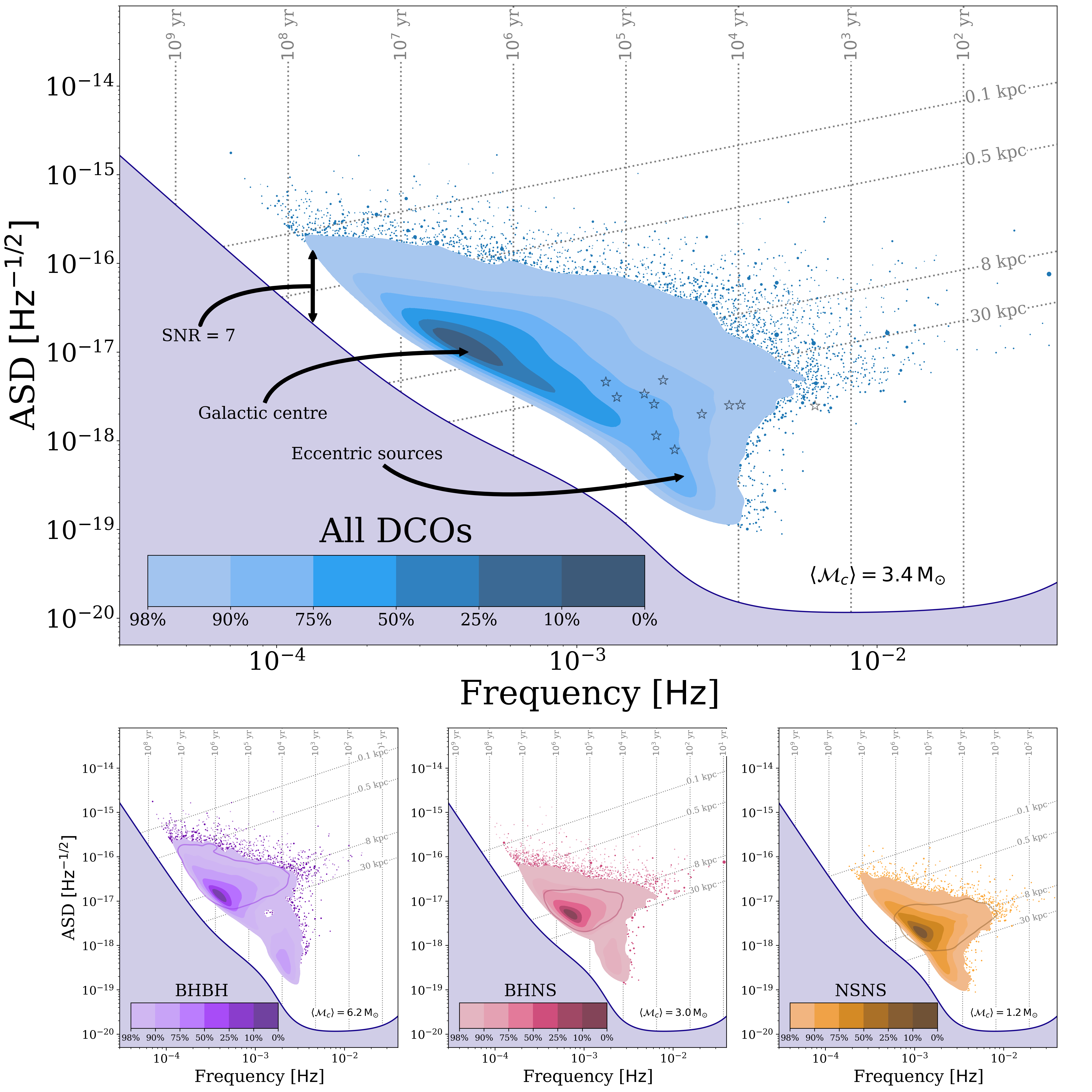}
    \caption{The LISA sensitivity curve is shown together with the density distribution of the characteristic strain and the dominant frequency for all detectable sources in our simulations (top) and separated by type (bottom). Contours show the percentage of the population enclosed. The remaining 2\% of the population is shown as dots with a size that scales with the sampling weight. To guide the interpretation, we show reference lines marking where a circular binary would reside for a given distance (diagonal line) and remaining inspiral time (vertical lines), assuming an average chirp mass $\avg{\mathcal{M}_c}$, orientation and sky location. The coloured lines in the bottom panels show a contour that encloses 90\% of the population that is circular. LISA verification binaries are overplotted in the top panel (star symbols).  See also Fig.~\ref{fig:dcos_on_sc_ecc_col} and Sec.~\ref{sec:dcos_on_sc} for a discussion. \href{https://github.com/TomWagg/detecting-DCOs-in-LISA/blob/main/paper/figures/fig2_dcos_on_sc.png}{\faFileImage} \href{https://github.com/TomWagg/detecting-DCOs-in-LISA/blob/main/paper/figure_notebooks/sensitivity_curve.ipynb}{\faBook}.}
    \label{fig:dcos_on_sc}
\end{figure*}
We show the expected LISA sensitivity curve \citep{Robson+2019} in Fig.~\ref{fig:dcos_on_sc}, which includes the confusion noise arising from the Galactic WDWD population, and overplot our predictions for the distribution of detectable sources. Eccentric systems emit gravitational waves in multiple harmonic frequencies ($n f_{\rm orb}$, with $n = 2, 3, \dots $). We choose to plot them at the $x$-coordinate that corresponds to the frequency of the harmonic that individually accumulates the largest SNR. For circular systems, the $x$-coordinate simply corresponds to $2 f_{\rm orb}$. The $y$-coordinate indicates the strength of the signal (or to be more precise, the amplitude spectral density, ASD), including the contribution from \textit{all} harmonics.

For reference, we show dotted lines to indicate where a hypothetical binary system would reside assuming a given distance from earth (diagonal lines) and a fixed remaining inspiral time (vertical lines). For each line we assume the binary is circular and has a chirp mass equal to the average of the sample ($\avg{\mathcal{M}_c}$, annotated in each panel). We also overplot the LISA verification binaries (star symbols, \citealt{Kupfer+2018}).

\vspace{1em}

We observe several features in Fig.~\ref{fig:dcos_on_sc} that are worth discussing and explaining. We note that some of these have also been described in earlier studies (see Sec.~\ref{sec:compare_studies}). Firstly, we note the empty band that separates the LISA sensitivity curve and the detectable population. This is the result of the criteria for detection where we require $\rm{SNR} > 7$.

We further note that the detectable population is concentrated on the left side of LISA's sensitivity window. The peak is located at a frequency of about $0.4 \unit{mHz}$, which is much lower than the frequency where LISA will be most sensitive (about $10 \unit{mHz}$). This can be understood from the acceleration of inspiraling DCOs as they evolve towards higher frequencies. DCOs typically form with wide orbits (low frequencies) that would not be detectable yet. Their orbits shrink as they lose angular momentum in the form of gravitational waves leading to an increase of their orbital frequency until they become detectable. These systems are increasingly rare because they evolve faster and faster towards higher frequency as the inspiral accelerates, even though the signal emitted by a more compact binary is stronger \citep{Peters+1964}. The vertical grid lines show these rapidly decreasing inspiral times at increasing frequencies. Most of the population is thus expected to reside at low frequencies.

In the lower three panels, we show the contributions of the different types of sources. Comparing them, one can observe the shift in the frequency at which the peak is located, at $0.3 \unit{mHz}$, $0.7 \unit{mHz}$ and $1 \unit{mHz}$ for BHBH, BHNS and NSNS systems respectively. This is a result of the difference in chirp mass. A higher mass system can emit at lower frequency and still produce the minimum SNR needed for detection. We note that this effect can be used to distinguish the heavier DCOs that we discuss in this work from WDWD systems, at least probabilistically (see Sec.~\ref{sec:WDWD_distinguish}). In the same way, this also explains the offset in frequency between the population of sources we predict and the LISA verification binaries.

Inspecting the dotted reference lines, we note that the peak of the density distribution of observable sources coincides with the location expected for circular systems at $8 \unit{kpc}$, which is the distance to the centre of the Milky Way. As can be seen best in the lower panels, the reference lines for $0.1$ and $30 \unit{kpc}$ enclose the majority of systems, as expected given the dimensions of the Milky Way.

There is a distinct subpopulation of binaries, most clearly visible in the lower panels as an offshoot that extends downwards to ASDs of $10^{-19} \unit{Hz}^{-1/2}$, especially around $2 \unit{mHz}$. This offshoot is almost uniquely composed of eccentric binaries, as can be seen in Fig.~\ref{fig:dcos_on_sc_ecc_col}, which shows a similar figure but colouring individual systems by their eccentricity. This can also be seen, albeit indirectly, from the contour lines shown in the bottom panels of Fig.~\ref{fig:dcos_on_sc}, which encompass 90\% of the \textit{circular} sources in each population. This contour does not include the offshoot. We conclude that eccentric sources occupy a very different region in this diagram.

\subsection{Properties of the detectable systems}\label{sec:fiducial_distributions}
\begin{figure*}[t]
    \centering
    \includegraphics[width=\textwidth]{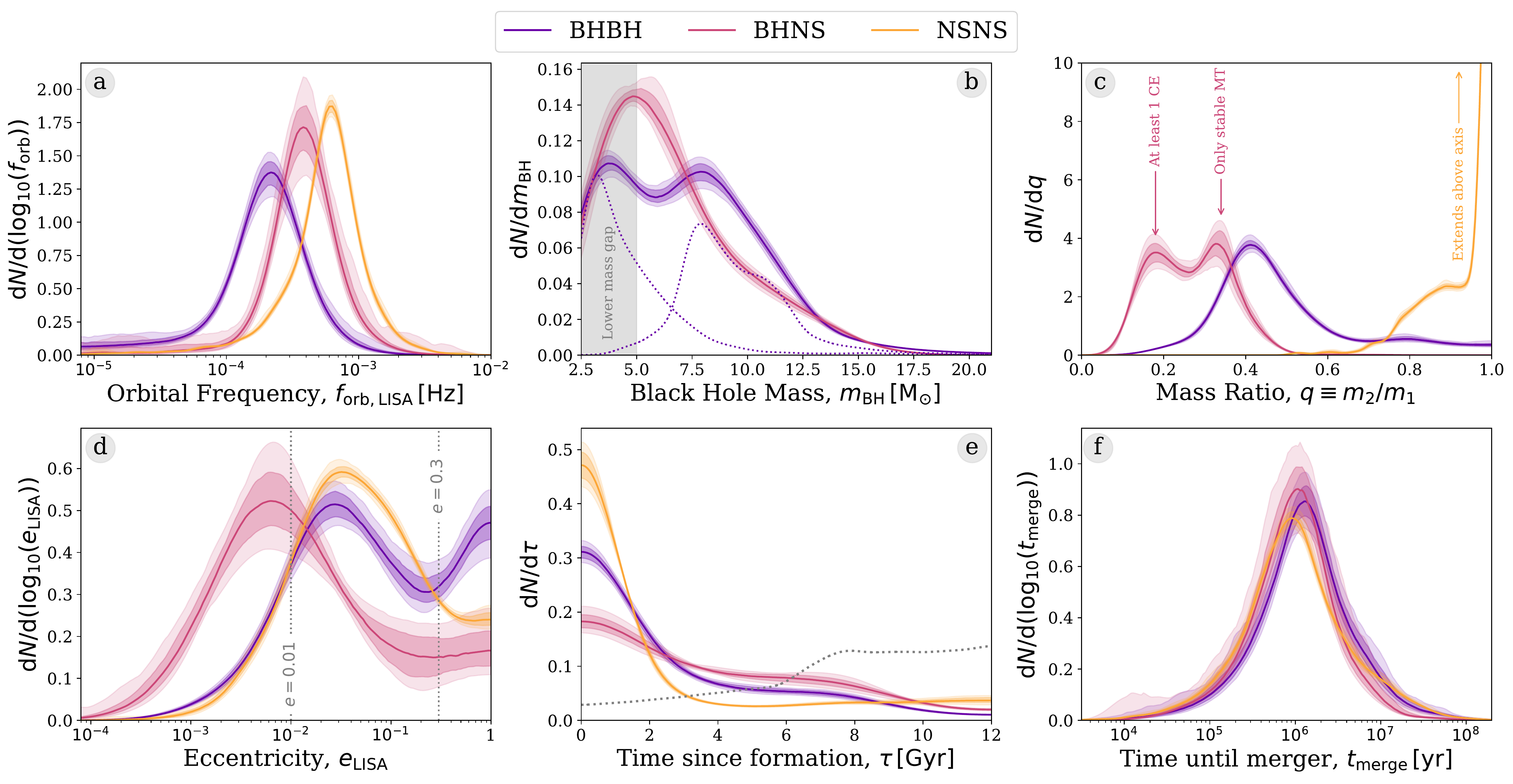}
    \caption{Properties of detectable systems for a 4-year LISA mission in our fiducial model. Each panel shows a kernel density estimator for a single property, coloured by DCO type. Each curve has been individually normalised. The shaded areas show the 1- and 2-$\sigma$ sampling uncertainties (obtained via bootstrapping). The dotted lines in panel b show the individual primary and secondary mass distributions. The dotted line in panel e shows the star formation history we have assumed in our Milky Way model. See Sec.~\ref{sec:fiducial_distributions} for a discussion. \href{https://github.com/TomWagg/detecting-DCOs-in-LISA/blob/main/paper/figures/fig3_detectable_properties_4yr.pdf}{\faFileImage} \href{https://github.com/TomWagg/detecting-DCOs-in-LISA/blob/main/paper/figure_notebooks/fiducial.ipynb}{\faBook}.}
    \label{fig:fiducial_pdf_distributions}
\end{figure*}

In Fig.~\ref{fig:fiducial_pdf_distributions}, we show the 1D distribution of several individual parameters of the population of detectable binaries together with the 1- and 2-$\sigma$ uncertainties obtained via bootstrapping. These uncertainties represent the fluctuations in our results over different random instances of the Milky Way. The distributions shown here are approximated by kernel density estimators, corrected for edge effects by mirroring the sample \citep{Schuster+1985}.

\paragraph{Orbital Frequency} The orbital frequency distributions for BHBHs, BHNSs and NSNSs (Fig.~\ref{fig:fiducial_pdf_distributions}a) peak at progressively increasing frequencies. As mentioned in Sec.~\ref{sec:dcos_on_sc}, this is because a higher mass DCO at the same distance and eccentricity requires a lower frequency to produce the same signal-to-noise ratio and thus be detected. The distributions appear nearly symmetric, but closer inspections shows that the left hand side is more populated, which can be seen most clearly in the curve for the BHBHs. This is due to the contribution of highly eccentric binaries, which are most abundant in the BHBH population. These systems are still detectable by LISA, despite their low orbital frequency, as the high eccentricity means that the majority of the GW signal is emitted at higher harmonics, where LISA is more sensitive.

\paragraph{Black Hole Mass}
In Fig.~\ref{fig:fiducial_pdf_distributions}b, we show the distribution masses of individual black holes that are part of BHNS and BHBH systems. 
The distribution for BHBH shows a bimodality. This results from the two contributions of the more and less massive BHs in BHBH systems, which peak at around $8 \unit{M_{\odot}}$ and $3.5 \unit{M_{\odot}}$ respectively as shown by the dotted lines (see also our discussion of the mass ratios below).

For both the BHBHs and BHNSs, we see that the black hole mass distribution favours low masses. About $90\%$ of BHs have masses below $11 \unit{M_{\odot}}$, in our fiducial simulations shown here. This is in stark contrast with observations from ground-based GW detectors, where heavy BHs with masses of about $30\unit{M_{\odot}}$ and higher have been common. There are two main reasons for this discrepancy. First, the population of BHs in the Milky Way (and, in particular, those detectable by LISA due to their recent formation times, see Fig.~\ref{fig:fiducial_pdf_distributions}e) primarily come from progenitors that formed from high metallicity gas according to our simulations. Stellar winds are stronger at high metallicity leading to increased mass loss. This affects the mass of the most massive black holes that can be formed \citep{Belczynski+2010}. Secondly, ground-based detectors are strongly biased towards high mass systems, since they can be detected out to larger distances and thus a greater volume is probed. In contrast, LISA has no such bias and is, in principle, sensitive to inspiraling BHBHs throughout the entire Milky Way regardless of their mass, as long as we catch them when they are emitting in the LISA band. For this reason, LISA is more likely to detect the more common lower mass BHBHs. 

We also note that our BH mass distribution extends down below $5\unit{M_{\odot}}$ to $2.5 \unit{M_{\odot}}$ which is our fiducial maximum neutron star mass. The BHs in this simulation fill the so-called ``lower mass gap'' marked as a grey band \citep{Ozel+2010,Farr+2011}, see also \citet{Shao+2021} who recently also pointed this out. This prediction is sensitive to adopted model for fallback during the SN explosion as we discuss this further in Section~\ref{sec:lower_mass_gap}.

\paragraph{Mass Ratio} The mass ratio distributions for each DCO type are very distinct from one another, as can be seen in Fig.~\ref{fig:fiducial_pdf_distributions}c. The majority of NSNSs have a mass ratio close to unity, with $\NSNSqAbovePointEight{}$ of systems having $q > 0.8$, where $q \equiv m_2 / m_1$. The reason for the concentration around equal masses is that most NSs are formed either through electron-capture supernovae (ECSN) or from low mass stars in our simulations. We assume a remnant mass for any NS formed through ECSN of $1.26 \unit{M_{\odot}}$ (see Sec.~\ref{app:fiducial_physics}). The remnant mass prescription that we use assumes a fixed fallback mass for any star with a CO core mass less than $2.5 \unit{M_\odot}$, such that many NSs end up with an identical mass of $1.278 \unit{M_\odot}$ \citep[see][Eq.~19]{Fryer+2012}. This means that many NSs are formed with equal masses and hence we see a mass ratio distribution peaked around unity.

In contrast, only $\BHBHqAbovePointEight{}$ of detectable BHBHs are formed with $q > 0.8$ and the distribution peaks around $q = 0.4$. We find that the strong stellar winds in our (typically high-metallicity) progenitors are the reason behind this. 

BHBHs with unequal masses typically come from progenitors that also had more extreme mass ratios at birth ($90$ and $30\unit{M_{\odot}}$ are typical for the progenitors of detectable BHBH systems in our simulations. The primary in such systems experiences strong mass loss by winds before filling its Roche lobe. This mostly happens during its early hydrogen-shell burning phase. The wind mainly reduces the mass of the envelope, but does not have a very strong effect on the core. By the time the primary fills its Roche lobe, it has become less massive and the mass ratio is closer to one. This favours stable mass transfer. The massive core of the primary star typically becomes the more massive BH. Accretion on the secondary star is limited and the secondary eventually provides the less massive black hole.

At the same time, stellar wind mass loss disfavours the formation of black holes with comparable masses. Such systems would have originated from progenitors that also started with comparable masses. The rather massive secondaries in these systems (especially after they accreted from the primary) experience very strong stellar wind mass loss due to LBV-like eruptions. This limits the amount by which they can expand \citep[e.g.][]{vanSon+2021}. At the same time, wind mass loss (and also stable and more conservative mas transfer in these systems) lead to widening of the orbit. Both effects, the reduced expansion and increased widening of the orbit, tend to prevent the secondaries from being able to fill their Roche lobe. This thus limits the number of systems that experience the reverse mass transfer or common-envelope phase needed shrink the orbits and make them tight enough to be detected as gravitational-wave sources.

We find that detectable BHNSs have even more unequal mass ratios. Moreover, the mass ratio distribution is bimodal, where the two peaks arise from two distinct formation scenarios. Around two thirds of detectable BHNSs experience at least one common-envelope event, whilst the last third are formed through only stable mass transfer. The first peak at $q = 0.18$ is from systems that experience at least one common-envelope phase and occurs at the expected mass ratio, which approximately follows the mean BH mass (${\sim}6.5 \unit{M_\odot}$) and NS mass (${\sim}1.2 \unit{M_\odot}$). Yet we also see a second peak at higher mass ratios around $q = 0.34$, which arises from the fraction of the population that underwent only stable mass transfer phases. The stability of mass transfer depends on the mass ratio, preferentially forming systems with more equal masses, i.e.\ at higher $q$.

\paragraph{Eccentricity} In Fig.~\ref{fig:fiducial_pdf_distributions}d we show the eccentricity distributions. We find that most systems (73\%) will have eccentricities larger than $0.01$ during the LISA mission, which should in principle be detectable according to \citet{Nishizawa+2016}. This means that we will potentially be able to use eccentricity to distinguish these sources from WDWDs, which are expected to have little to no eccentricity (see Sec.~\ref{sec:WDWD_distinguish}). We note that several previous studies assumed all systems to be circular when calculating the detection rates \citep[e.g.][]{Liu+2014, Lamberts+2018, Sesana+2020}. We discuss the impact of this assumption in Section~\ref{sec:compare_studies}.

For systems with eccentricities higher than $e\gtrsim0.3$, most gravitational wave energy is emitted in higher harmonics. Such systems are more rare, but we find them to be significant among the BHBH population, where they account for \BHBHHighlyEccentric{} of systems. Detectable BHBHs in our simulation (and, in particular, those that are eccentric) are primarily systems formed through the stable mass transfer channel (see Fig.~\ref{fig:formation_channels}). These systems are still relatively wide (compared to those formed through the CE channel) immediately prior to formation of the second BH, which makes them more easily affected by kicks. If the kick is oriented roughly in opposite direction to the orbital motion and has a velocity that is of similar magnitude as the orbital velocity, it will lead to the formation of a highly eccentric system. 

It is rare to get such a ``lucky kick'', but there are a few effects that favour this for BHBHs. The kicks of BHs are reduced by fallback and they are thus less likely to disrupt the system. Moreover, because of their higher masses, BHBHs can be observed already at lower orbital frequencies. This means that they have not had as much time to circularise and so still have significant eccentricity by the time of the LISA mission. Finally, LISA favours the detection of eccentric systems, if all other properties are held fixed. This is because the gravitational-wave emission is stronger (Eq.~\ref{eq:peters_f}) and the energy is emitted at higher frequencies \citep[Eq.~20]{Peters+1963} where LISA is more sensitive.

The lower abundance of highly eccentric systems among the NSNS and BHNS systems may seem counter-intuitive since neutron stars are lower mass and would be more strongly affected by natal kicks, which one may expect to lead to more eccentric systems. However, the majority of NSs in our simulations are formed through ECSN and USSN and for these types of supernovae we draw from a Maxwellian with $\sigma_{\rm rms}^{\rm 1D} = 30 \unit{km}{s^{-1}}$. Thus the kicks received by NSs in our simulations are often much smaller than for BHs.

\paragraph{Time since formation} 
In Fig.~\ref{fig:fiducial_pdf_distributions}e we show how long ago the LISA detectable DCOs formed. Star formation was highest at early times $6-12\unit{Gyr}$ ago, after which it declined. In contrast, the LISA detectable DCOs primarily formed in the relatively \textit{recent} history of our Milky Way, about $2 \unit{Gyr}$ ago. This reflects the fact that binaries in our simulation typically take about a Gyr to merge.

When comparing the distribution of formation times for the three different DCO types we see that NSNSs are most strongly concentrated at recent times, followed by BHBHs and then BHNSs. To understand this it is helpful to consider how the inspiral time scales with various parameters \citep{Peters+1964}
\begin{equation}
    t_{\rm inspiral} \propto \frac{a^4}{(m_1 + m_2)^{3}} \cdot \frac{q}{(1+q)^2} \cdot (1 - e^2)^{-7/2}.
\end{equation}
The inspiral time depends most strongly on the separation at DCO formation, $a$, and this is where the three types also differ most strongly (see Fig.~\ref{fig:dco_formation_properties}). The detectable NSNS systems have the tightest orbits at DCO formation. The median of the distribution of separations at DCO formation, $\langle a_{\rm DCO} \rangle_{\rm med}$, relate as 8:3:1 for detectable BHBH:BHNS:NSNS in our simulations. This results in increase of the inspiral time by a factor of about 4000:80:1. The total masses affect the inspiral time to the third power and this where the heavier BHBH systems are favoured. The median total masses differ by ratios of 6:4:1 for detectable BHBH:BHNS:NSNS in our simulations, impacting the inspiral times such that they are a factor of 200:60:1 shorter, partially counteracting the effect of the separations. The term depending on the mass ratio $q$ only varies by about 30\% for the mass ratio ranges considered here and so is not of interest. The eccentricity term is not of importance for mildly eccentric systems, $f(e_{\rm DCO} \le 0.3)  \le 1.4$ but of large importance for the very eccentric  $f(e_{\rm DCO} \ge 0.9) \ge 300$. The fraction of highly eccentric systems with $e_{\rm DCO}>0.9$ is 33\%, 16\%, and 8\% of for BHBH, BHNS and NSNS respectively, see also Fig.~\ref{fig:dco_formation_properties}. 

We conclude that the shorter median separations at DCO formation are the main reason why NSNS are most strongly peaked at short delay times. They are followed by BHBHs rather than BHNSs due to the high masses and substantial eccentricities of BHBHs.

\paragraph{Time until merger} Fig.~\ref{fig:fiducial_pdf_distributions}f shows the remaining time until merger for each of the DCO types at the start of LISA mission. The distributions are strikingly similar and peak with merger times of around a Myr.

The merger time is a function of the mass, frequency and eccentricity of the sources, such that more massive, higher frequency and more eccentric sources merge faster \citep[][Eq.~5.14]{Peters+1964}. So, despite the fact that each DCO type often has higher values in any one of these properties, the convolution of all three tends to negate the differences. For example, NSNSs have the highest orbital frequencies and are mildly eccentric whilst BHNSs have moderate orbital frequencies and are more circular. However, BHNSs are more massive in general and so the overall merger times are distributed very similarly for both DCO types.

\subsection{Distribution in the Milky Way}\label{sec:mw_detectable_distribution}
For our Milky way model we consider three different components, a low-\achem~(``thin'') disc, an older high-\achem~(``thick'') disc and a bulge/bar (see Sec.~\ref{sec:galaxy_synthesis}). In Table~\ref{tab:component_fractions} we summarise the number of detections originating from each of these components. Despite the fact that only $42.5\%$ of systems are formed in the low-\achem~disc, we find that $86\%$ of the detections originate from this component. This is because most detectable systems were formed relatively recently (see Fig.~\ref{fig:fiducial_pdf_distributions}e) and so the high-\achem~disc and bulge are effectively too old to contribute many detectable systems. Nevertheless, we do find a significant fraction of detectable systems originate in the high-\achem~disc and bulge, indicating that it is still important to include these components, as was ignored in some earlier works (see Sec.~\ref{sec:compare_studies}).

\begin{table}[tb]
    \centering
    \begin{tabular}{l|c|cccc}
        \hline
        \multirow{2}{*}{Component} & Formation & \multicolumn{4}{c}{Detectable} \\ \cline{2-6}
        & {\scriptsize All} & {\scriptsize All} & {\scriptsize BHBH} & {\scriptsize BHNS} & {\scriptsize NSNS} \\
        \hline
        Low-$\alpha$ disc & 42.5\% & 86\% & 89\% & 82\% & 85\%\\
        High-$\alpha$ disc & 42.5\% & 6\% & 5\% & 8\% & 12\%\\
        Bulge/bar & 15.0\% & 8\% & 6\% & 10\% & 3\%\\
    \end{tabular}
    \caption{Percentage of systems in each Galactic component.}
    \label{tab:component_fractions}
\end{table}
\begin{figure}[tb]
    \centering
    \includegraphics[width=\columnwidth]{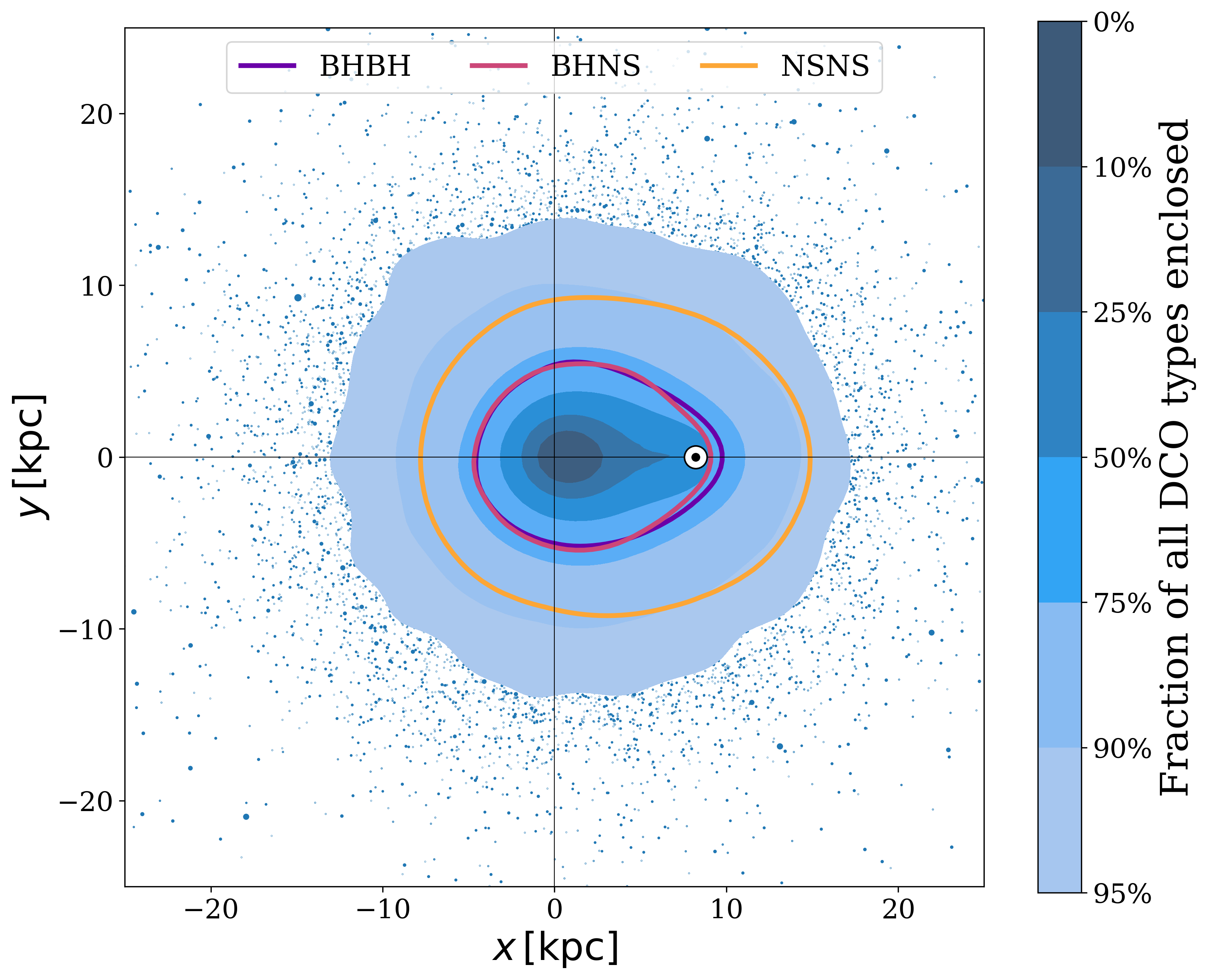}
    \includegraphics[width=\columnwidth]{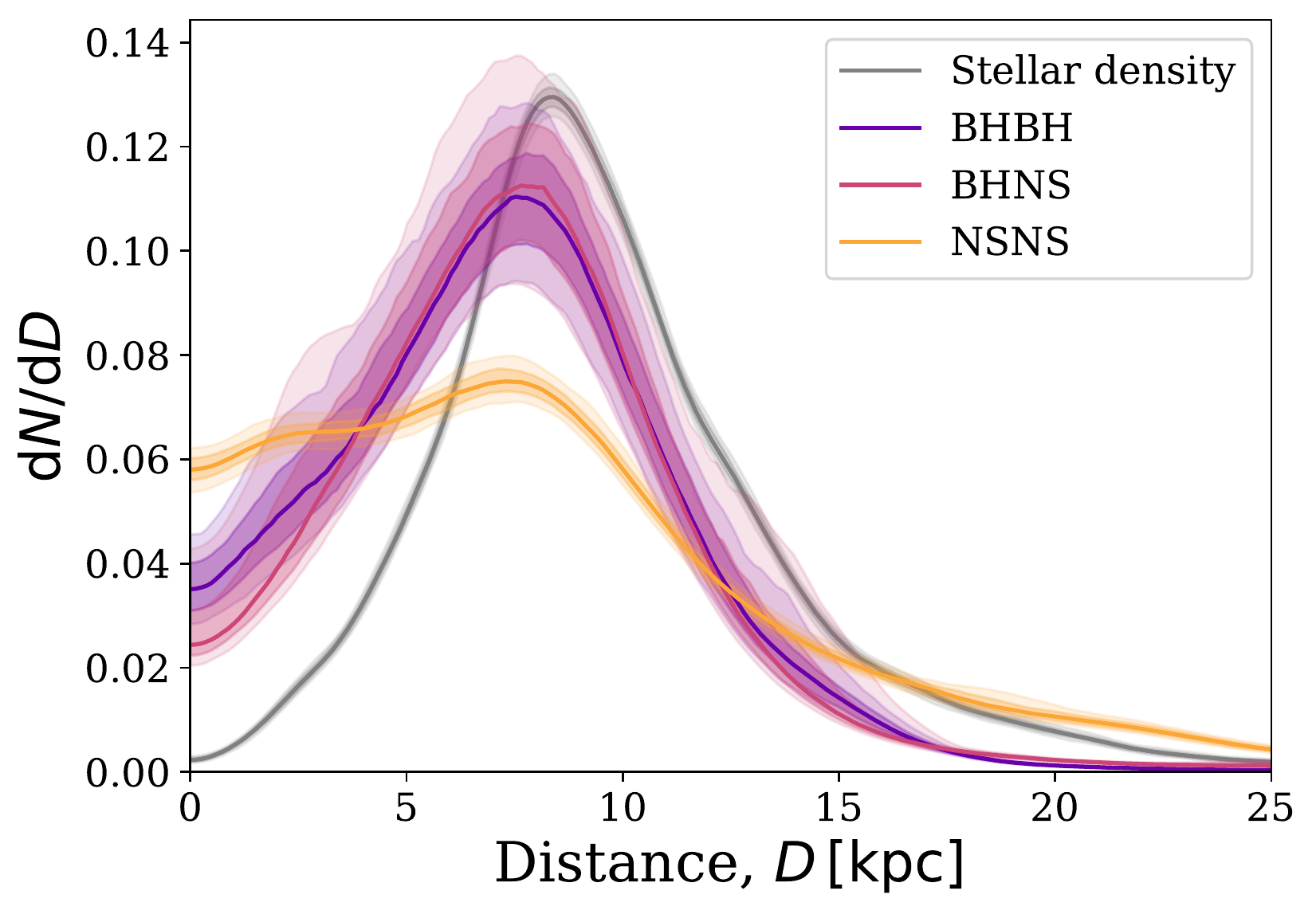}
    \caption{
    \textbf{Top:} A face-on view of the Galactic density distribution for detectable DCOs. We show the density distribution for the top 95\% of the sources, the rest are indicated by scatter points whose sizes correspond to their sampling weights. The coloured lines show the 75\% contour for each of the individual DCO types. The large cross passes through (0, 0) and helps to highlight the bias towards the position of the sun, which is indicated by the $\odot$.
    \textbf{Bottom:} As Fig.~\ref{fig:fiducial_pdf_distributions}, but for the luminosity distance. \href{https://github.com/TomWagg/detecting-DCOs-in-LISA/blob/main/paper/figures/fig4_1_galaxy_density_distribution.png}{\faFileImage} \href{https://github.com/TomWagg/detecting-DCOs-in-LISA/blob/main/paper/figure_notebooks/galaxy_creation_station.ipynb}{\faBook} \href{https://github.com/TomWagg/detecting-DCOs-in-LISA/blob/main/paper/figures/fig4_2_detectable_distance_distribution_4yr.pdf}{\faFileImage} \href{https://github.com/TomWagg/detecting-DCOs-in-LISA/blob/main/paper/figure_notebooks/fiducial.ipynb}{\faBook}.}
    \label{fig:detectable_distance_dist}
\end{figure}

In the top panel of Fig.~\ref{fig:detectable_distance_dist}, we show the density distribution for detectable DCOs in the galaxy. We see that most detectable sources are concentrated towards the Galactic centre, with a strong bias towards sources that are on our side of the Milky Way in the vicinity of the solar system (indicated with the $\odot$ symbol). In principle, systems are detectable out to large distances of about $20\unit{kpc}$ and more, although they become increasingly rare, as can be seen from the 95\% contour. 

The differences between different DCO types can be seen more clearly in the bottom panel of Fig.~\ref{fig:detectable_distance_dist} where we show the distribution of the distances, $D$, from earth to the detectable systems. Each distribution peaks around $8\unit{kpc}$, which is the distance to the centre of the Milky Way. The distribution for BHBH and BHNS systems follow a very similar shape, favouring the detection of systems with distance $<8\unit{kpc}$, but with a tail extending out to about $20\unit{kpc}$.

The distribution for NSNS stands out by being flatter, making them more common nearby and, surprisingly, also at larger distances compared to the stellar density. This may seem counter intuitive as one might naively expect the less massive NSNS systems would not be observable out to larger distances than the more massive BHNS and NSNS systems. To understand the differences we need to consider not only the mass distribution of binaries, but also their eccentricity and frequency distributions. Since, each parameter contributes to the calculation of the SNR (and thus affects the maximum distance at which systems can be detected).

The reason that the NSNSs are favoured at higher distances is that the NSNS population has the highest fraction of ``mildly'' eccentric systems ($0.01 < e < 0.03$). In contrast, the BHNS population has a much higher fraction of effectively circular systems ($e < 0.01$), which emit weaker gravitational waves compared to equivalent eccentric systems. Therefore, despite their typically higher masses, the distance at which a BHNS source is detectable is generally lower than for the mildly eccentric NSNS.

Conversely, the BHBH population has the highest fraction of highly eccentric systems ($e > 0.3$). Although one may naively expect that this would result in stronger signals (and so further distances), for a system to have these high eccentricities in LISA, it must still be early in its evolution (otherwise it would have circularised) and thus have a low orbital frequency. The result of this is that highest eccentricity systems tend to have lower SNRs and so cannot be detected at large distances. 

Overall we see that the eccentricity distribution of NSNSs occupies a ``sweet spot'' where the gravitational wave power is increased compared to circular systems, but it isn't too high that the frequency is significantly impacted. This means that NSNSs can be seen out to the largest distances of the three DCO types.

\subsection{Formation channels}\label{sec:progenitors_and_formation}
In our fiducial model, approximately two thirds of detectable BHBHs are formed through the `only stable mass transfer' channel, whilst the remaining third are primarily formed through the `classic' CE channel. Detectable BHNSs follow an inverse pattern, such that around two thirds are formed through the classic channel and the rest are mainly formed through only stable mass transfer.

In contrast, detectable NSNSs are very rarely formed through only stable mass transfer. Approximately half of systems are formed through the `classic' channel and the rest are formed through a double-core common-envelope event \citep{Brown+1995} where both progenitors evolve on a similar timescale and initiate a double-core common-envelope event whilst they are on the giant branch. All detectable DCOs show a small fraction of systems are formed through a channel that does not fit into the other categories and hence are labelled `other'. These systems tend to be formed through `lucky' supernova kicks that happen to shrink the binary significantly by chance.

The fraction of detectable DCOs that are formed through different formation channels in the other model variations are shown Fig.~\ref{fig:formation_channels}, where the first column in the plot corresponds to the fiducial model that we described above.

\subsection{How accurately will we be able to infer the parameters of detected systems?}\label{sec:measurement_uncertainties}
So far we have discussed the properties of the all detectable sources. However, only for a subset of systems we expect to get high enough SNR to obtain accurate and useful measurements of these parameters. Below we discuss the expected SNR distribution and the typical uncertainties expected for the most relevant parameters, namely, the chirp mass and sky localisation.

\paragraph{Signal-to-noise ratio}

\begin{figure}[tb]
    \centering
    \includegraphics[width=\columnwidth]{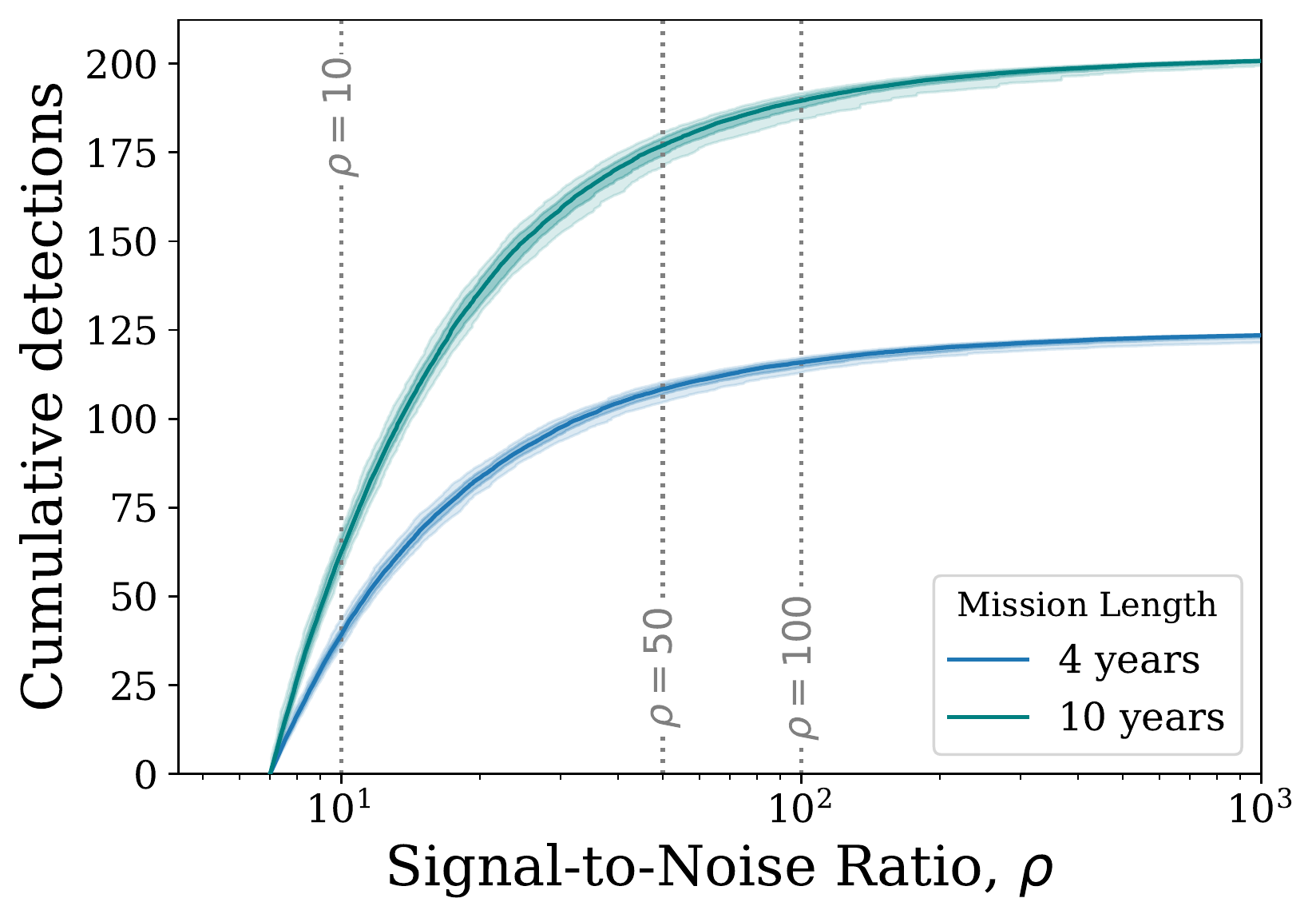}
    \caption{Cumulative number of LISA detections with a given signal-to-noise ratio. Colours indicate LISA mission length and shading shows 1- and 2-$\sigma$ uncertainties (obtained via bootstrapping). \href{https://github.com/TomWagg/detecting-DCOs-in-LISA/blob/main/paper/figures/fig5_snr_cumulative_all.pdf}{\faFileImage} \href{https://github.com/TomWagg/detecting-DCOs-in-LISA/blob/main/paper/figure_notebooks/fiducial.ipynb}{\faBook}.}
    \label{fig:snr_dist}
\end{figure}

In Fig.~\ref{fig:snr_dist} we show the cumulative number of detections with a given SNR. Although many a large fraction of sources have SNRs around our assumed detection threshold of 7, many systems are detected with very high SNRs. We find that on average for a 4(10)-year LISA mission, of the 124(202) detections, 85(138), 16(27) and 9(14) systems have SNRs greater than 10, 50 and 100, respectively. These high SNR systems are typically, but not only, the more massive BHBH systems.

\paragraph{Chirp mass}

The chirp mass is important for identifying the type of the source of a detected GW signal. The uncertainty of the chirp mass depends on the uncertainty in the measured orbital frequency, the time derivative of the orbital frequency and the eccentricity as detailed in Appendix~\ref{app:chirp_mass_uncertainty}.

\begin{figure}[tb]
    \centering
    \includegraphics[width=\columnwidth]{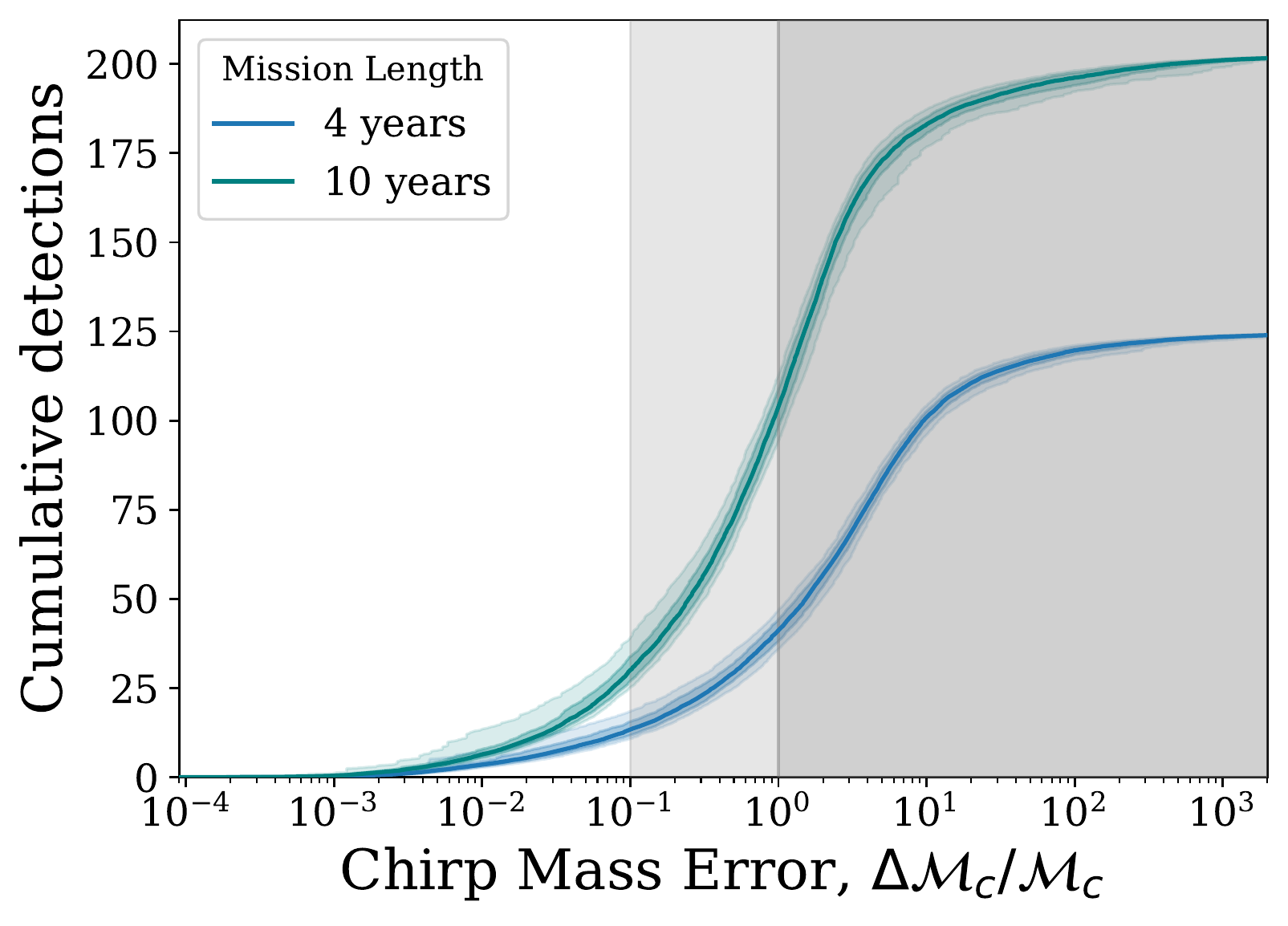}
    \caption{As Fig.~\ref{fig:snr_dist}, but for the chirp mass uncertainty. The shaded areas indicate regions with more than 10\% and 100\% uncertainty. \href{https://github.com/TomWagg/detecting-DCOs-in-LISA/blob/main/paper/figures/fig6_chirp_mass_uncertainty_all.pdf}{\faFileImage} \href{https://github.com/TomWagg/detecting-DCOs-in-LISA/blob/main/paper/figure_notebooks/fiducial.ipynb}{\faBook}.}
    \label{fig:m_c_unc}
\end{figure}

We find that for a 4(10)-year LISA mission, approximately 41(105) detections have measurable chirp masses ($\Delta \mathcal{M}_c / \mathcal{M}_c < 1$, indicated by the dark shaded region in Fig.~\ref{fig:m_c_unc}) whilst 13(31) have chirp mass uncertainties below 10\% ($\Delta \mathcal{M}_c / \mathcal{M}_c < 0.1$, indicated by the light shaded region in Fig.~\ref{fig:m_c_unc}). This uncertainty is generally dominated by the uncertainty on the time derivative of the frequency, since most of the binaries are too early in their inspiral for LISA to measure a strong chirp.
Note from Fig.~\ref{fig:m_c_unc} that increasing the mission length significantly increases the number of detections for which the chirp mass uncertainty is below 100\%. The total number of detections only scales as $\sqrt{T_{\rm obs}}$, yet we find that the the total number of detections with a chirp mass uncertainty is below 100\% and 10\% both scale approximately as $T_{\rm obs}$.

\paragraph{Sky localisation}

An accurate sky localisation will be essential to possibly identify electromagnetic counterparts or distinguish sources that come from different components of our Milky Way. 

We quantify the sky localisation of a source by estimating the angular resolution for the detectable sources. Since all potential sources are effectively stationary on the timescale of the LISA mission, we can follow \citet{Mandel+2018} and use the timing accuracy of LISA and the effective detector baseline to calculate the angular resolution, $\sigma_{\theta}$, as
\begin{equation}
    \sigma_{\theta} = 16.6^\circ \left(\frac{7}{\rho}\right) \left(\frac{5 \times 10^{-4} \unit{Hz}}{f_{\rm dom}}\right) \left( \frac{2 \unit{AU}}{L} \right),
\end{equation}
where $L$ is the effective detector baseline, which for LISA is 2 AU in the frame of the solar system, since it will orbit the Sun.

\begin{figure}[htb]
    \centering
    \includegraphics[width=\columnwidth]{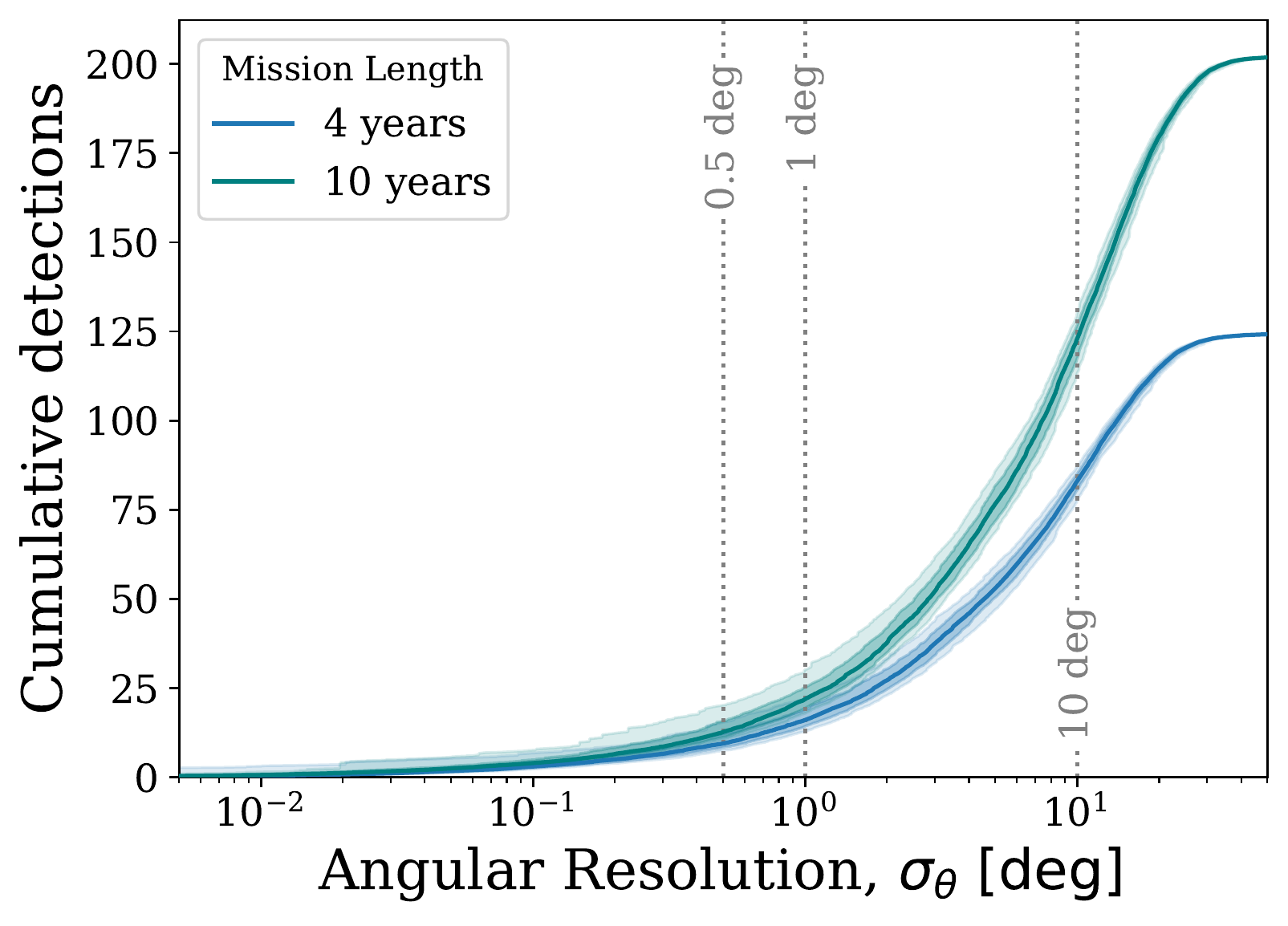}
    \caption{As Fig.~\ref{fig:snr_dist}, but for the angular resolution. \href{https://github.com/TomWagg/detecting-DCOs-in-LISA/blob/main/paper/figures/fig7_angular_resolution_all.pdf}{\faFileImage} \href{https://github.com/TomWagg/detecting-DCOs-in-LISA/blob/main/paper/figure_notebooks/fiducial.ipynb}{\faBook}.}
    \label{fig:ang_res}
\end{figure}

We plot the distribution of expected angular resolutions in Fig.~\ref{fig:ang_res}. We see that, for a 4(10)-year LISA mission, approximately 82(123) sources can be resolved to an angular resolution better than 10 degrees and only 16(23) better than 1 degree. 

For comparison, the size of a pencil beam for a $15 \unit{m}$ diameter SKA dish observing at 1.4 GHz is roughly 0.67 square degrees \citep{Smits+2009}, corresponding to an angular resolution of $\sigma_\theta = \sqrt{(0.67 / \pi)} = 0.46^\circ$ (similar to the angular size of the moon). We will further discuss the prospects of matching LISA detections to radio pulsars with SKA in Sec.~\ref{sec:pulsar_matching}.

\section{Results II - Impact of physics assumptions} \label{sec:variations}
In this section we explore the effect of varying the uncertain assumptions governing the evolution of binary system and the formation of compact objects. For this we use the population synthesis simulations and model variations presented in \citet{Broekgaarden+2021} and Broekgaarden et al.\ 2021b (in prep.).

We first discuss the robustness of our predictions for the number of detectable systems (Section~\ref{sec:detection_rate_analysis}). We then discuss examples of how the observable properties of the detectable systems, such as the distribution of masses and eccentricities, are affected and can potentially be used to probe the physics of double compact object formation (Section~\ref{sec:property_variations}).

\subsection{Detection rates}\label{sec:detection_rate_analysis}
We predict approximately \rangeFourYear{} detections in a 4-year LISA mission, across all our simulations for varying physics assumptions. This increases to about \rangeTenYear{} for a 10-year LISA mission. Although the number of detections per type can vary by about 2 orders of magnitude, we find that the total detection rate is fairly robust, among the variations we have considered (see Table~\ref{tab:detection_rates}).

\begin{figure*}[p]
    \centering
    \includegraphics[width=\textwidth]{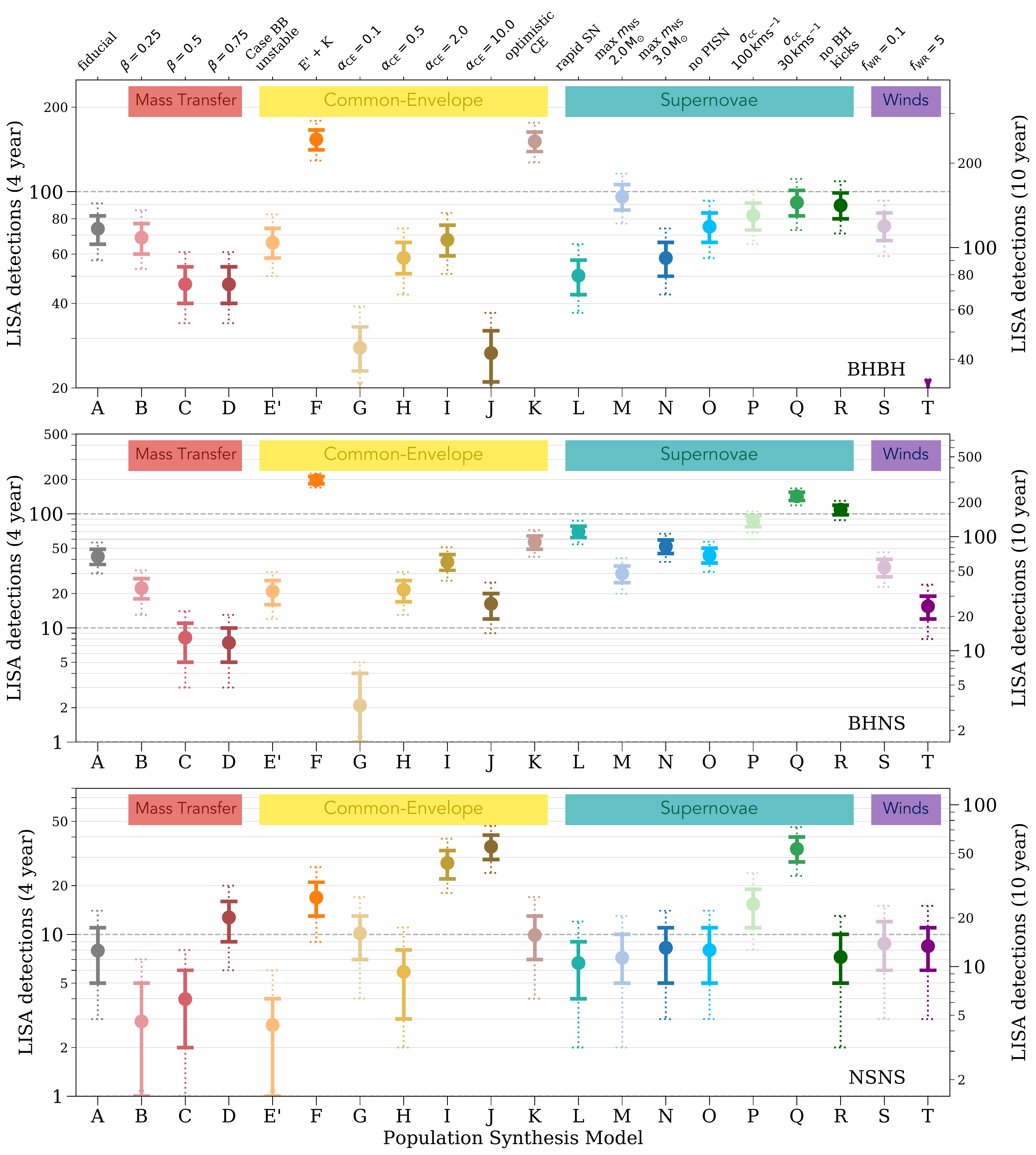}
    \caption{The number of expected detections in the LISA mission for different DCO types and model variations. Error bars show the 1- (solid) and 2-$\sigma$ (dotted) Poisson uncertainties. An arrow indicates that the error bar extends to zero. The left axis and grid lines show the number of detections in a 4-year LISA mission and the right axis shows an approximation of the number of detections in a 10-year mission (we scale the axis by $\sqrt{T_{\rm obs}}$, see Table~\ref{tab:detection_rates} for exact rates). Each model is described in further detail in Table~\ref{tab:physics_variations} and details of the fiducial assumptions are in Section~\ref{app:fiducial_physics}. See Sec.~\ref{sec:detection_rate_analysis} for a discussion. \href{https://github.com/TomWagg/detecting-DCOs-in-LISA/blob/main/paper/figures/fig8_dco_detections.pdf}{\faFileImage} \href{https://github.com/TomWagg/detecting-DCOs-in-LISA/blob/main/paper/figure_notebooks/detections.ipynb}{\faBook}.}
    \label{fig:detection_rates}
\end{figure*}

In Fig.~\ref{fig:detection_rates}, we show the expected number of LISA detections based on our simulations considering variations in the physical assumptions. We show the expected number of detections for BHBH, BHNS and NSNS systems in the top, middle and bottom panel respectively. All the rates and their uncertainties plotted in this figure are also provided in Table~\ref{tab:detection_rates}. 
In the sections that follow, we briefly explain the variations considered and discuss the most prominent trends. 

\subsubsection{Efficiency of mass transfer}

The efficiency of mass transfer, i.e.\ the fraction of mass lost by the donor through Roche-lobe overflow that is accreted by the companion, is poorly constrained and is considered as one of the main uncertainties in binary evolution \citep[e.g.][]{deMink+2007}. In our fiducial model A, we use a prescription in which the accretion rate onto stellar companions is regulated by their thermal timescale, i.e.\ the timescale on which a star can react to changes and restore thermal equilibrium \citep[see e.g.][]{Schneider+2015}. 

In models \modBetaLow{}-\modBetaHigh{}, we instead adopt a fixed value for the mass transfer efficiency, $\beta$, from $\beta=0.25$ up to 0.75, in cases of stable mass transfer onto a stellar companion. For accretion onto NS and BH we still assume that their accretion is limited to the Eddington rate.

Nearly all systems that can be detected form through channels where the very first interaction is stable mass transfer. Generally, higher mass transfer efficiencies lead to higher masses for the accreting stars, but also leads to wider orbits \citep{Soberman+1997, vanSon+2020}. Changing $\beta$ thus already affects the masses and orbital separation after the first interaction phase, which in turn changes the starting conditions and outcome of all subsequent interactions phases. This makes it complicated to fully understand the impact of varying $\beta$ in simple terms, but we can distinguish two main patterns for higher and lower mass systems respectively.   

For the most massive progenitors, increasing $\beta$ leads to secondary stars that are so massive and luminous that they experience strong wind mass loss. This leads to further widening of the orbit. In addition the more massive secondaries may not be able to fully expand as mass loss may prematurely remove their hydrogen envelope. Both of these effects tend to prevent the most massive secondaries from filling their Roche lobe. This means that they cannot initiate the reverse interaction needed to shrink the binary system and eventually produce a detectable double compact object. We indeed see that increasing $\beta$ leads to a decrease of the expected number of detections for BHBHs and BHNSs, which originate from the most massive progenitors.  

For lower mass progenitors, which primarily produce NSNS systems, we find the opposite: increasing $\beta$ leads to an \textit{increase} in the number of detectable NSNS systems. This is in part because  the changes in secondary mass and the orbital widening, affect the number of systems for which the reverse interaction successfully ejects the envelope and shrinks the orbit. Furthermore, the increased mass of the secondary stars allows stars that would have otherwise ended their life as a WD to instead become massive enough to form a NS \citep[e.g.][]{Zapartas+2017}. The same effect also allows some NS progenitors to become massive enough to become BH progenitors, which partially cancels the extra progenitors that would have originally been destined to become WDs.

\subsubsection{Common-envelope evolution}\label{sec:detection_rate_CE_trends}

The common-envelope phase constitutes a highly uncertain phase in the evolution of interacting binary systems \citep[e.g.][]{Ivanova+2013}. The uncertainties concerns the conditions required for the onset of a common-envelope phase and, if a common-envelope phase occurs, what the outcome is. Rapid population synthesis simulations such as ours approximate both questions in a crude way. We therefore consider several model variations. 

\paragraph{The efficiency parameter $\alpha_{\rm CE}$} To estimate the outcome of a CE phase, we use a simple consideration of the binding energy and orbital energy \citep{Webbink+1984, deKool+1990}. Our fiducial model assumes a common-envelope efficiency parameter $\alpha_{\rm CE} = 1$ which can be interpreted as the case where all the energy liberated by shrinking the orbit is used in an optimal way to unbind the envelope. There have been many attempts to constrain this parameter using observations and more recently also using 3D simulations \citep[e.g.][]{DeMarco+2011, Law-Smith+2020, Lau+2021}, but no consistent picture has emerged. Therefore we consider large variations in this parameter. 

In models \modAlphaLowest{}-\modAlphaHighest{} we alter the common-envelope efficiency parameter $\alpha_{\rm CE}$ to $0.1, 0.5, 2.0$ and $10.0$ respectively. Values smaller than 1 may represent cases where not all energy is used efficiently to unbind the envelope, for example when part of the energy escapes in the form of radiation or if part is used to impart additional kinetic energy in the ejecta \citep[e.g.][]{Ivanova+2013, Nandez+2016}. Values larger than 1 may represent cases where additional energy sources can be tapped into, such as for example jets powered by accretion \citep[e.g.][]{Schreier+2021}. The variations can also be seen as a way to cover uncertainties in estimates for the binding energy itself.

Increasing $\alpha_{\rm CE}$ makes common-envelope ejection more efficient or, in other words, less orbital shrinkage is needed to successfully eject the envelope. This has two consequences. (1) A larger fraction of systems avoids merging during the CE phase. This increases the overall number of DCOs. (2) The systems that survive the CE phase are wider, possibly too wide to become detectable as gravitational wave sources \citep[e.g.][]{Klencki+2021}. So, while the first consequence favours the formation of DCOs, the second consequence disfavours the formation of DCOs that are tight enough to be detected.

These two opposing effect result in a fine tuning situation. Only a very small subset of progenitor systems have the right orbital parameters prior to the CE phase to successfully produce detectable systems. Changing $\alpha_{\rm CE}$ moves and changes this window in the parameter space that successfully leads to the formation of detectable systems. How the number of the detectable systems changes depends on whether the relevant parameter space grows or shrinks and on how well the relevant part of the parameters space is populated. Fully unravelling these effects and how they interplay with the assumed star formation history is beyond our scope (and possibly not even of large relevance given the severe simplifications). We will limit the further discussions to simply stating the trends we observe.

We find that the BHBH rate peaks for $\alpha_{\rm CE} = 1$ (model \modFid{}) and reduces whether we increase \textit{or} decrease $\alpha_{\rm CE}$. The BHNS rate follows the same pattern as BHBHs although the value of $\alpha_{\rm CE}$ which maximises BHNSs seems to be between 1 and 2. In contrast, for NSNSs we find that increasing $\alpha_{\rm CE}$ (models \modAlphaHigh{} and \modAlphaHighest{}) results in significantly higher rates. 

\paragraph{The ``optimistic'' CE treatment} We further explore a model variation introduced by \citet{Belczynski+2007} often referred to as the ``optimistic'' CE scenario. This variation (model \modOpt{}) relaxes our restriction that donor stars that are on the Hertzsprung cannot survive common-envelope events.

In agreement with other studies, we find that this treatment leads to a significant increase in the formation rate of BHBHs, by a factor of two. This is because the progenitors expand significantly during the Hertzsprung gap phase in our simulations. In our fiducial simulation, all progenitors that initiate unstable interaction during this phase would end as stellar mergers, while in this variation they will survive. The progenitors of BHNSs and NSNSs are less strongly affected with an increase about 30\% increase.

\paragraph{Case BB mass transfer} In models \modCaseBB{} and \modCaseBBOpt{} we consider uncertainties in case BB mass transfer. This is a phase of mass transfer where the donor star has already lost its hydrogen envelope in a prior interaction, but fills its Roche lobe again as it expands during helium shell burning phase \citep[e.g.][]{Dewi+2002, Tauris+2015, Tauris+2017}. This is of particular interest for the formation of NSs, as their lower mass progenitors are swelling the most during this phase \citep[e.g.][and references therein]{Laplace+2020}. Population synthesis studies find that nearly all NSNS systems form through a phase of case BB mass transfer \citep{Vigna-Gomez+2018}.

In model \modCaseBB{} we enforce that case BB mass transfer is always unstable, such that it always leads to a CE. Note that this is slightly different from model E described in  \citet{Broekgaarden+2021}. In their work the pessimistic approach to CE evolution is implemented such that all HG stars are excluded including helium stars in the helium shell burning phase. This, in combination with the assumption that case BB mass transfer is always unstable, effectively leads to the exclusion of all systems that originate through this channel. We are interested in NSNS systems, which are frequently formed through this channel in our simulations. Therefore, we adapted this model to only exclude H-rich HG donors, but allow systems with donors that are helium stars in the helium shell burning phase to survive a CE phase.

As expected, in model \modCaseBB{}, we find that case BB systems form through a common-envelope phase rather than only stable mass transfer (see Fig.~\ref{fig:formation_channels}). This model gives the lowest NSNS rate of all of our variations, which is a factor of 3 lower than our fiducial rate. We also find a reduction of the BHNS systems by a factor 2. The BHBH systems are not significantly affected, as expected, since case BB mass transfer does not play a role for high mass progenitors.

Finally, in model \modCaseBBOpt{}, we again enforce that case BB mass transfer is always unstable, but in combination with the optimistic treatment for CE (essentially combining models \modCaseBB{} and \modOpt{}). This allows the systems that have HG donors for common-envelopes (as well as those formed through case BB mass transfer) to survive the CE phase. We find that this model leads to the highest predictions for the detections among all variations that we have considered.

\subsubsection{Supernovae and compact remnants}

The formation of compact remnants and their associated natal kicks also constitute important uncertainties.  

In model \modRapid{} we consider the so-called ``rapid'' remnant mass function by \citet{Fryer+2015} as an alternative to the ``delayed'' description used in our fiducial simulations. This affects the mass distribution (see Sec.~\ref{sec:lower_mass_gap}), but the effects on the number of detectable systems is modest for BHBH and BHNS, and negligible for NSNS.  

The same is true for the impact of changing the assumed maximum neutron star mass, $m_{\rm NS, max}$ (\modNSLow{} and \modNSHigh{}). Lowering $m_{\rm NS, max}$ increases the number of detections involving BHs (since more stars form BHs instead of NSs) and vice versa, but has no significant effect on the number of NSNS detections since the vast majority are formed from low mass NSs.

We find that not implementing pair-instability supernovae (PISN) or pulsational pair-instability supernovae (PPISN) in model \modNoPISN{} has no effect on the number of detections with LISA. This is because the average metallicity of the Milky Way is high enough such that no progenitor retains enough mass to initiate a PISN or PPISN.

Decreasing the natal kicks for all core-collapse supernovae (models \modSigLow{}-\modSigLower{}) increases the detection rates for each DCO type, since lower kicks result in fewer disrupted binaries and hence a more numerous detectable population. The BHNS and NSNS systems are strongly affected whilst the impact on BHBH systems is insignificant. The reason that BHBHs are relatively unaffected is that, in our models, the natal kicks for BHs are scaled down with the amount of mass that falls back. In the case of BHBHs, the black holes have very massive cores and thus low kicks.

In model \modNoBH{} we assume BHs form without any kick, while using our fiducial assumption for the natal kicks of neutron stars. This increases the predictions for BHNS by a factor 3 but the impact on BHBH systems is much smaller for the same reason as for models \modSigLow{}-\modSigLower{}. As expected, the NSNS population is not affected. 

\subsubsection{Stellar winds}
Mass loss in the form of stellar winds or eruptions is also a main uncertainty. It affects by how much stars can grow in size and it affects the final core masses. We consider variations in the mass loss by naked helium (Wolf-Rayet like) stars and choose to vary the efficiency of these winds between $0.1$ and $5$, to account for uncertainties in the derived rates \citep[e.g.][]{Vink+2017, Shenar+2019, Hamann+2019, Sander+2020}.

We find that a reduction of the wind mass loss has very little effect on our predictions. This is the consequence of several effects that cancel each other. Firstly, the decreased Wolf-Rayet like winds mean that the DCOs (particularly those containing BHs) are generally more massive and so more detectable in LISA. Secondly, one may expect that LISA sources in model \modWRLow{} would be higher frequency than in our fiducial model as decreased winds generally result in tighter binaries. However, though this is the case at DCO formation, we find that by the time the sources have evolved until they are observable by the LISA mission, they have lower orbital frequencies than in our fiducial model. This is because the reduced winds allow DCOs to be formed at higher metallicity and, therefore, at more recent times. This means that most DCOs do not evolve for as long before the LISA mission and so remain at lower frequencies (wider separations) thus making them less detectable.

In addition, we find that NSNSs are more eccentric and BHBHs are less eccentric than our fiducial model (with BHNSs relatively unchanged). The increase in eccentricity for NSNSs comes from the same reason as the lower frequency, more recent birth times mean that binaries have less time to circularise. The same is not true for BHBHs as the more massive systems are less affected by supernova kicks and so fewer high eccentricity systems are formed. Overall, despite the large differences in the system properties, these three effects in combination leave the detection rates relatively unchanged.

In model \modWRHigh{} we instead \textit{increase} the efficiency of Wolf-Rayet winds by a factor of 5. In this model the detection rate of BHBHs decreases by over a factor of 10 and BHNSs by over a factor of 2 whilst the NSNS rate is relatively unchanged. Increasing the efficiency of WR winds widens the orbit and decreases the final masses of DCOs. This means that some progenitors that would have formed LISA sources under our fiducial assumptions would not have enough mass to produce a DCO, or produce a NS instead of a BH. The effect is strongest for the progenitors of black holes, since these are most strongly affected by Wolf-Rayet winds. This effect is less pronounced in NSs since the rate of mass loss for the progenitors of these systems is low enough that changing by a factor of 5 still does not impact the final fate. Moreover, DCOs that are formed tend to be less massive and therefore less detectable. 

\subsection{Properties of detectable systems}\label{sec:property_variations}

In this section, we consider how varying underlying physics assumptions changes the properties of detectable systems. We focus on several key differences across physics variations rather than showing the differences in every model and thus this section is by no means exhaustive.

\subsubsection{Using LISA to investigate the lower mass gap}\label{sec:lower_mass_gap}

\begin{figure}[tb]
    \centering
    \includegraphics[width=\columnwidth]{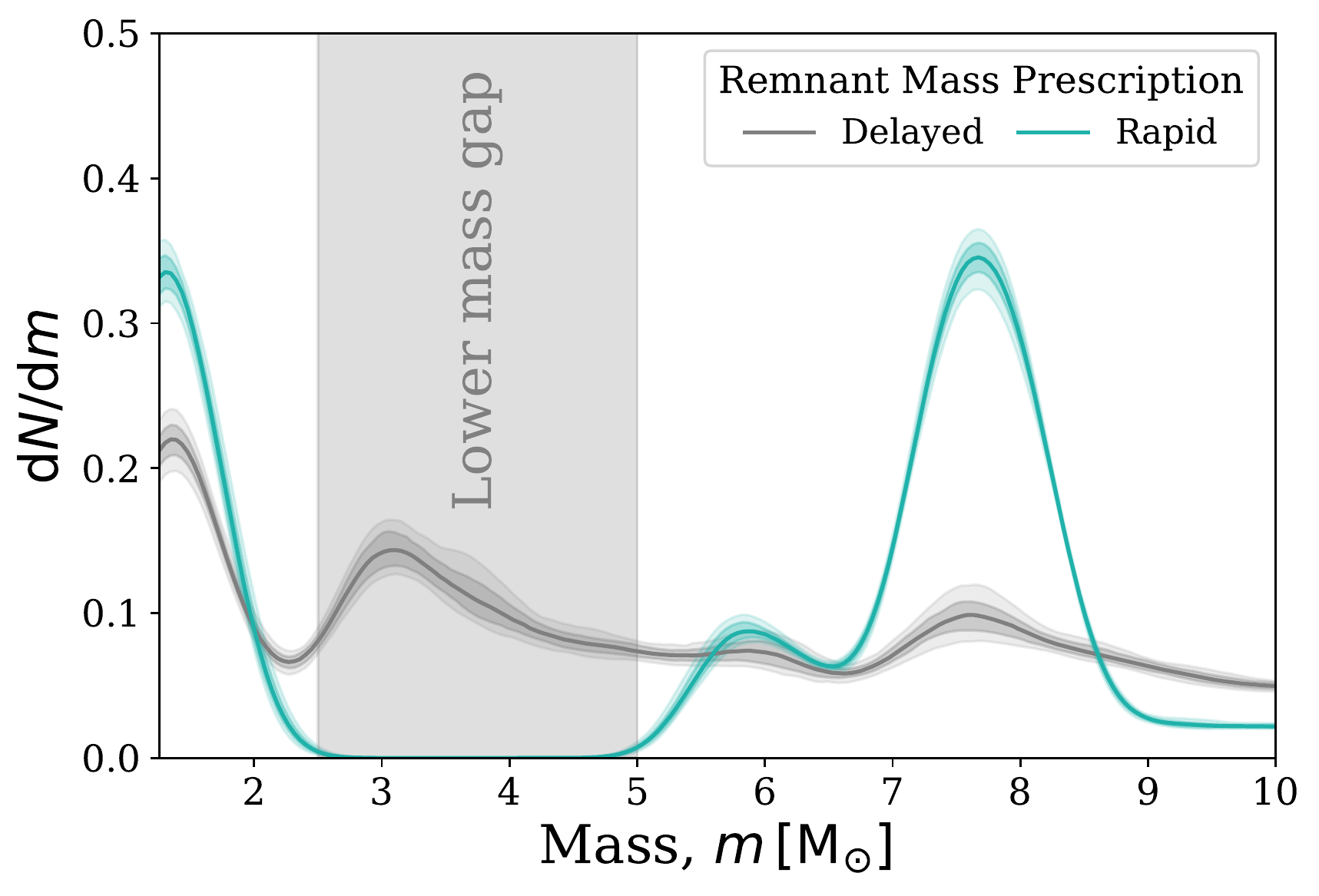}
    \caption{Comparison of the component mass distribution of LISA detectable DCOs when using the Fryer delayed (model \modFid{}) and rapid (model \modRapid{}) remnant mass prescriptions. Distributions are plotted in the same way as Fig.~\ref{fig:fiducial_pdf_distributions}, except all DCOs types are shown in one curve and each type is weighted by its detection rate in the respective model. \href{https://github.com/TomWagg/detecting-DCOs-in-LISA/blob/main/paper/figures/fig9_lower_mass_gap_rapid_variation.pdf}{\faFileImage} \href{https://github.com/TomWagg/detecting-DCOs-in-LISA/blob/main/paper/figure_notebooks/variations.ipynb}{\faBook}.}
    \label{fig:lower_mass_gap_variation}
\end{figure}

In Fig.~\ref{fig:lower_mass_gap_variation}, we show the component mass distribution for all LISA detectable DCOs (BHBHs, BHNSs and NSNSs) for two different remnant mass prescriptions. The grey distribution uses the Fryer \textit{delayed} remnant mass prescription \citep{Fryer+2012}, which is our fiducial assumption (model \modFid{}). This prescription produces compact objects in the lower mass gap ($2.5 \unit{M_{\odot}} \le m \le 5 \unit{M_{\odot}}$) and indeed we find that, of the LISA detectable DCOs, approximately \BHBHatLeastOneLowerMassGapPerc{}\% of BHBHs, \BHNSatLeastOneLowerMassGapPerc{}\% of BHNSs and \NSNSatLeastOneLowerMassGapPerc{}\% of NSNSs have at least one component in the lower mass gap. Overall, weighting by the relative detection rates, this gives that, in our fiducial model, 55\% of our predicted LISA DCO detections would have at least one component in the lower mass gap when using this remnant mass prescription. This equates to approximately 69 systems being detected in the lower mass gap. Alternatively, the blue curve in Fig.~\ref{fig:lower_mass_gap_variation} shows the same distribution but for the \textit{rapid} remnant mass prescription \citep{Fryer+2012}, which we use in model \modRapid{}. In this case, no compact objects are formed (and therefore, detected) in the lower mass gap.

From the stark difference between these models, it is clear that it is difficult at this point to say with any certainty what fraction of systems LISA will detect in the lower mass gap given the highly uncertain formation rate of systems in this mass range. Indeed, we find that the percentage of detectable systems with at least one component in the lower mass gap varies between approximately 30-70\% (or, in terms of detections, from 15 to 156) for the different model variations. However, it is important to highlight that \textit{if} DCOs are formed with components in the lower mass gap, LISA \textit{will} be able to detect them. And thus, LISA could be a useful instrument for providing constraints on the existence or non-existence of a lower mass gap based on the mass distribution of detected DCOs.

\subsubsection{Effect of natal kicks on eccentricity distribution}

\begin{figure}[tb]
    \centering
    \includegraphics[width=\columnwidth]{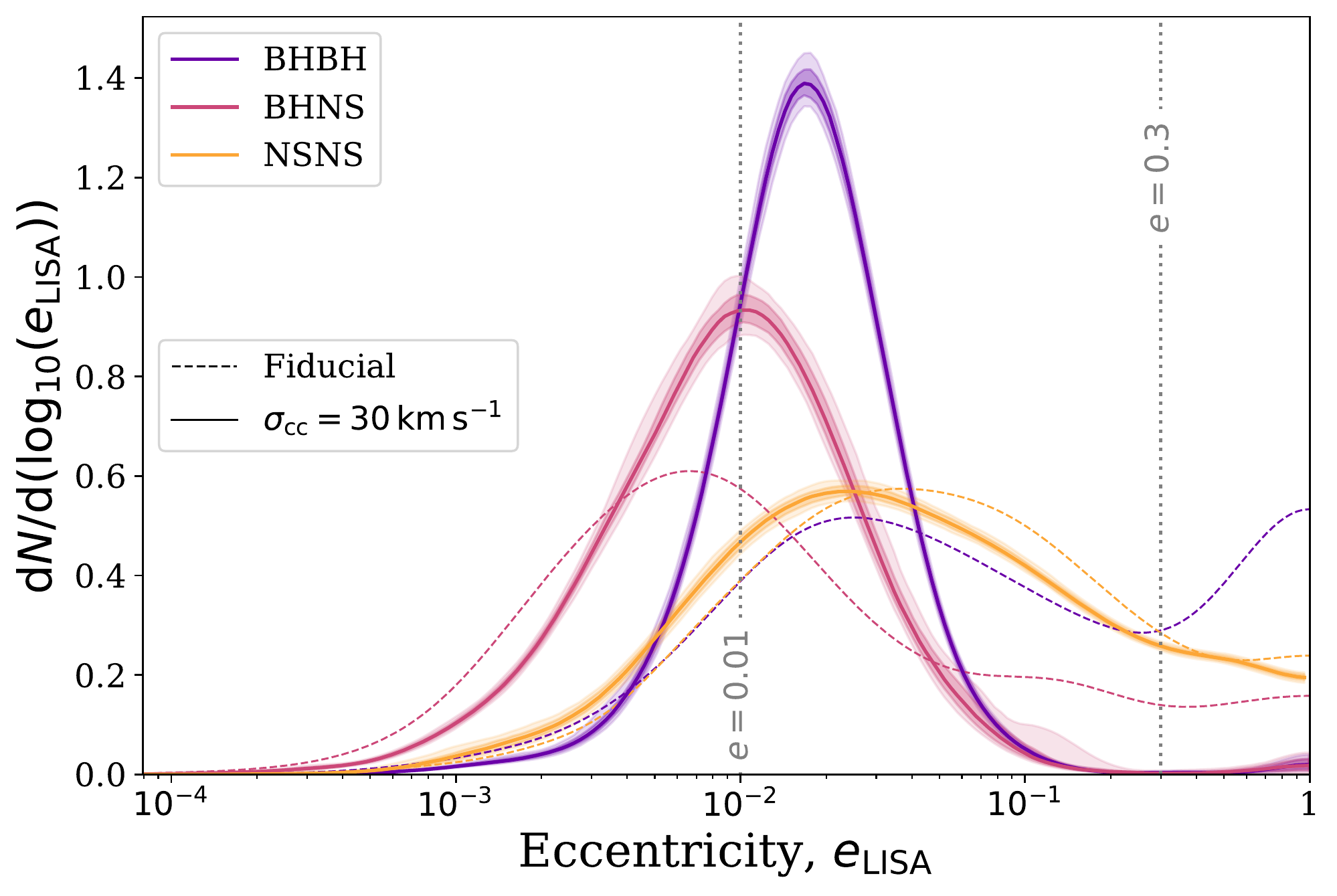}
    \caption{As Fig.~\ref{fig:fiducial_pdf_distributions}d, but for model \modSigLower{}. For comparison, we show the mean distribution for the fiducial model (model \modFid{}) as dashed lines. \href{https://github.com/TomWagg/detecting-DCOs-in-LISA/blob/main/paper/figures/fig10_ecc_low_kick_variation.pdf}{\faFileImage} \href{https://github.com/TomWagg/detecting-DCOs-in-LISA/blob/main/paper/figure_notebooks/variations.ipynb}{\faBook}.}
    \label{fig:ecc_low_kick_variation}
\end{figure}

In Fig.~\ref{fig:ecc_low_kick_variation}, we investigate how decreasing the magnitude of natal kicks from core-collapse supernovae affects the eccentricity distribution of LISA detectable DCOs. For reference, we show the mean fiducial distributions (model \modFid{}) as dashed lines (see Fig.~\ref{fig:fiducial_pdf_distributions}d for full comparison). In the main curves, we reduce the velocity dispersion for core-collapse supernovae from $265 \unit{km}{s^{-1}}$ to $30 \unit{km}{s^{-1}}$ (model \modSigLower{}).

We find that the LISA detectable BHBHs are significantly less eccentric with weaker kicks, such that the population above $e = 0.2$ is nearly completely eliminated. This is because BHBHs are often massive enough to withstand strong natal kicks without disrupting and these kicks tend to impart significant eccentricity. In model \modSigLower{}, very few systems are ever given such strong kicks and thus very few BHBHs are detected with significant eccentricity.

Since BHNSs are less massive than BHBHs and have more unequal mass ratios, they are more vulnerable to disruption during supernova kicks. BHNSs can only withstand strong kicks when they are aimed in the correct direction and so only a small `lucky' fraction of the fiducial population is highly eccentric. Therefore in model \modSigLower{}, although we see that the population of highly eccentric BHNS systems is eliminated (similar to BHBHs), the peak of the distribution actually shifts to \textit{higher} eccentricity. This is because a larger fraction of systems are given weaker kicks that BHNSs can withstand and these impart much more moderate eccentricities.

Finally, we find that the NSNS distribution is relatively unchanged between model \modFid{} and \modSigLower{}. This is not surprising however since the majority of NSNSs are formed through electron-capture supernovae and ultra-stripped supernovae and for these types of supernovae we use $\sigma^{\rm 1D}_{\rm rms} = 30 \unit{km}{s^{-1}}$ already (see App.~\ref{app:fiducial_physics}) and thus there is little difference between the models.

Overall, compared to our fiducial model, we find that decreasing supernova natal kicks, though it strongly increases the number of detections (see Fig.~\ref{fig:detection_rates}), strongly \textit{decreases} the fraction of highly eccentric systems that are detected.

\subsubsection{Effect of Wolf-Rayet winds on mass distribution}

\begin{figure}[tb]
    \centering
    \includegraphics[width=\columnwidth]{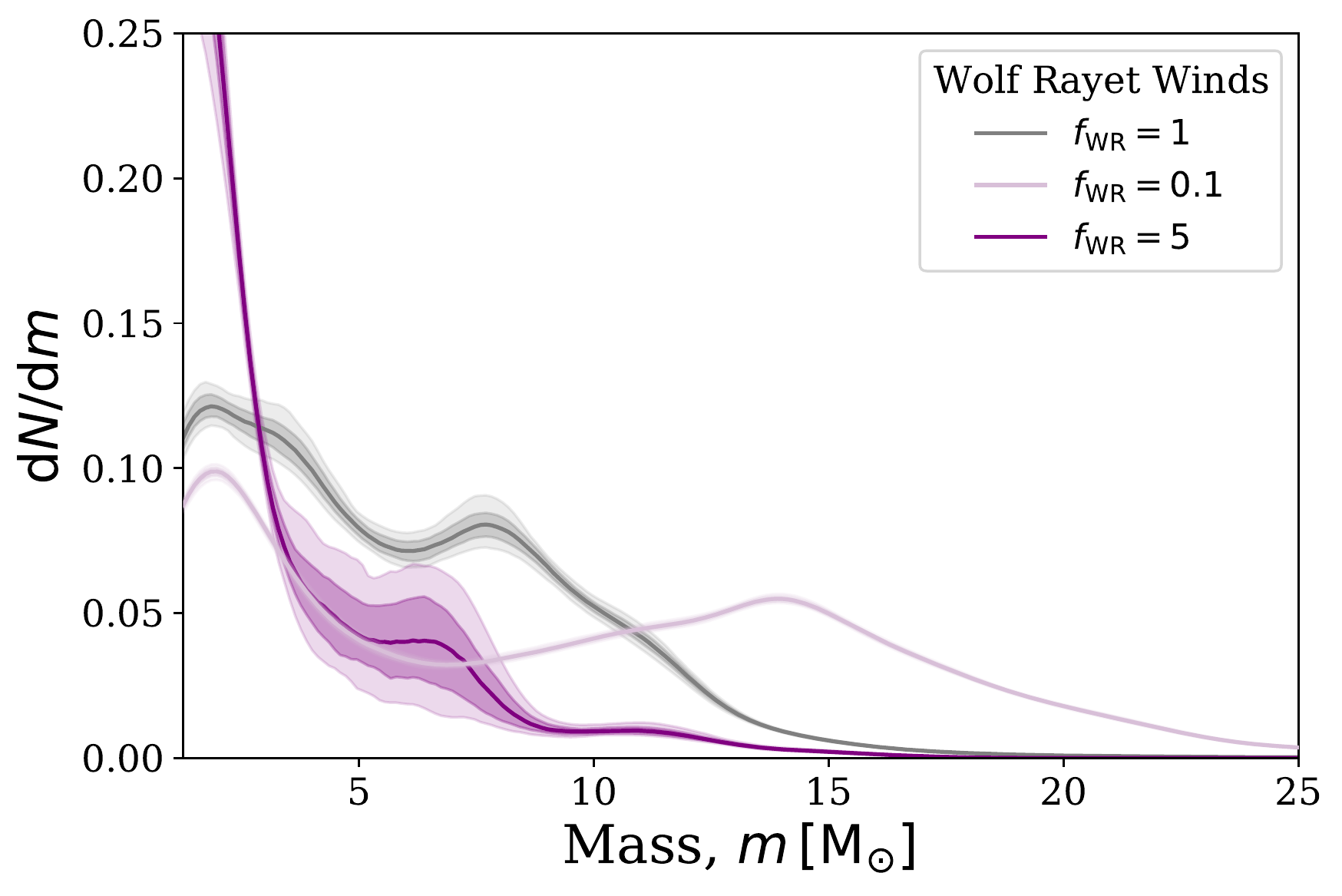}
    \caption{As Fig.~\ref{fig:lower_mass_gap_variation}, but instead varying Wolf-Rayet wind efficiency (models \modWRLow{} and \modWRHigh{}). The curve for $f_{\rm WR} = 5$ has much higher uncertainties as there are many fewer DCO systems formed in this model (the inverse reasoning also explains the lower uncertainties for $f_{\rm WR} = 0.1$). \href{https://github.com/TomWagg/detecting-DCOs-in-LISA/blob/main/paper/figures/fig11_wr_wind_mass_variations.pdf}{\faFileImage} \href{https://github.com/TomWagg/detecting-DCOs-in-LISA/blob/main/paper/figure_notebooks/variations.ipynb}{\faBook}.}
    \label{fig:wr_wind_mass_variations}
\end{figure}

In Fig.~\ref{fig:wr_wind_mass_variations} we show the effect that changing the efficiency of Wolf-Rayet winds has on the individual component mass distribution. Decreasing the Wolf-Rayet wind efficiency allows the formation of more massive DCOs in the Milky Way and, indeed, we see that the distribution extends to $25 \unit{M_{\odot}}$ and relatively fewer detectable systems are formed at low masses. By contrast, increasing the Wolf-Rayet efficiency by a factor of 5 strongly disfavours the formation of systems at high masses and approximately 85\% of detectable systems have masses below $5 \unit{M_{\odot}}$. These three distributions are very distinct and so it is possible that the mass distribution of LISA could help to constrain the efficiency of Wolf-Rayet winds.

\section{Discussion} \label{sec:discussion}
In this section we discuss the prospects of (and methods for) identifying LISA sources (Sec.~\ref{sec:identify_sources}), the possibility of matching LISA signals to SKA detections (Sec.~\ref{sec:pulsar_matching}), the main caveats for this study (Sec.~\ref{sec:caveats}) and the possible contribution from other formation channels (Sec.~\ref{sec:other_formation_channels}). All predictions quoted in each subsection are derived for the fiducial model (model \modFid{}).

\subsection{Identification of GW sources}\label{sec:identify_sources}
It is important to note that, though we present predictions for the detection rates of specific DCO types, the nature of the source may not be immediately apparent from the gravitational wave signal. LISA can detect a variety of sources, from exoplanets \citep[e.g.][]{Tamanini+2019} to common-envelopes \citep[e.g.][]{Ginat+2020, Renzo+2021} that may cause confusion. However, by far the most prominent will be the population of Galactic WDWDs detectable with LISA, which will be several orders of magnitude larger than the population of the more massive DCOs that we focus on in this paper \citep[e.g.][]{Korol+2017}. It is therefore imperative that we consider how to distinguish NS and BH binaries from this much more numerous population of sources. In addition to distinguishing them from WDWDs, we must consider how to discriminate between BHBHs, BHNSs and NSNSs themselves.

\subsubsection{Distinguishing from WDWD population}\label{sec:WDWD_distinguish}
The simplest way to check whether a source is a WDWD is to evaluate its chirp mass. The mass of a non-rotating white dwarf cannot be larger than the Chandrasekhar limit of $1.4 \unit{M_\odot}$ \citep{Chandrasekhar+1931, Hamada+1961}, so we can take the maximum chirp mass of a WDWD to be ${\sim}1.2 \unit{M_{\odot}}$. Therefore, any DCO with a chirp mass that satisfies $\mathcal{M}_c > 1.2 \unit{M_{\odot}} + \Delta \mathcal{M}_c$ must not be a WDWD (where $\Delta \mathcal{M}_c$ is the error on the chirp mass, estimated using Eq.~\ref{eq:chirp_mass_uncertainty}). We find that for the detectable population of a 4(10)-year LISA mission, \BHBHAboveMaxWDWDFourPerc{}(\BHBHAboveMaxWDWDTenPerc{})\% of BHBHs, \BHNSAboveMaxWDWDFourPerc{}(\BHNSAboveMaxWDWDTenPerc{})\% of BHNSs and \NSNSAboveMaxWDWDFourPerc{}(\NSNSAboveMaxWDWDTenPerc{})\% of NSNSs satisfy this condition. This method is not particularly effective for NSNSs since their average chirp mass, $1.17 \unit{M_\odot}$, is below the Chandrasekhar limit.

Another discriminator between WDWDs and other DCOs is eccentricity. WDWDs formed in the disc are thought to be formed mainly through isolated binary formation and have little to no eccentricity (e.g.\ \citealt{Nelemans+2001}, see however \citealt{Dosopoulou+2016a, Dosopoulou+2016b, Gosnell+2019}). This is because WDWDs formed through isolated binary evolution all experience a phase of mass transfer or a common envelope, which typically circularises the binary \citep[e.g.][]{Marsh+2004}. However, in contrast to the more massive DCOs that we study, WDWDs do not experience strong natal kicks which we find to be the main source of eccentricity. Therefore, if any system is detected with anything other than one detectable harmonic, this suggests that the system is unlikely to be a WDWD. We find that for a 4(10)-year LISA mission, \BHBHMultipleHarmonicsFourPerc{}(\BHBHMultipleHarmonicsTenPerc{})\% of BHBHs, \BHNSMultipleHarmonicsFourPerc{}(\BHNSMultipleHarmonicsTenPerc{})\% of BHNSs and \NSNSMultipleHarmonicsFourPerc(\NSNSMultipleHarmonicsTenPerc{})\% of NSNSs are detected with multiple harmonics (see also Sec.~\ref{sec:fiducial_distributions}). Both the absolute percentage and the relative improvement with an extended LISA mission is lower for the BHNSs with respect to other DCOs as we find that these BHNSs are less eccentric on average (see Fig.~\ref{fig:fiducial_pdf_distributions}d and discussion in Sec.~\ref{sec:fiducial_distributions}).

However, we should also consider that eccentric WDWDs could be formed through dynamical formation in Milky Way globular clusters \citep[e.g.][]{Willems+2007, Kremer+2018}, or with third companions \citep[e.g.][]{Antonini+2017}. This means that we cannot assume that eccentric binaries are not WDWDs unless they are detected in the Galactic plane (though even then there is a chance they were formed dynamically). We can use the sky localisation, scale height of the disc and distance to the source to estimate what fraction of eccentric sources can be localised to the Galactic plane. This condition can be written as $\sigma_\theta < \arcsin(z_{\rm plane} / D_L)$ or $D_L < z_{\rm plane}$, where we set the height of the Galactic plane, $z_{\rm plane}=0.95 \unit{kpc}$, to the scale height of the high-$\alpha$ disc. We apply this condition to find that the fraction of sources that are eccentric and localised within the disc for a 4(10)-year LISA mission are \BHBHEccInDiscFourPerc{}(\BHBHEccInDiscTenPerc{})\% for BHBHs, \BHNSEccInDiscFourPerc{}(\BHNSEccInDiscTenPerc{})\% for BHNSs and \NSNSEccInDiscFourPerc{}(\NSNSEccInDiscTenPerc{})\% for NSNSs. Note that although the fractions are the same for the 10-year mission, the absolute number of detections is still greater.

Overall, combining these methods (chirp mass, eccentricity and sky localisation) we find that for a 4(10)-year mission, LISA will detect at least \BHBHNotWDWDFour{}(\BHBHNotWDWDTen{}) BHBHs, \BHNSNotWDWDFour{}(\BHNSNotWDWDTen{}) BHNSs and \NSNSNotWDWDFour{}(\NSNSNotWDWDTen{}) NSNSs that are distinguishable from the WDWD population. Thus we will be able to confidently distinguish approximately half of all detected sources from WDWDs. This increases to roughly 60\% for a 10-year mission. We highlight that, though the overall number of LISA detections in an extended mission only increases by a factor of $\sqrt{T_{\rm obs}}$, the number of distinguishable detections increases by a greater factor since each of the more numerous sources are better measured. This further underlines the benefits of extending the LISA mission to 10 years.

\subsubsection{Discriminating between BHBHs, BHNSs and NSNSs}

The problem of discriminating between the BHBH, BHNS and NSNS populations can be more difficult than distinguishing them from WDWDs. For NSNSs, we can follow a similar method to the WDWDs (see Sec.~\ref{sec:WDWD_distinguish}) by applying our knowledge of the maximum mass of a neutron star. Following our fiducial assumption, we can take the maximum mass of a neutron star as $2.5 \unit{M_{\odot}}$ and thus the maximum chirp mass that a system can attain without one of the components being a black hole is $\mathcal{M}_{c} = 2.2 \unit{M_\odot}$. For a 4(10)-year LISA mission, the fraction of systems that are above or below this limit (and thus \textit{must} respectively contain or not contain a BH component) by more than $\Delta \mathcal{M}_c$ is \BHBHEitherBHOrNSFourPerc{}(\BHBHEitherBHOrNSTenPerc{})\% for BHBHs, \BHNSEitherBHOrNSFourPerc{}(\BHNSEitherBHOrNSTenPerc{})\% for BHNSs and \NSNSEitherBHOrNSFourPerc{}(\NSNSEitherBHOrNSTenPerc{})\% of NSNSs, which in terms of absolute detections is \BHBHEitherBHOrNSFour{}(\BHBHEitherBHOrNSTen{}) for BHBHs, \BHNSEitherBHOrNSFour{}(\BHNSEitherBHOrNSTen{}) for BHNSs and \NSNSEitherBHOrNSFour{}(\NSNSEitherBHOrNSTen{}) for NSNSs.

For separating the BHBH and BHNS population one could do so probabilistically given the properties that are measured, particularly the orbital frequency, mass ratio and eccentricity, since these distributions are fairly different for the two DCO types (see Fig.~\ref{fig:fiducial_pdf_distributions}). This method would pose a challenge, however, as it would likely only indicate which type was more likely rather than discriminate between them with strong evidence.

Another possible solution would be the existence of electromagnetic counterparts to the gravitational wave signal. In Section~\ref{sec:pulsar_matching} we consider the possibility of detecting a pulsar within a BHNS or NSNS system. This could be used to identify the type of the source.

\subsection{Matching LISA detections to pulsars with the SKA}\label{sec:pulsar_matching}
Since the vast majority of the LISA detectable population of DCOs will not merge for many years, the main type of electromagnetic counterpart for this population is pulsars. Therefore, for this section we focus only on BHNSs and NSNSs since no BHBH system will contain a pulsar. The joint detection of a binary pulsar with LISA and the Square Kilometre Array (SKA, \citealt{Dewdney+2009}) would not only help to constrain the parameters of the binary, but also enable investigation of other compact object physics. A pulsar(PSR)+BH can provide stringent tests of theories of gravity, in particular the ``No-hair theorem'' \citep{Keane+2015}. Alternatively, an ultrarelativstic PSR+NS system could be used to measure the neutron star equation of state up to an order of magnitude more accurately than other proposed observational constraints \citep{Kyutoku+2019, Thrane+2020}.

We estimate that on average, given the number of detectable pulsars and the SKA sky area, each pulsar in the SKA occupies a region with an angular resolution of $\sigma_{\theta} < 1.3^\circ$ or $0.7^\circ$ for SKA-1 and SKA-2 respectively (see Appendix~\ref{app:ska_area}). Therefore, any DCOs containing NSs localised by LISA with an angular resolution lower than these values can be unambiguously matched to the radio signal in the SKA. By considering Fig.~\ref{fig:ang_res}, approximately $11$ and $6$ (for SKA-1 and SKA-2) DCOs will satisfy this constraint.

If there is more than one pulsar in the region given by the LISA sky localisation, one can compare the measured parameters of the system in LISA and the SKA. Both the SKA and LISA will measure the orbital frequency to high precision, as well as the time derivative of the frequency and chirp mass to a lesser precision, of each of these systems. Therefore, one could perform a targeted search with the SKA that checks the sky location given by LISA, only looking for binary pulsars with orbital frequencies within the uncertainties. If there was \textit{still} more than one possible pulsar one could also check against the chirp mass. In this way, we expect it will be possible to get a joint detection between the SKA and LISA even when the sky area implied by the LISA detection contains more than one pulsar.

In order to assess the efficacy of this method, we would need to know the probability that two random binary pulsars would have orbital frequencies and chirp masses close enough that one could not tell which pulsar matches the LISA detection. This would require simulating the SKA population of pulsars with a code such as PSRPOPPy \citep{Bates+2014} to find the frequency and chirp mass distribution, which is beyond the scope of this paper. However, the uncertainty in the orbital frequency of a binary on the detection threshold (${\rm SNR} = 7$) for a 4-year LISA mission is $2.5 \times 10^{-9} \unit{Hz}$ and $1.0 \times 10^{-9} \unit{Hz}$ for a 10-year mission (calculated using Eq.~\ref{eq:f_orb_unc}). Therefore, we expect that the SKA could likely isolate the correct binary pulsar to match to a LISA detection even when several are present in the sky localisation region.

\subsection{Caveats}\label{sec:caveats}

Our predictions are subject to various uncertainties which can be broadly divided into two different categories: those arising from the progenitor models for the population of DCOs and those arising from the choices we have made when placing these DCOs in our model for the Milky-Way. Although we are unable, at present, to evaluate the impact of all these uncertainties, the reader should nevertheless keep in mind that they are likely very substantial. Most of these concerns are not unique to these study, but apply to most of the predictions available in present literature. We highlight a few main concerns. 

\paragraph{Progenitor models} Our binary-star progenitors models have been computed with a rapid population synthesis code (see Sec.~\ref{sec:COMPAS_explained}). This code relies on approximate parametric prescriptions for the stellar evolutionary tracks of single tracks and simple algorithms to mimic the effects of evolutionary and binary interaction  processes. Even though we explicitly consider the impact of some of the main physics uncertainties (see Sect.~\ref{sec:variation_assumptions}) the list of variations that we considered is far from exhaustive. Moreover, it is by no means guaranteed that the parametric prescriptions used in this code lead to realistic results, even when varying the values of the parameters to their extremes. We stress in particular the uncertainties affecting our most massive progenitor models. Observational constraints are scarce for high mass stars and practically non-existent for the rapid evolutionary phases \citep[e.g.][]{Langer2012, Mapelli+2021}. This is even more true for the evolution of massive stars at low metallicity. In addition to our limited understanding of massive stars, we note that the rapid population synthesis code, such as the one employed to compute the models used in this study, rely on extrapolations of the original fitting formulae to approximate the evolutionary tracks for these higher mass progenitors \citep{Hurley+2000,Hurley+2002}. 
 
\paragraph{Populating the Milky Way} Our Milky Way model is semi-empirical and has been calibrated based on observations. Unfortunately, the early evolution of the (metallicity dependence of the) star formation history is poorly constrained. We do not expect this to be a major concern, as most of the double compact objects have relatively short delay times of less than $2 \unit{Gyr}$ (see Fig.~\ref{fig:fiducial_pdf_distributions}e), but this is a caveat that should be kept in mind. Furthermore, to estimate the rate of detectable systems, we rely on normalisation choices (e.g.\ how many detectable double compacts are formed per unit of star formation). This depends heavily on the initial mass function, as low mass stars account for most of the mass while high mass stars are the progenitors of double compact objects. Further choices, such as the binary fraction and the initial distributions of binary parameters also play a lesser but probably still significant role \citet[e.g.][]{deMink+2015, Chruslinska+2017, Klencki+2018}. 

We also note that, for reasons of computational efficiency, we have not accounted for the spatial velocities resulting from the Blaauw-Boersma kick \citep{Blaauw+1961,Boersma1961}. In test simulations we find that accounting for this spreads out the population (increasing the typical height above the Galactic plane and Galactocentric radius), but we find that the impact on the rate is limited. In light of the other much larger uncertainties, we felt that this was justified (see however, e.g., \citealt{Brandt+1995, Abbott+2017_GW170817_progenitor}). We have further ignored a possible contribution coming from the Galactic halo, as \citet{Lamberts+2018} estimates this not be significant. However, this may not be true for other formation channels other than those we have considered here.

\subsection{Other formation channels}\label{sec:other_formation_channels}

In this paper we considered the formation of NS and BH binaries formed via isolated binary evolution, through the classical CE channel, the stable mass transfer channel and variations on these (see Fig.~\ref{fig:formation_channels}). We did not consider further possible contributions from other formation channels, which may play a role.

We highlight the possible role of dynamical formation in globular clusters. \citet{Kremer+2018} predict, for a nominal 4-year LISA mission, that 21 sources will have SNR $> 7$, of which 7 are BHBHs, 0 are BHNSs and 1 is a NSNS \citep[see Table~1][]{Kremer+2018}. This is significantly lower than the rates we predict for nearly every model variation. If true, this would mean that formation through isolated binary formation will dominate the LISA detections. 

\citet{Banerjee+2020} investigates formation of LISA detectable BHBHs in young massive and open stellar clusters and estimates approximately 128 BHBHs with SNR $>5$ in a 5-year LISA mission \citep[see Table~1, Column 9][]{Banerjee+2020}. Although this is similar to the number we predict for our fiducial model, we note these authors adopt a threshold SNR required for a detection that is lower and a mission length is slightly longer than what is typically assumed (i.e. SNR $>7$ and 4 years, as we have also adopted in our work). We expect that, after correcting for this and making a fair comparison, our fiducial model predicts about twice as many detections.

The contribution of triples systems \citep[e.g.][]{Antonini+2017}, or even higher order multiple systems \citep[e.g.][]{Vynatheya+2021} will likely also be of interest, in particular for the formation of eccentric sources. We are, however, not aware of specific predictions for the detection rates that we can compare to directly.

\section{Comparison with previous studies}\label{sec:compare_studies}
\begin{figure*}[p]
    \centering
    \includegraphics[width=\textwidth]{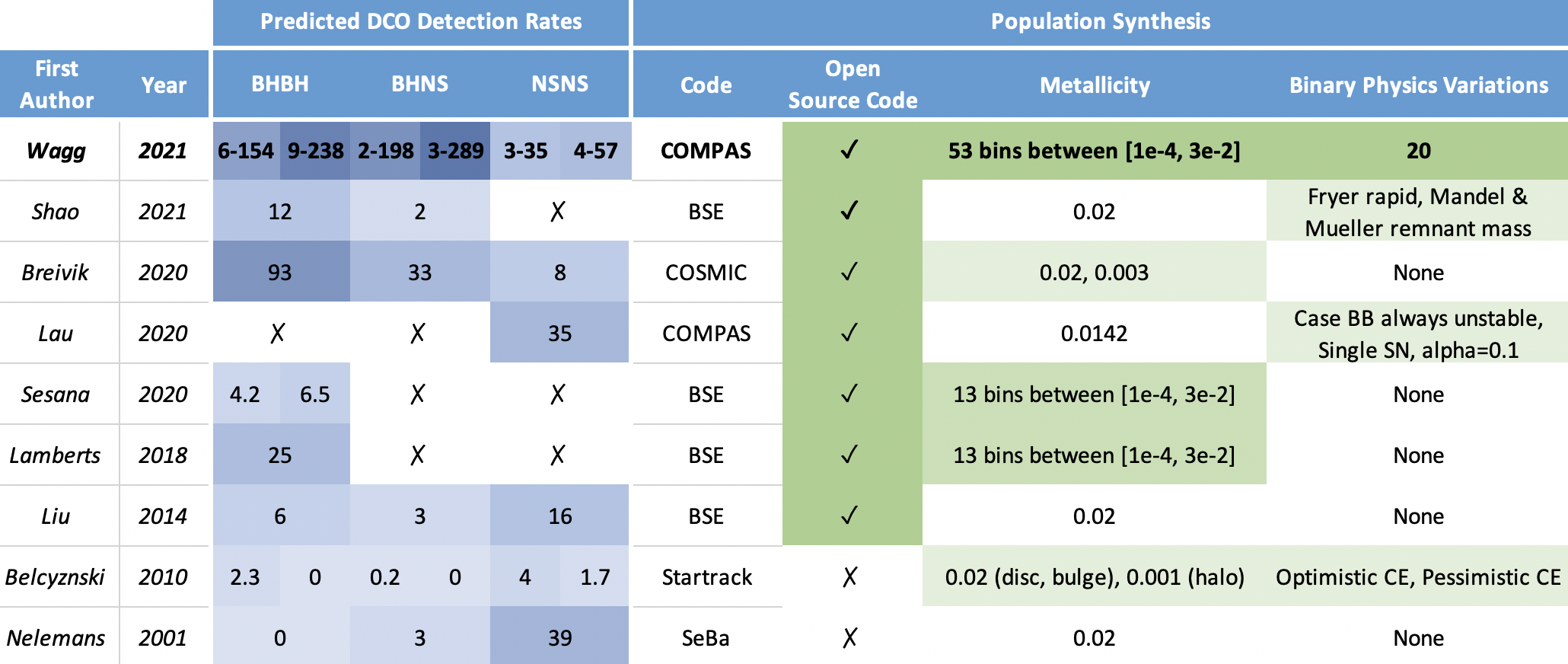}

    \vspace{0.5cm}

    \includegraphics[width=\textwidth]{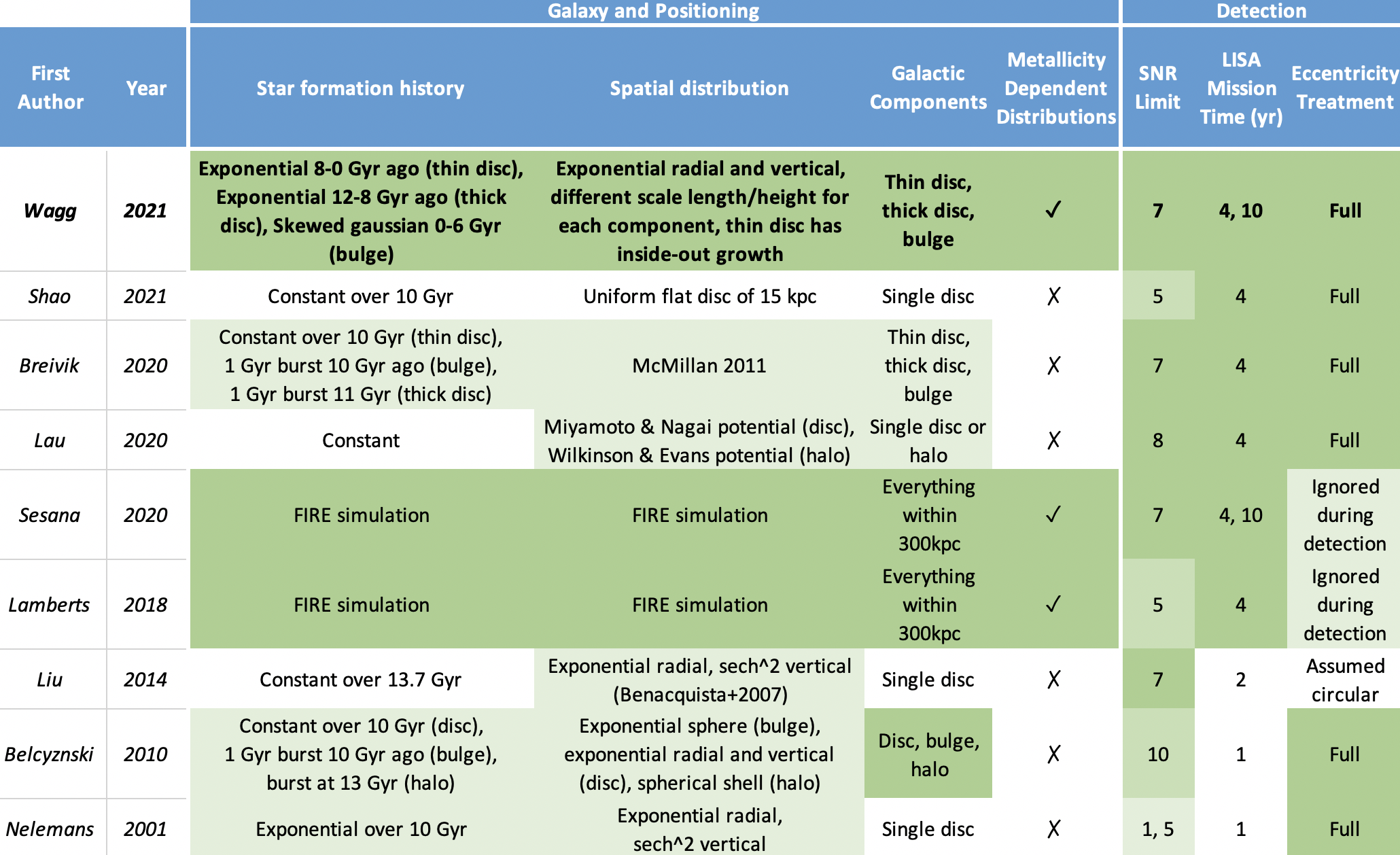}
    \caption{A table comparing previous studies of a similar nature to this work. In the detection rate section darker shades of blue indicate higher detection rates. In other sections, darker shades of green indicate more detailed models or more accurate choices. Entries in detection rates with two values listed correspond to the different mission lengths (penultimate column). The works listed in the table are \citet{Nelemans+2001}, \citet{Belczynski+2010}, \citet{Liu+2014}, \citet{Lamberts+2018}, \citet{Sesana+2020}, \citet{Lau+2020}, \citet{Breivik+2020} and \citet{Shao+2021}. \href{https://github.com/TomWagg/detecting-DCOs-in-LISA/blob/main/paper/figures/fig12_1_compare_dco.png}{\faFileImage} \href{https://github.com/TomWagg/detecting-DCOs-in-LISA/blob/main/paper/figures/fig12_2_compare_dco.png}{\faFileImage} \href{https://github.com/TomWagg/detecting-DCOs-in-LISA/blob/main/paper/previous_BH_NS_studies.xlsx}{\faFileExcel}.}
    \label{fig:previous_studies}
\end{figure*}

In Fig.~\ref{fig:previous_studies}, we compare our results to similar previous studies that investigate the population of stellar-mass BHBHs, BHNSs and NSNSs that are detectable with LISA. Fig.~\ref{fig:previous_studies} details the expected detection rates predicted by each paper as well as their assumptions regarding their Milky Way galaxy model, binary population synthesis simulation and LISA mission specifications. We only include papers that are similar to our work, such that they use population synthesis and simulate sources in the Galactic plane. Moreover, Fig.~\ref{fig:previous_studies} does not include the numerous papers on the LISA WDWD population as we do not make predictions for WDWDs.
 
\paragraph{\citet{Nelemans+2001}} were the first to investigate the population of LISA detectable stellar-mass double compact objects. We find a significantly higher detection rate for BHBHs and BHNSs, as well as a slightly lower rate for NSNSs. We can understand this difference from changes both to the specifications of LISA (such as the mission length and SNR threshold for detection) and our understanding of massive star evolution since the publication of their paper, which both strongly affect the expected detections rates.

\paragraph{\citet{Belczynski+2010}} built upon the work of \citet{Nelemans+2001}, by using a different population synthesis code with two model variations and a multi-component model for the Milky Way. They find a much lower detection rate for BHNSs and NSNSs (and agreed on zero BHBHs) when compared to \citet{Nelemans+2001}. They state that this discrepancy from \citet{Nelemans+2001} comes from differences in their population synthesis and an overall lower formation rate rather than any changes to LISA detectability. The low total detection rate for all DCOs in this paper compared to our work is unsurprising given the relatively high SNR threshold of 10 and short mission length of 1 year. The reduced mission length means that the source signal has much less time to accumulate, whilst also fewer WDWDs can be resolved in this time, leading to a weaker signal and an increased Galactic confusion noise relative to our work.

\paragraph{\citet{Liu+2014}} performed a similar investigation using a different population synthesis code and find higher rates than earlier works. Their lower detection threshold and longer mission length compared to \citet{Belczynski+2010} likely explains the relatively increased rates. Yet their rates are still significantly below what we find. This could be for several reasons; they assume all binaries are circular both in their evolution and for detection. This means that systems may not have inspiralled as far before the LISA mission or may appear to have weaker gravitational waves when eccentricity is not accounted for. They also use a simplified model for the Milky Way with a single disc of one metallicity and constant star formation, whilst also using a mission length half what we assume. Each of these factors likely contributes to the lower overall detection rates.

\paragraph{\citet{Lamberts+2018}} presented a new approach to the problem by using the FIRE simulation \citep{Hopkins+2014} to distribute their sources rather than an analytical model of the Milky Way and were the first paper in this area to incorporate metallicity dependence into their Milky Way model. \citet{Sesana+2020} followed up on this paper using the same simulated BHBH population and presented updated results for the number of expected BHBH detections. They find significantly fewer BHBHs than our fiducial model despite using the same SNR threshold and LISA mission length.
The discrepancy between the results of \citet{Sesana+2020} and those presented in this work could be caused by different treatments of eccentricity. Unlike our work, \citet{Sesana+2020} assume that all binaries are circular for the purpose of detection in LISA, which could result in a lower number of detections by missing eccentric binaries that appear as weaker signals when assumed to be circular. This is especially relevant as we find that around $\BHBHNotCirc{}$ of LISA detectable BHBHs are not circular and around $\BHBHHighlyEccentric{}$ have significant eccentricity (see Fig.~\ref{fig:fiducial_pdf_distributions}d). We also improve upon this work by using a larger number of metallicity bins compared to \citet{Sesana+2020}, since a low number of metallicity bins can produce artificial features in the mass distribution of DCOs and possibly affect the detection rate \citep[see][ and also appendix~\ref{app:mw_changes} for further discussion]{Dominik+2015, Neijssel+2019, Kummer_thesis}. Finally, it could be that different implicit assumptions in their population synthesis code lead to differences in our results \citep{Toonen+2014}.

\paragraph{\citet{Lau+2020}} focussed on the number of Galactic NSNS binaries that could be detected by LISA. Their study uses the same population synthesis code, COMPAS, as this work, though an earlier version. Despite this, their study finds a much larger number of detections. They make several different physical assumptions in their population synthesis, using the \citet{Fryer+2012} \textit{rapid} remnant mass prescription, assuming the optimistic CE scenario, limiting the maximum neutron star mass to $2 \unit{M_{\odot}}$ and not implementing PISN. However, we note that none of these assumptions strongly affect the NSNS LISA detection rate (see bottom panel of Fig.~\ref{fig:detection_rates}, models \modOpt{}, \modRapid{}, \modNSLow{} and \modNoPISN{}) and so this is unlikely to entirely account for the differences. It is also important to highlight that COMPAS has received several improvements and bug fixes since \citet{Vigna-Gomez+2018} (which contains the simulations used by \citealt{Lau+2020}) and these could possibly have affected the formation rate of NSNSs.
Yet it is most likely that the remaining difference between our results is due to way in which we simulate the Milky Way. \citet{Lau+2020} use a model for the Milky Way similar to that of \citet{Breivik+2020}, which we use to estimate the impact of the choice of MW model in Appendix~\ref{app:mw_changes}. The Milky Way model by \cite{Breivik+2020} applies only two metallicity bins, while we consider a range of metallicities between $10^{-4}$ and $0.03$. When applying the simpler model for the Milky Way, we find that the NSNS detection rate is increased by at least a factor of two. \citet{Lau+2020} only uses a single metallicity, and so assuming that all star formation happens at a single high metallicity (which has a high efficiency of producing NSNS), could lead to an even greater overestimate of the detection rate.  Hence, we expect the low number of metallicity bins in their Milky Way model to be the main driver behind the discrepancy between our results.

\paragraph{\citet{Breivik+2020}} introduced the population synthesis code COSMIC and presented detections for many different DCO types in LISA using this code. They find that LISA will detect 93 BHBHs, 33 BHNSs and 8 NSNSs in the Milky Way over a 4 year mission. \citet{Breivik+2020} make many physical assumptions that differ from our fiducial model, the most notable being that they assume the optimistic CE scenario and that case BB mass transfer is always unstable, whilst also using a simpler model for the Milky Way (see Appendix~\ref{app:mw_changes}). Thus for better comparison we ran our simulation using model \modCaseBBOpt{} and the Milky Way model from \citet{Breivik+2020}. This results in 97, 101 and 43 detections for BHBHs, BHNSs and NSNSs respectively. Therefore, though we are in very good agreement for BHBHs, we predict much higher rates for BHNSs and NSNSs. These differences are likely due to using a different population synthesis code (COSMIC), which has different underlying physics assumptions from COMPAS. Given our strong agreement for BHBHs, it is possible that COSMIC and COMPAS handle NSs differently and so lead to different detection rates for DCOs containing NSs. However checking this would require a more in-depth study of the intrinsic formation rate of DCOs containing NSs in the two codes.

\paragraph{\citet{Shao+2021}} most recently investigated the detectability binaries containing BHs in LISA using BSE and a relatively simple model for the Milky Way (assuming a uniform flat disc, constant star formation and a single metallicity). They assume that kicks for NSs formed through ECSN are slightly higher than our work ($50 \unit{km}{s^{-1}}$ instead of $30 \unit{km}{s^{-1}}$). This may account for their particularly low BHNS rate (as the binaries would be more likely to disrupt), which is a factor of 20 lower than ours, but we expect their assumption of the optimistic CE scenario, reduced Wolf-Rayet winds and lower SNR detection threshold could partially offset this. As we show in Appendix~\ref{app:mw_changes}, their use of a simpler Milky Way model, especially with only a single metallicity, would lead to an underestimate of the BHBH and BHNS rates, which may explain the discrepancy in our results.\\

\noindent{}Overall, since the work of \citet{Nelemans+2001}, in addition to the LISA mission specifications, the methods that we use to simulate binaries and the Milky Way have all changed significantly. We now predict that LISA detections of these massive DCOs are dominated by BHBHs, rather than NSNSs, whilst the absolute detection rates for BHBHs and BHNSs are much higher. Further studies in this area could improve on this work by including the effects of systemic kicks on the position of systems in the Milky Way and accounting for contributions from other formation channels.
\section{Conclusion \& Summary} \label{sec:conclusion}
We provide predictions for the detection rate and population properties of LISA detectable BHBH, BHNS and NSNS.
We use a novel empirically-informed analytical model for the metallicity dependent star formation history of the Milky Way, calibrated against the APOGEE stellar spectroscopic survey. We use this to model Monte-Carlo realisations of the present-day BHBH, BHNS and NSNS populations in our Milky Way. 
For the binary population, we use the results of a large grid of simulations performed with the rapid population synthesis code COMPAS \citefloorp{}. These simulations have been optimised with the adaptive sampling algorithm STROOPWAFEL to preferentially sample NS and BH binaries. In total these comprise over two billion massive binaries that span 20 physics variations, which represent the most common uncertainties in binary physics.
To determine the detectability of sources with LISA we use the LEGWORK package \citep{Wagg+2021}, that was specifically developed for this purpose and is publicly available. 
We investigate the results expected for a 4- and 10-year LISA mission. Our main conclusions can be summarised as:
\begin{enumerate}
    \item \textbf{Total detections:} We predict \rangeFourYear{} detections in a 4-year LISA mission, across all our simulations for varying physics assumptions. This increases to about \rangeTenYear{} for a 10-year LISA mission. Although the number of detections per type can vary by about 2 orders of magnitude, we find that the total detection rate is fairly robust, among the variations we have considered (see Table~\ref{tab:detection_rates}).
    
    \item \textbf{Detections by type:} For our fiducial model, we predict a total of $124 \pm 11$ detections and out of these we find about $\BHBHFourYear{}\pm 9$ BHBHs, $\BHNSFourYear{}\pm 6$ BHNSs and $\NSNSFourYear{}\pm 3$ NSNSs. The errors quoted here are the $1$-$\sigma$ Poisson uncertainties resulting from the random initialisation of the Milky Way (see Table~\ref{tab:detection_rates}).
    
    \item \textbf{Physics variations:} Among the model variations we consider, we find that our predictions for the rates for the different DCO types are robust within a factor of 2 of the fiducial rate, with the following exceptions. For BHBHs, the rate is most sensitive to the treatment of common envelope phases or an increase of the WR wind mass loss. For BHNSs and NSNSs, the assumptions regarding the assumed common envelope ejection efficiency, treatment of case BB mass transfer and the kicks are most important. In addition, the assumed mass transfer efficiency impacts the BHNS (see Fig.~\ref{fig:detection_rates}).
    
    \item \textbf{Probing the black hole mass distribution and the lower mass gap:} We expect LISA to predominantly detect lower mass BHs (with 90\% of BHBH and BHNSs having BH masses lower than $11 \unit{M_\odot}$ in our fiducial simulations) in stark contrast to current ground-based detectors which are heavily biased towards high mass systems. For our fiducial simulation, we predict that approximately 69 systems with a component with a mass between $2.5$-$5 \unit{M_\odot}$ would be detected by a 4-year LISA. This implies that LISA can potentially make important contributions to the debate about the existence of a lower mass gap (see Fig.~\ref{fig:lower_mass_gap_variation}).
    
    \item \textbf{Eccentricity distribution:} We find that for all DCO types a large fraction of detectable systems still have nonzero eccentricities ($e > 0.01$) when entering the LISA band, which can be used to distinguish them from the more numerous WDWD binaries, which are largely expected to be circular. In particular, for our fiducial model, we find that this is the case for around three quarters of detectable binaries. Furthermore, around 16\% of detectable binaries have eccentricities that are so high ($e > 0.3$) that the emission at frequencies corresponding to higher order harmonics start to dominate (see Fig.~\ref{fig:fiducial_pdf_distributions}).
    
    \item \textbf{Distinguishing from WDWD sources:} For about half of all detections we expect that we will be able to confidently determine the type of compact objects involved and this increases to 60\% for a 10-year LISA mission (see Sec.~\ref{sec:WDWD_distinguish}).
    
    \item \textbf{Chirp mass determinations:} For about 10\% of systems we expect to be able to determine the chirp mass better than 10\% and this increases to 15\% for a 10-year LISA mission (see Fig.~\ref{fig:m_c_unc}).

    \item \textbf{Prospects for finding EM counterparts:} We expect about 13\% of detections with a sky localisation better than 1 degree for our fiducial model (though the fraction remains roughly constant among model variations). This fraction remains the same for a 10-year LISA mission, meaning that the number increases proportionally. This will be of interest for electromagnetic searches for counterparts, in particular for radio pulsar searches with SKA (see Fig.~\ref{sec:pulsar_matching}).
 
   \item \textbf{Benefits of extending the LISA mission:} The number of detections scale approximately as $\sqrt{T_{\rm obs}}$, where $T_{\rm obs}$ is the mission length. Therefore, extending the LISA mission from 4- to 10-years increases the number of detections by about 60\% for each model variation. A further important benefit is the improvement of the characterisation of the sources, since the relative error on the frequency derivative (which dominates the relative error in the chirp mass) scales as $T_{\rm obs}^{-2.5}$ for stationary sources (Eq.~\ref{eq:f_orb_dot_unc}). We find that the number of systems with chirp masses that can be measured better than 10\% increases by a factor of 2.4 for each model variation.
   In addition, the number of systems with a sky localisation better than one degree increases by a factor of 1.5. Overall, the number of sources that can be unambiguously distinguished from WDWDs increases by almost a factor of 2 (see Section~\ref{sec:WDWD_distinguish}).
\end{enumerate}
\acknowledgments{We thank Katie Brevik, Will Farr, Rob Farmer, Valeriya Korol, Floris Kummer, Eva Laplace, Mike Lau, Tyson Littenberg, Ilya Mandel, Mathieu Renzo, Alberto Sesana, Katie Sharpe, Simon Stevenson, the CCA GW group and COMPAS collaboration for insightful discussions. TW also thanks Terence Lovatt for his advice and help on improving an earlier draft of this project. Portions of this study were inspired by the LISA Sprint at the Center for Computational Astrophysics of the Flatiron Institute, supported by the Simons Foundation. This project was funded in part by the National Science Foundation under Grant No. (NSF grant number 2009131), the European Union’s Horizon 2020 research and innovation program from the European Research Council (ERC, Grant agreement No. 715063), and by the Netherlands Organization for Scientific Research (NWO) as part of the Vidi research program BinWaves with project number 639.042.728. We further acknowledge the Black Hole Initiative funded by a generous contribution of the John Templeton Foundation and the Gordon and Betty Moore Foundation.}

\software{We used LEGWORK to evolve sources over time and calculate signal-to-noise ratios \citep{Wagg+2021}. It is freely available at \url{https://legwork.readthedocs.io/en/latest/}. Simulations in this paper made use of the COMPAS rapid binary population synthesis code \citep{COMPAS:2021methodsPaper}. The simulations performed in this work were simulated with a COMPAS version that predates the publicly available code. Our version is most similar to v02.13.01 of the publicly available COMPAS code. Requests for the original code can be made to Floor Broekgaarden. The authors used {\sc{STROOPWAFEL}} from \citep{Broekgaarden+2019}, publicly available at \url{https://github.com/FloorBroekgaarden/STROOPWAFEL}\footnote{For the latest pip installable version of STROOPWAFEL please contact Floor Broekgaarden.}.
The authors also made use of Python (v3.8), available at \url{http://www.python.org}. In addition the following Python packages were used: \texttt{matplotlib} \citep{Hunter+2007}, \texttt{NumPy} \citep{2020NumPy-Array}, \texttt{Astropy} \citep{AstropyCollaboration+2013,AstropyCollaboration+2018}, \texttt{Seaborn} \citep{Waskom+2021}, \texttt{SciPy} \citep{2020SciPy-NMeth}, \texttt{h5py} \citep{Collette+2021} and \texttt{Jupyter Lab} \citep{Kluyver+2016}. This research has made use of NASA’s Astrophysics Data System Bibliographic Services. We also made use of the computational facilities from the Harvard FAS Research Computing cluster.}

\bibliographystyle{aasjournal}
\bibliography{paper}

\restartappendixnumbering

\allowdisplaybreaks
\appendix
\section{Population Synthesis}\label{app:pop_synth}

In this section we summarise the main assumptions and settings from the population synthesis simulation from Broekgaarden et al.\ (2021a, 2021b).

\subsection{Initial conditions}

Broekgaarden et al.\ (2021a, 2021b) simulate between 1 and 100 million massive binaries for each of 50 metallicities equally spaced in log space between $Z \in [0.0001, 0.022]$, where $Z$ is the mass fraction of heavy elements. They simulate more binaries for higher metallicities so that large enough sample of DCOs at each metallicity (since DCOs are formed at a lower rate at higher metallicities). These metallicities span the allowed metallicity range for the original fitting formulae on which COMPAS is based \citep{Hurley+2000}. This is repeated for \nMinusOneModels{} physics variations (see Section \ref{sec:variation_assumptions}) and so in total over two billion binaries were simulated.

Each binary is sampled from initial distributions for the primary and secondary masses as well as the separation. The primary mass, that is the mass of the initially more massive star, is restricted to $m_1 \in [5, 150] \unit{M_{\odot}}$, which spans the range of interest for NS and BH formation in binary systems, and drawn from the \citet{Kroupa+2001} initial mass function (IMF), $p(m_1) \propto m_1^{-2.3}$. The secondary mass, $m_2$, is drawn using the initial mass ratio of the binary, $q \equiv m_2 / m_1$, which Broekgaarden et al.\ (2021a, 2021b) assume to be uniform on $[0, 1]$, therefore $p(q) = 1$ \citep[e.g.\ consistent with][]{Sana+2012}. They additionally restrict the secondary masses $m_2 \ge 0.1 \unit{M_{\odot}}$, which is approximately the minimum mass for a main sequence star. They assume that the initial separation follows a flat in the log distribution with $p(a_i) \propto 1 / a_i$ and $a_i \in [0.01, 1000] \unit{AU}$ \citep{Opik+1924, Abt+1983}. They assume that all binary orbits are circular at birth to reduce the dimensions of initial parameters. Since they focus on post-interaction binaries which will have circularised after mass transfer they argue this is an reasonable assumption (as many studies have in the past) and is likely not critical for predicting detection rates \citep{Hurley+2002, deMink+2015}.

Broekgaarden et al.\ (2021a, 2021b) apply the adaptive importance sampling algorithm STROOPWAFEL \citep{Broekgaarden+2019} to improve the yield of their sample. This algorithm increases the prevalence of target DCOs (BHBHs, BHNSs and NSNSs in this case) in the sample and assigns each a weight, $w$, which represents the probability of drawing the DCO without STROOPWAFEL in effect.

\subsection{Physical assumptions in the fiducial model}\label{app:fiducial_physics}

\textit{Stellar Evolution:} To follow the evolution of massive stars, COMPAS relies on fitting formulae by \citet{Hurley+2000} to detailed single star models by \citet{Pols+1998}. COMPAS models the evolution of stars that lose or gain mass closely following the algorithms originally described in \citet{Tout+1996} and \citet{Hurley+2002}.

\textit{Wind mass loss:} Broekgaarden et al.\ (2021a, 2021b) follow the wind prescription from \citet{Belczynski+2008}, which was based on results from Monte Carlo radiative transfer simulation of \citet{Vink+2000, Vink+2001}. They use the wind mass loss rates from \citet{Vink+2001} for stars above $12500 \unit{K}$ and the rates from \citet{Hurley+2000} for cooler stars. Additionally, they use a separate, higher wind mass loss rate for luminous blue variable (LBV) stars, following \citet{Belczynski+2008}, to mimic observed LBV eruptions for stars with luminosities and effective temperatures above the Humphreys-Davidson limit. They use the Wolf-Rayet-like mass loss rate from \citet{Hamann+1998} with an additional metallicity scaling from \citet{Vink+2005} for helium stars, and set $f_{\rm WR} = 1$. See \citet{COMPAS:2021methodsPaper}, Section 3 for the explicit equations.

\textit{Mass Transfer:} In determining the stability of mass transfer Broekgaarden et al.\ (2021a, 2021b) use the $\zeta$-prescription, which compares the radial response of the star with the response of the Roche lobe radius to the mass transfer \citep[e.g.][]{Hjellming+1987}. The mass transfer efficiency, $\beta \equiv \Delta M_{\rm acc} / \Delta M_{\rm don}$, is the fraction of the mass transferred by the donor that is actually accreted by the accretor. They limit the maximum accretion rate for stars to $\Delta M_{\rm acc} / \Delta t \le 10 M_{\rm acc} / \tau_{\rm KH}$, where $\tau_{\rm KH}$ is the Kelvin-Helmholtz timescale of the star \citep{Paczynski+1972, Hurley+2002}. The maximum accretion rate for compact objects is limited to the Eddington accretion rate. If more mass than these rates is accreted then they assume that the excess is lost through isotropic re-emission in the vicinity of the accreting star \citep[e.g.][]{Massevitch+1975, Soberman+1997}. They assume that all mass transfer from a stripped post-helium-burning-star (case BB) onto a neutron star or black hole is unstable \citep{Tauris+2015}.

\textit{Common-Envelope:} A common-envelope phase follows dynamically unstable mass transfer and Broekgaarden et al.\ (2021a, 2021b) parameterise this using the $\alpha$-$\lambda$ prescription from \citet{Webbink+1984} and \citet{deKool+1990}. They assume $\alpha = 1$, such that all of the gravitational binding energy is available for the ejection of the envelope. For $\lambda$ they use the fitting formulae from \citet{Xu+2010, Xu+2010a}. They assume that any Hertzsprung gap donor stars that initiate a common-envelope phase will not survive this phase due to a lack of a steep density gradient between the core and envelope \citep{Taam+2000, Ivanova+2004, Klencki+2021}. This follows the `pessimistic' common-envelope scenario \citep[c.f.][]{Belczynski+2007}. They remove any binaries where the secondary immediately fills its Roche lobe upon the conclusion of the common-envelope phase as they treat these as failed common-envelope ejections, likely leading to a stellar merger.

\textit{Supernovae:} Broekgaarden et al.\ (2021a, 2021b) draw the remnant masses and natal kick magnitudes from different distributions depending on the type of supernova that occurs. For stars undergoing a general core-collapse supernova, they use the \textit{delayed} supernova remnant mass prescription from \citet{Fryer+2012}. The \textit{delayed} prescription does not reproduce a neutron star black hole mass gap and they use this as their default as it has been shown to provide a better fit for observed populations of DCOs \citep[e.g.][]{Vigna-Gomez+2018}. They draw the natal kick magnitudes from a Maxwellian velocity distribution with a one-dimensional root-mean-square velocity dispersion of $\sigma_{\rm rms}^{\rm 1D} = 265 \unit{km}{s^{-1}}$ \citep{Lyne+1994, Hobbs+2005}. They assume that stars with helium core masses between $1.6$--$2.25 \unit{M_{\odot}}$ \citep{Hurley+2002} experience electron-capture supernovae (ECSN) \citep{Nomoto+1984, Nomoto+1987, Ivanova+2008}. They set all remnant masses to $1.26 \unit{M_{\odot}}$ in this case as an approximation of the solution to Equation 8 of \citet{Timmes+1996}. For these supernovae, they set $\sigma_{\rm rms}^{\rm 1D} = 30 \unit{km}{s^{-1}}$ \citep[e.g.][]{Pfahl+2002, Podsiadlowski+2004}. They assume that stars that undergo case BB mass transfer \citep{Dewi+2002} experience extreme stripping which leads to an ultra-stripped supernova \citep{Tauris+2013, Tauris+2015}. For these supernovae they calculate the remnant mass using the \citet{Fryer+2012} prescription and use $\sigma_{\rm rms}^{\rm 1D} = 30 \unit{km}{s^{-1}}$ (as with ECSN). Stars with final helium core masses between $35$-$135 \unit{M_{\odot}}$ are presumed to undergo a pair-instability, or pulsational pair-instability supernova \citep[e.g.][]{Woosley+2007, Farmer+2019}. They follow the prescription from \citet{Marchant+2019} as implemented in \citep{Stevenson+2019} for these supernovae. They assume that kicks are isotropic in the frame of the collapsing star. They adopt a maximum neutron star mass of $2.5 \unit{M_{\odot}}$ \citep[e.g.][]{Kalogera+1996, Fryer+2015, Margalit+2017} for the fiducial model and change the \citet{Fryer+2012} prescription accordingly.

\subsection{Model variations} \label{sec:variation_assumptions}
In addition to their fiducial model for the formation of DCOs, Broekgaarden et al.\ (2021a, 2021b) explore \nMinusOneModels{} other models in which they change various aspects of the mass transfer, common-envelope, supernova and wind mass loss physics assumptions in order to assess the effect of their uncertainties on the overall double compact object detection rates and distributions. Each of the models varies a single physics assumption (fiducial assumptions are outlined in Section~\ref{app:fiducial_physics}) and these models are outlined in Table~\ref{tab:physics_variations}.

Their fiducial model is labelled model \modFid{}. Models \modRangeMT{} focus on changes to the mass transfer physics assumptions. They explore the effect of fixing the mass transfer efficiency $\beta$ to a constant value, rather than allowing it to vary based on the maximum accretion rate. In models \modBetaLow{}, \modBetaMed{}, \modBetaHigh{}, in which they set the value of $\beta$ to $0.25$, $0.5$ and $0.75$ respectively.

Models \modRangeCE{} focus on altering the common-envelope physics. In model \modCaseBB{} we modify model E from Broekgaarden et al.\ (2021a, 2021b) to investigate the consequence of assuming that case BB mass transfer is always unstable, whilst allowing Helium HG donors to survive CE events. They change the common-envelope efficiency parameter to $\alpha_{\rm CE} = 0.1, 0.5, 2.0, 10.0$ in models \modAlphaLowest{}, \modAlphaLow{}, \modAlphaHigh{} and \modAlphaHighest{} respectively. In model \modOpt{}, they relax their restriction that Hertzsprung gap donor stars cannot survive common-envelope events, thereby following the `optimistic' common-envelope scenario. They combine this with model \modCaseBB{} in model \modCaseBBOpt{}.

In models \modRangeSN{} they consider changes related to their assumptions about supernova physics. Model \modRapid{} uses the alternate \textit{rapid} remnant mass prescription from \citet{Fryer+2012} instead of the \textit{delayed} prescription. They change the maximum neutron star mass in models \modNSLow{} and \modNSHigh{} to $2$ and $3 \unit{M_{\odot}}$ respectively to account for the range of predicted maximum neutron star masses. Model \modNoPISN{} removes the implementation of pair-instability and pulsational pair-instability supernovae. In models \modSigLow{} and \modSigLower{} they decrease the root-mean-square velocity dispersion for core-collapse supernovae to explore the effect of lower kicks. Model \modNoBH{} removes the natal kick for all black holes.

Finally, in models \modRangeML{} Broekgaarden et al.\ (2021a, 2021b) investigate the effect of changing their assumption about wind mass loss rates, specifically for Wolf-Rayet winds. They vary $f_{\rm WR}$ to $0.1$ and $5.0$ in models \modWRLow{} and \modWRHigh{} respectively. These values approximately span the current range of possible Wolf-Rayet wind efficiencies suggested from observations \citep[e.g.][]{Vink+2017, Hamann+2019, Shenar+2019, Miller-Jones+2021, vanSon+2021}.

\begin{table}[htb]
    \centering
    \begin{tabular}{cl}
        \hline \hline
        Model & Physics Variation \\
        \hline \hline
        \modFid & Fiducial (see Section~\ref{app:fiducial_physics}) \\
        \hline
        \modBetaLow & Fixed mass transfer efficiency of $\beta=0.25$ \\ 
        \modBetaMed & Fixed mass transfer efficiency of $\beta=0.5$  \\ 
        \modBetaHigh & Fixed mass transfer efficiency of $\beta=0.75$ \\
        \hline
        \modCaseBB & Case BB mass transfer always unstable \\
        \modCaseBBOpt & Case BB always unstable + Optimistic CE \\
        \modAlphaLowest & CE efficiency parameter $\alpha = 0.1$ \\
        \modAlphaLow & CE efficiency parameter $\alpha = 0.5$ \\
        \modAlphaHigh & CE efficiency parameter $\alpha = 2$   \\
        \modAlphaHighest & CE efficiency parameter $\alpha = 10$   \\
        \modOpt & HG donor stars initiating a CE survive CE \\
        \hline
        \modRapid & Fryer rapid SN remnant mass prescription \\
        \modNSLow & Maximum NS mass is fixed to $2\unit{M_{\rm \odot}}$ \\
        \modNSHigh & Maximum NS mass is fixed to $3\unit{M_{\rm \odot}}$ \\
        \modNoPISN & PISN and pulsational-PISN not implemented \\
        \modSigLow & $\sigma_{\rm{rms}}^{\rm{1D}}=100 \unit{km}{s^{-1}}$ for core-collapse supernova \\  
        \modSigLower & $\sigma_{\rm{rms}}^{\rm{1D}}=30  \unit{km}{s^{-1}}$ for core-collapse supernova \\ 
        \modNoBH & Black holes receive no natal kick \\
        \hline
        \modWRLow & Wolf-Rayet wind factor $f_{\rm WR} = 0.1$ \\
        \modWRHigh & Wolf-Rayet wind factor $f_{\rm WR} = 5.0$ \\
        \hline \hline
    \end{tabular}%
    \caption{A description of the \nModels{} binary population synthesis models used in this study. \modFid{} is the fiducial model, \modRangeMT{} change mass transfer physics, \modRangeCE{} change common-envelope physics , \modRangeSN{} change supernova physics and \modRangeML{} change wind mass loss \citep[c.f.][Table 2]{Broekgaarden+2021}.}
    \label{tab:physics_variations}
\end{table}

\section{Detection Rate Normalisation}\label{app:rate_normalisation}
In this section we explain the normalisation process that we refer to in Section~\ref{sec:gw_detection}. From each simulated instance of the Milky Way we extract the fraction of targets that are detectable, where we define a target as one of BHBH, BHNS or NSNS that merges in a Hubble time. To convert the detectable fraction to a detection rate for the Milky Way, we write that the \textit{number} of detectable targets in the Milky Way is
\begin{equation}\label{eq:norm_frame}
    N_{\rm detect} = f_{\rm detect} \cdot N_{\rm target, MW},
\end{equation}
where $f_{\rm detect}$ is the fraction of targets in the instance that were detectable and $N_{\rm target, MW}$ is the total number of targets that have been formed in the Milky Way's history. We can further break this total down into
\begin{equation}
    N_{\rm target, MW} = \avg{ \mathcal{R}_{\rm target} } \cdot M_{\rm SF, MW},
\end{equation}
where $\avg{ \mathcal{R}_{\rm target} }$ is the average number of targets formed per star forming mass and $M_{\rm SF, MW}$ is the star forming mass of the Milky Way, meaning the total mass of every star ever formed in the Milky Way.

\subsection{Average target formation rate}
Double compact object formation is metallicity dependent, so we find the average rate as the integral over metallicity, which is given by
\begin{equation}\label{eq:norm_avg_target_formation}
    \avg{ \mathcal{R}_{\rm target} } = \int_{Z_{\rm min}}^{Z_{\rm max}} p_{Z} \mathcal{R}_{\rm target, Z} \dd{Z},
\end{equation}
where $Z_{\rm min}, Z_{\rm max}$ are the minimum and maximum sampled metallicities, $p_Z$ is the probability of forming a star at the metallicity $Z$ (which can be found using the distribution in \citealp{Frankel+2018}) and $\mathcal{R}_{\rm target, Z}$ is the number of targets formed per star forming mass,
\begin{equation}
    \mathcal{R}_{\rm target, Z} =  \frac{N_{\rm target, Z}}{ M_{\rm SF, Z} }.
\end{equation}
In practice, this integral is instead approximated as a sum over the metallicity bins that we use in our simulation. The number of targets in our sample at a metallicity $Z$, $N_{\rm target, Z}$, can be written simply as the sum of the targets' weights:
\begin{equation}
    N_{\rm target, Z} = \sum_{i=1}^{N_{\rm binaries, Z}} w_i \theta_{\rm target, i},
\end{equation}
where $w_i$ is the binary's adaptive importance sampling weight assigned, $N_{\rm binaries, Z}$ is the number of binaries at metallicity $Z$ in our sample and $\theta_{\rm target, i}$ is only $1$ when the binary is a target and otherwise $0$.

The total star forming mass at a metallicity $Z$, $M_{\rm SF, Z}$, can be written as
\begin{equation}
    M_{\rm SF, Z} = \frac{\avg{m}_{\rm COMPAS, Z}}{f_{\rm trunc}} N_{\rm binaries, Z},
\end{equation}
where $\avg{m}_{\rm COMPAS}$ is the average star forming mass of a binary in a simulation using our cutoffs (discussed in Section~\ref{sec:COMPAS_explained}) and $f_{\rm trunc}$ is the fraction of the total stellar mass from which our COMPAS simulations sample, given our truncated mass and separation ranges (see Section~\ref{sec:COMPAS_explained}). These truncations mean that only $f_{\rm trunc} \approx 0.17$ of the stellar mass in the galaxy is sampled from.

\subsection{Total star forming mass in the Milky Way}
It is important to distinguish between the \textit{total} mass of every star formed over the entire history of the Milky Way and the \textit{current} stellar mass in the Milky Way. Many stars born in the Milky Way are no longer living and have lost much of their mass to stellar winds and supernovae, thus the current stellar mass in the Milky Way is an underestimate of the total star forming mass.

\citet{Licquia+2015} find that the total stellar mass today in the Milky Way is $6.08 \pm 1.14 \times 10^{10} \unit{M_{\odot}}$. This total includes all stars and stellar remnants (white dwarfs, neutrons stars and black holes) but \textit{excludes} brown dwarfs. We can write that the total mass of every star every formed in the Milky Way is
\begin{equation}\label{eq:m_SF_MW}
    M_{\rm SF, MW} = (6.08 \pm 1.14) \times 10^{10} \unit{M_{\rm \odot}} \cdot \frac{\avg{m}_{\rm SF, total}}{\avg{m}_{\rm SF, today}},
\end{equation}
where $\avg{m}_{\rm SF, total}$ is the average mass of a star over the history of the Milky Way and is defined as
\begin{equation}
    \avg{m}_{\rm SF, total} = \int_{0}^{t_{\rm MW}} p_{\rm birth}(\tau) \int_{0.01}^{200} \zeta(m)\ m \dd{m} \dd{\tau},
\end{equation}
where $t_{\rm MW}$ is the age of the Milky Way, $\zeta(m)$ is the \citet{Kroupa+2001} IMF function and $p_{\rm birth}(\tau)$ is the probability of a star being formed at a lookback time $\tau$ (Eq.~\ref{eq:thin_disc_tau}). $\avg{m}_{\rm SF, today}$ is the average mass of all stars and stellar remnants (excluding brown dwarfs) present in the Milky Way today is defined as follows (note that we integrate from $0.08$ not $0.01$ since observations of today's Milky Way mass exclude brown dwarfs)
\begin{equation}
    \avg{m}_{\rm SF, today} = \int_{0}^{t_{\rm MW}} p_{\rm birth}(\tau) \int_{0.08}^{200} \zeta(m)\ m_{\rm today} \dd{m} \dd{\tau},
\end{equation}
where $m_{\rm today}(m, Z, \tau)$ is the current mass of a star that was formed $\tau$ years ago at a metallicity $Z$. We calculate $m_{\rm today}(m, Z, \tau)$ by interpolating the final masses given by COMPAS for a grid of single stars over different masses and metallicities using the \citet{Fryer+2012} delayed prescription and default wind mass loss settings. For $Z$, we use the average star forming metallicity in the Milky Way at a lookback time $\tau$ using our galaxy model. Evaluating Equation~\ref{eq:m_SF_MW}, we find that the total mass of every star that has ever formed in the Milky Way is
\begin{align}
    M_{\rm SF, MW} &= (6.1 \pm 1.1) \times 10^{10} \unit{M_{\odot}} \cdot \frac{0.378 \unit{M_{\odot}}}{0.221 \unit{M_{\odot}}}, \nonumber \\
    &= (10.4 \pm 1.1) \times 10^{10} \unit{M_{\odot}},
\end{align}
an increase of approximately 70\% from the value still in stars today!

\subsection{Normalisation summary}
Finally, we can substitute Equations~\ref{eq:norm_avg_target_formation} and \ref{eq:m_SF_MW} into \ref{eq:norm_frame} and write that the overall normalisation of the detection rate is calculated as
\begin{align}
    N_{\rm detect} &= f_{\rm detect} \cdot 10.4 \times 10^{10} \unit{M_{\odot}} \nonumber \\
    &\times \sum_{Z=Z_{\rm min}}^{Z_{\rm max}} p_{Z} \qty(\sum_{i=1}^{N_{\rm binaries, Z}} w_i \theta_{\rm target, i}) \nonumber \\
    &\times \qty(\frac{\avg{m}_{\rm COMPAS, Z}}{f_{\rm trunc}} \sum_{i=1}^{N_{\rm binaries, Z}} w_i)^{-1}.
\end{align}

\section{Calculation of the uncertainties in the chirp mass for detectable sources}\label{app:chirp_mass_uncertainty}

How accurately the chirp mass of a detected binary can be determined depends on the signal to noise ratio, duration of the mission, its orbital frequency and the time derivative of the orbital frequency. 

Here we describe how we estimate the uncertainty of the chirp mass . First, consider the chirp mass, which can be expressed as
\begin{equation}
    \mathcal{M}_c = \frac{c^3}{G} \left( \frac{5 \pi}{48 n} \frac{\dot{f}_{n}}{F(e)} \right)^{3/5} \frac{1}{(2 \pi f_{\rm orb})^{11/5}},
\end{equation}
where ${f}_{n}$ is the frequency of the n-th harmonic,  $f_{\rm orb}$ is the orbital frequency, $\mathcal{M}_{c}$ is the chirp mass (defined in Eq.~\ref{eq:chirp_mass}), $e$ is the eccentricity and
\begin{equation}\label{eq:peters_f}
    F(e) = \frac{1 + \frac{73}{24} e^2 + \frac{37}{96} e^4}{(1 - e^2)^{7/2}},
\end{equation}
is the enhancement factor of gravitational wave emission for an eccentric binary over an otherwise identical circular binary \citep[][Eq.~17]{Peters+1963}. 
In practice, we will use the dominating harmonic, with $n=n_{\rm dom}$ and $f_n = n_{\rm dom} f_{\rm orb} = f_{\rm dom}$. The dominating harmonic for circular binaries is $n_{\rm dom} = 2$ and so the dominating frequency is twice the orbital frequency. 

Therefore the chirp mass uncertainty can be estimated as
\begin{equation}\label{eq:chirp_mass_uncertainty}
    \frac{\Delta \mathcal{M}_c}{\mathcal{M}_c} = \frac{11}{5} \frac{\Delta f_{\rm orb}}{f_{\rm orb}} + \frac{3}{5} \frac{\Delta \dot{f}_{\rm dom}}{\dot{f}_{\rm dom}} + \frac{3}{5} \frac{\Delta F(e)}{F(e)},
\end{equation}
%where $f_{\rm dom}$ is the harmonic frequency with the strongest SNR ($f_{\rm dom} = n_{\rm dom} f_{\rm orb}$ and $n_{\rm dom} = 2$ for circular binaries) as this will provide the best measurement.

We estimate the frequency uncertainties using \citet{Takahashi+2002}, such that
\begin{align}\label{eq:f_orb_unc}
    \frac{\Delta f_{\rm orb}}{f_{\rm orb}} &= 4 \sqrt{3} \cdot \frac{1}{\rho} \frac{1}{T_{\rm obs}} \frac{1}{f_{\rm orb}}, \\
    \frac{\Delta \dot{f}_{\rm dom}}{\dot{f}_{\rm dom}} &= 6 \sqrt{5} \cdot \frac{1}{\rho} \left(\frac{1}{T_{\rm obs}} \right)^2 \frac{1}{\dot{f}_{\rm dom}},\label{eq:f_orb_dot_unc}
\end{align}
where $\rho$ is the signal-to-noise ratio and $T_{\rm obs}$ is the LISA mission length. We estimate the eccentricity certainty, $\Delta e$, following the methods of \citet{Lau+2020} and \citet{Korol+2021}, which use the relative SNRs of different harmonics to work out the eccentricity. We propagate this uncertainty such that
\begin{equation}
    \frac{\Delta F(e)}{F(e)} = \Delta e \cdot \frac{(1256 + 1608 e^2 + 111 e^4) e}{96 + 196 e^2 - 255 e^4 - 37 e^6}.
\end{equation}

We use Eq.~\ref{eq:chirp_mass_uncertainty} to calculate the chirp mass uncertainty for each DCO type in our sample and plot it in Fig.~\ref{fig:m_c_unc}.

\section{Assessing the impact of Milky Way model choices}\label{app:mw_changes}
The model that we use for the Milky Way adds several layers of complexity, accounting for the inside-out growth of the thin disc, using empirically informed star formation histories that are a function of time and assigning metallicities based on the position and age of binaries. In this section, we repeat our main analysis but instead apply a simpler model for the Milky Way in order to assess the effect of these added features. For this purpose, we use model for the Milky Way used in \citet{Breivik+2020} as this is representative of the models used in most previous works.

Their model can be summarised as follows: the Milky Way is assumed to comprise of three components, a thin disc, a thick disc and a bulge. The spatial distributions and relative masses for these components are given in \citet{McMillan+2011}. \citet{Breivik+2020} assume constant star formation over 10 Gyr for the thin disc, a 1 Gyr burst of star formation 11 Gyr ago for the thick disc and a 1 Gyr burst of star formation 10 Gyr ago for the bulge. A major difference is that only two metallicities are used and they are assigned to binaries independent of age or position. Binaries formed in the thin disc and bulge are assumed to have a metallicity of $Z = 0.02$ and those formed in the thick disc are assumed to have $Z = 0.003$.

\begin{figure}[htb]
    \centering
    \includegraphics[width=\columnwidth]{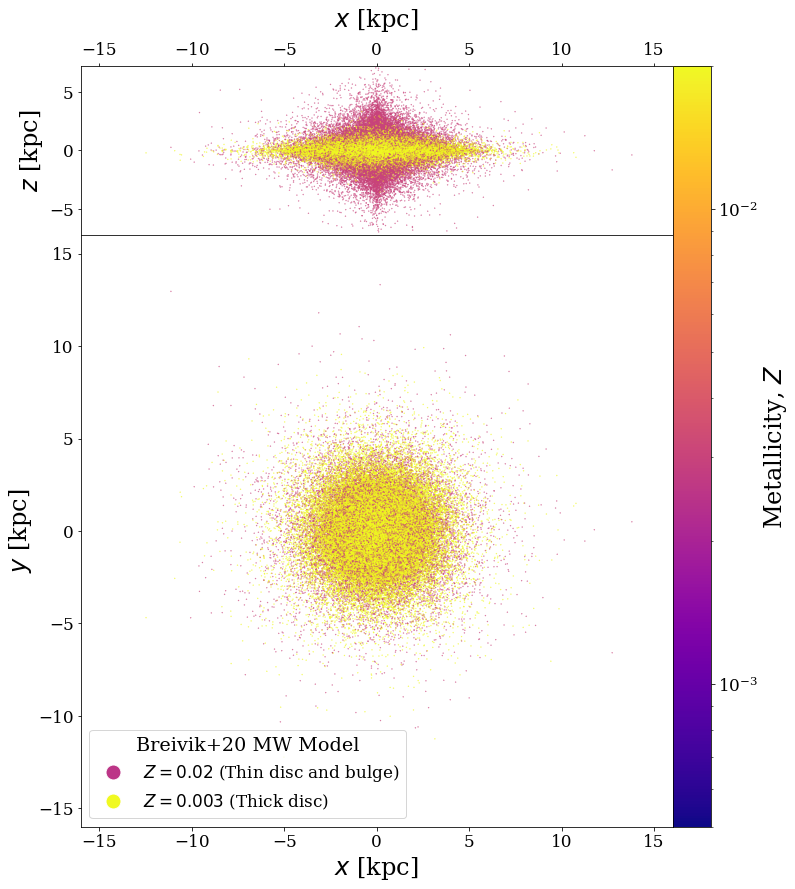}
    \caption{As Fig.~\ref{fig:galaxy_schematic} (right panel), but for the Milky Way model used in \citet{Breivik+2020}.  \href{https://github.com/TomWagg/detecting-DCOs-in-LISA/blob/main/paper/figures/figD1_random_simple_galaxy.png}{\faFileImage} \href{https://github.com/TomWagg/detecting-DCOs-in-LISA/blob/main/paper/figure_notebooks/galaxy_creation_station.ipynb}{\faBook}.}
    \label{fig:simple_mw}
\end{figure}

We show the spatial metallicity distribution for this model in Fig.~\ref{fig:simple_mw} in the same form as Fig.~\ref{fig:galaxy_schematic} for ease of comparison between our models. The two main differences we can see between Fig.~\ref{fig:galaxy_schematic} and \ref{fig:simple_mw} are that the \citet{Breivik+2020} model is more centrally concentrated and only has two fixed metallicity populations.

When applying this simpler Milky Way model in combination with our fiducial binary physics assumptions (model \modFid{}), we find that the expected number of detections for BHBHs, BHNSs and NSNSs for a 4-year LISA mission is $52$, $25$ and $17$ respectively. Thus the BHBH detection has decreased slightly compared to our main findings, whilst for BHNSs and NSNSs the rate has approximately halved and doubled respectively.

\begin{figure}[htb]
    \centering
    \includegraphics[width=\columnwidth]{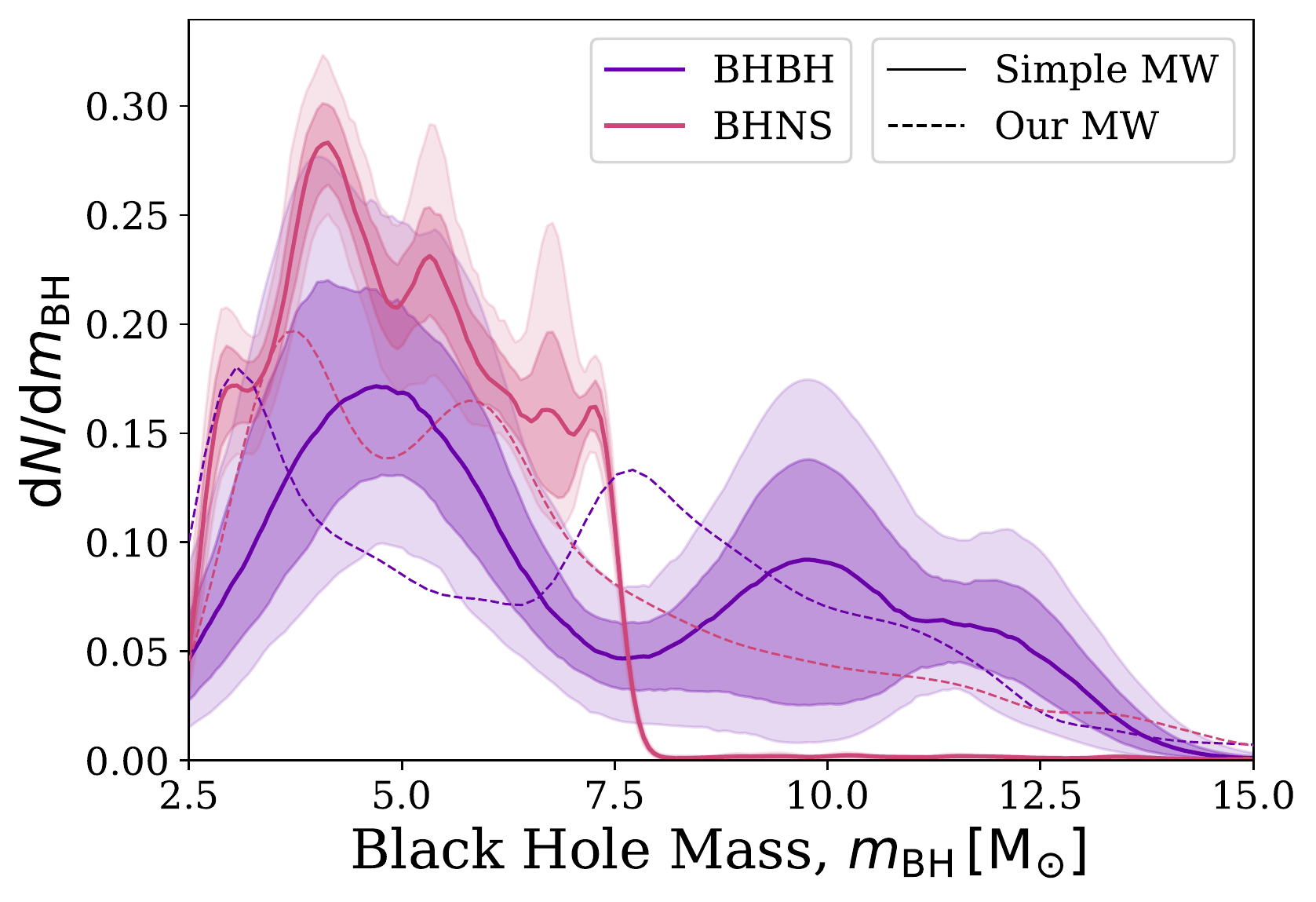}
    \caption{As Fig.~\ref{fig:fiducial_pdf_distributions}b, but for the Milky Way model used in \citet{Breivik+2020}. Dotted lines show the distribution from Fig.~\ref{fig:fiducial_pdf_distributions}b for comparison. \href{https://github.com/TomWagg/detecting-DCOs-in-LISA/blob/main/paper/figures/figD2_mBH_simple_mw_variation.pdf}{\faFileImage} \href{https://github.com/TomWagg/detecting-DCOs-in-LISA/blob/main/paper/figure_notebooks/fiducial.ipynb}{\faBook}.}
    \label{fig:bh_mass_simple_mw}
\end{figure}

Moreover, the distribution of parameters within the population, particularly the mass distributions, are notably disparate. By using only two fixed metallicity populations, unphysical artifacts are introduced into distribution of DCO masses \citep[e.g.][]{Dominik+2015, Neijssel+2019, Kummer_thesis}. For example, in Fig.~\ref{fig:bh_mass_simple_mw}, we show the black hole mass distribution produced by the simulation using the simple Milky Way model. Despite the fact that these KDEs use the same bandwidth as Fig.~\ref{fig:fiducial_pdf_distributions}, the distributions show many more sharp transitions, which is a result of pileups occurring at specific masses for specific metallicities. Moreover, the lack of lower metallicities systems means that higher mass systems are not formed and so we see the distributions do not include a high mass tail such as in our fiducial results.

The unphysical artifacts present in the mass distributions can have far-reaching effects since the masses of DCOs affect most other parameters. The inspiral time and SNR are directly dependent on the mass, whilst the uncertainty estimates depend on the SNR. This means that the artifacts can affect the predictions for most distributions of LISA detectable populations.

Overall, we find that previous studies that use Milky Way models analogous to this simpler model may significantly underestimate the LISA BHNS rate whilst overestimating the NSNS detection rate. They may also miss higher mass systems (particular for BHNSs) and contain unphysical artifacts in their parameter distributions.

\section{Estimating the number of pulsars for a given sky area in SKA}\label{app:ska_area}

In this section, we perform some back-of-the-envelope calculations in order to estimate the number of pulsars that SKA will observe within a given sky area.

First, we consider how many pulsars SKA is likely to detect. \citet{Keane+2015} uses PSRPOPPy \citep{Bates+2014} to simulate the Milky Way pulsar population. They find that for SKA-1, approximately $10000$ pulsars will be discovered. The second phase of SKA, which should be in operation by the time of the LISA mission, would yield a total of $35000$-$41000$ pulsars \citep{Keane+2015}. We use the average, $38000$, in further estimates below. Moreover, we are only interested in pulsars that are part of a binary system. We estimate this pulsar binary fraction as the fraction of known pulsars that are in binaries using the ATNF Pulsar Catalogue\footnote{\url{https://www.atnf.csiro.au/research/pulsar/psrcat}} \citep{Manchester+2005}. $290$ of the $2872$ currently known pulsars are in binary systems and thus we estimate the binary fraction of pulsars as $10\%$. Therefore, we expect that SKA-1 and SKA-2 will detect approximately $1000$ and $3800$ binary pulsars respectively.

Next, we can find the total number of pulsars SKA will detect in a patch on the sky. The total sky area that the SKA mission covers is approximately $5700 \unit{deg^2}$, which is calculated by integrating over the sky for all Galactic longitudes and Galactic latitudes limited to $\abs{b} < 10^\circ$ and $\delta < 45^\circ$, which are the limits on SKA-mid \citep{Keane+2015}. If we assume that the pulsars are found uniformly across the sky, this means that roughly $0.2$ and $0.7$ binary pulsars are expected per square degree for SKA-1 and SKA-2 respectively. Note that the assumption of a uniform distribution is not realistic as pulsars will tend to be far more concentrated in the Galactic centre but we use it to provide a slightly optimistic estimate.

Overall, we therefore expect a single pulsar per $5.7 \unit{deg^2}$ and $1.5 \unit{deg^2}$ for SKA-1 and SKA-2 respectively, which correspond to angular resolutions of $\sigma_\theta = 1.3^\circ$ and $\sigma_\theta = 0.7^\circ$.

\clearpage
\onecolumngrid

\section{Supplementary material}

\begin{table*}[htb]
    \centering
    \caption{The number of detectable binaries in a 4- and 10-year LISA mission for the \nModels{} different model variations and each DCO type. Each model variation is discussed in App.~\ref{sec:variation_assumptions} and the trends in detection rates are discussed in Sec.~\ref{sec:detection_rate_analysis}. The `All' column contains the total expected detections when summed over the three types. The final two rows show the minimum and maximum rates across all model variations. We embolden the corresponding rate for convenience of seeing which variation results in the minimum/maximum. Each value shows the mean and the 1-$\sigma$ Poisson uncertainty. \href{https://github.com/TomWagg/detecting-DCOs-in-LISA/blob/main/paper/figure_notebooks/detections.ipynb}{\faBook}.}
    \begin{tabular}{cl|cccc|cccc}
        \hline
        \multirow{2}{*}{Model} & \multirow{2}{*}{Description} & \multicolumn{4}{c|}{LISA detections (4 year)} & \multicolumn{4}{c}{LISA detections (10 year)} \\ \cline{3-10}
        & & {All} & {BHBH} & {BHNS} & {NSNS} & {All} & {BHBH} & {BHNS} & {NSNS} \\
        \hline
        A & Fiducial & \confinv{124}{11}{11} & \confinv{74}{9}{8} & \confinv{42}{6}{7} & \confinv{8}{3}{3} & \confinv{202}{15}{14} & \confinv{117}{10}{11} & \confinv{71}{8}{8} & \confinv{13}{4}{4}\\
        B & Fixed mass transfer efficiency of $\beta=0.25$ & \confinv{94}{10}{10} & \confinv{69}{9}{8} & \confinv{22}{4}{5} & \confinv{3}{2}{2} & \confinv{149}{12}{12} & \confinv{108}{11}{10} & \confinv{37}{6}{6} & \confinv{5}{2}{2}\\
        C & Fixed mass transfer efficiency of $\beta=0.5$ & \confinv{59}{8}{8} & \confinv{47}{7}{7} & \confinv{8}{3}{3} & \confinv{4}{2}{2} & \confinv{96}{10}{9} & \confinv{76}{9}{8} & \confinv{14}{4}{3} & \confinv{7}{3}{2}\\
        D & Fixed mass transfer efficiency of $\beta=0.75$ & \confinv{67}{8}{8} & \confinv{47}{7}{7} & \confinv{7}{2}{3} & \confinv{13}{4}{3} & \confinv{104}{10}{11} & \confinv{71}{8}{8} & \confinv{12}{3}{4} & \confinv{21}{4}{5}\\
        E$^\prime$ & Case BB mass transfer is always unstable & \confinv{90}{10}{9} & \confinv{66}{8}{8} & \confinv{21}{5}{5} & \boldconfinv{3}{2}{1} & \confinv{133}{11}{12} & \confinv{101}{10}{10} & \confinv{29}{5}{5} & \boldconfinv{4}{2}{2}\\
        F & Case BB Unstable + Optimistic CE & \boldconfinv{368}{19}{19} & \boldconfinv{154}{13}{12} & \boldconfinv{198}{14}{14} & \confinv{17}{4}{4} & \boldconfinv{553}{24}{23} & \boldconfinv{238}{15}{16} & \boldconfinv{289}{17}{17} & \confinv{25}{5}{5}\\
        G & CE efficiency $\alpha = 0.1$ & \confinv{40}{6}{6} & \confinv{28}{5}{5} & \boldconfinv{2}{1}{2} & \confinv{10}{3}{3} & \confinv{64}{8}{8} & \confinv{44}{7}{6} & \boldconfinv{3}{1}{2} & \confinv{17}{4}{4}\\
        H & CE efficiency $\alpha = 0.5$ & \confinv{86}{10}{9} & \confinv{58}{7}{8} & \confinv{22}{5}{4} & \confinv{6}{3}{2} & \confinv{136}{12}{11} & \confinv{91}{9}{10} & \confinv{35}{6}{6} & \confinv{10}{3}{3}\\
        I & CE efficiency $\alpha = 2.0$ & \confinv{133}{12}{11} & \confinv{67}{8}{9} & \confinv{38}{6}{6} & \confinv{28}{6}{5} & \confinv{218}{15}{14} & \confinv{109}{10}{11} & \confinv{62}{7}{8} & \confinv{46}{7}{7}\\
        J & CE efficiency $\alpha = 10.0$ & \confinv{78}{9}{9} & \confinv{27}{6}{5} & \confinv{16}{4}{4} & \boldconfinv{35}{6}{6} & \confinv{126}{11}{11} & \confinv{42}{6}{7} & \confinv{27}{6}{5} & \boldconfinv{57}{7}{8}\\
        K & HG donor stars initiating a CE survive CE & \confinv{218}{15}{14} & \confinv{151}{12}{12} & \confinv{57}{8}{7} & \confinv{10}{3}{3} & \confinv{340}{18}{19} & \confinv{229}{15}{15} & \confinv{96}{10}{10} & \confinv{16}{4}{4}\\
        L & Fryer rapid SN remnant mass prescription & \confinv{127}{11}{11} & \confinv{50}{7}{7} & \confinv{70}{8}{8} & \confinv{7}{3}{2} & \confinv{204}{14}{15} & \confinv{76}{8}{9} & \confinv{117}{11}{11} & \confinv{11}{3}{3}\\
        M & Maximum NS mass = 2.0 ${\rm M_{\odot}}$ & \confinv{133}{11}{12} & \confinv{96}{10}{10} & \confinv{30}{5}{5} & \confinv{7}{2}{3} & \confinv{214}{14}{15} & \confinv{153}{12}{13} & \confinv{50}{7}{7} & \confinv{12}{4}{3}\\
        N & Maximum NS mass = 3.0 ${\rm M_{\odot}}$ & \confinv{118}{11}{11} & \confinv{58}{8}{8} & \confinv{52}{7}{7} & \confinv{8}{3}{3} & \confinv{189}{13}{14} & \confinv{91}{9}{10} & \confinv{85}{10}{9} & \confinv{14}{4}{3}\\
        O & No PISN and pulsational-PISN & \confinv{126}{11}{12} & \confinv{75}{9}{9} & \confinv{43}{6}{7} & \confinv{8}{3}{3} & \confinv{205}{14}{14} & \confinv{120}{11}{11} & \confinv{72}{8}{9} & \confinv{13}{4}{3}\\
        P & $\sigma_{\rm RMS}^{\rm 1D} = 100 \ {\rm km\ s^{-1}}$ for CCSN & \confinv{184}{14}{14} & \confinv{82}{9}{9} & \confinv{86}{9}{10} & \confinv{15}{4}{4} & \confinv{300}{17}{17} & \confinv{130}{12}{11} & \confinv{145}{12}{12} & \confinv{26}{6}{5}\\
        Q & $\sigma_{\rm RMS}^{\rm 1D} = 30 \ {\rm km\ s^{-1}}$ for CCSN & \confinv{268}{16}{16} & \confinv{92}{10}{9} & \confinv{143}{12}{12} & \confinv{34}{6}{6} & \confinv{426}{20}{21} & \confinv{142}{11}{12} & \confinv{229}{16}{15} & \confinv{55}{7}{8}\\
        R & Black holes receive not natal kick & \confinv{205}{14}{15} & \confinv{89}{9}{10} & \confinv{109}{11}{10} & \confinv{7}{2}{3} & \confinv{332}{18}{18} & \confinv{140}{12}{12} & \confinv{180}{13}{13} & \confinv{12}{4}{3}\\
        S & Wolf-Rayet wind factor $f_{\rm WR} = 0.1$ & \confinv{118}{11}{11} & \confinv{75}{8}{9} & \confinv{34}{6}{6} & \confinv{9}{3}{3} & \confinv{182}{14}{13} & \confinv{112}{11}{11} & \confinv{56}{8}{7} & \confinv{14}{4}{4}\\
        T & Wolf-Rayet wind factor $f_{\rm WR} = 5.0$ & \boldconfinv{30}{6}{5} & \boldconfinv{6}{3}{2} & \confinv{15}{3}{4} & \confinv{8}{2}{3} & \boldconfinv{49}{7}{7} & \boldconfinv{9}{3}{3} & \confinv{26}{5}{5} & \confinv{13}{3}{4}\\
        \hline 
        - & Minimum rate & \confinv{30}{6}{5} & \confinv{6}{3}{2} & \confinv{2}{1}{2} & \confinv{3}{2}{1} & \confinv{49}{7}{7} & \confinv{9}{3}{3} & \confinv{3}{1}{2} & \confinv{4}{2}{2} \\
        - & Maximum rate & \confinv{368}{19}{19} & \confinv{154}{13}{12} & \confinv{198}{14}{14} & \confinv{35}{6}{6} & \confinv{553}{24}{23} & \confinv{238}{15}{16} & \confinv{289}{17}{17} & \confinv{57}{7}{8} \\
        \hline
    \end{tabular}
    \label{tab:detection_rates}
\end{table*}

\begin{figure}[p]
    \centering
    \includegraphics[height=0.85\textheight]{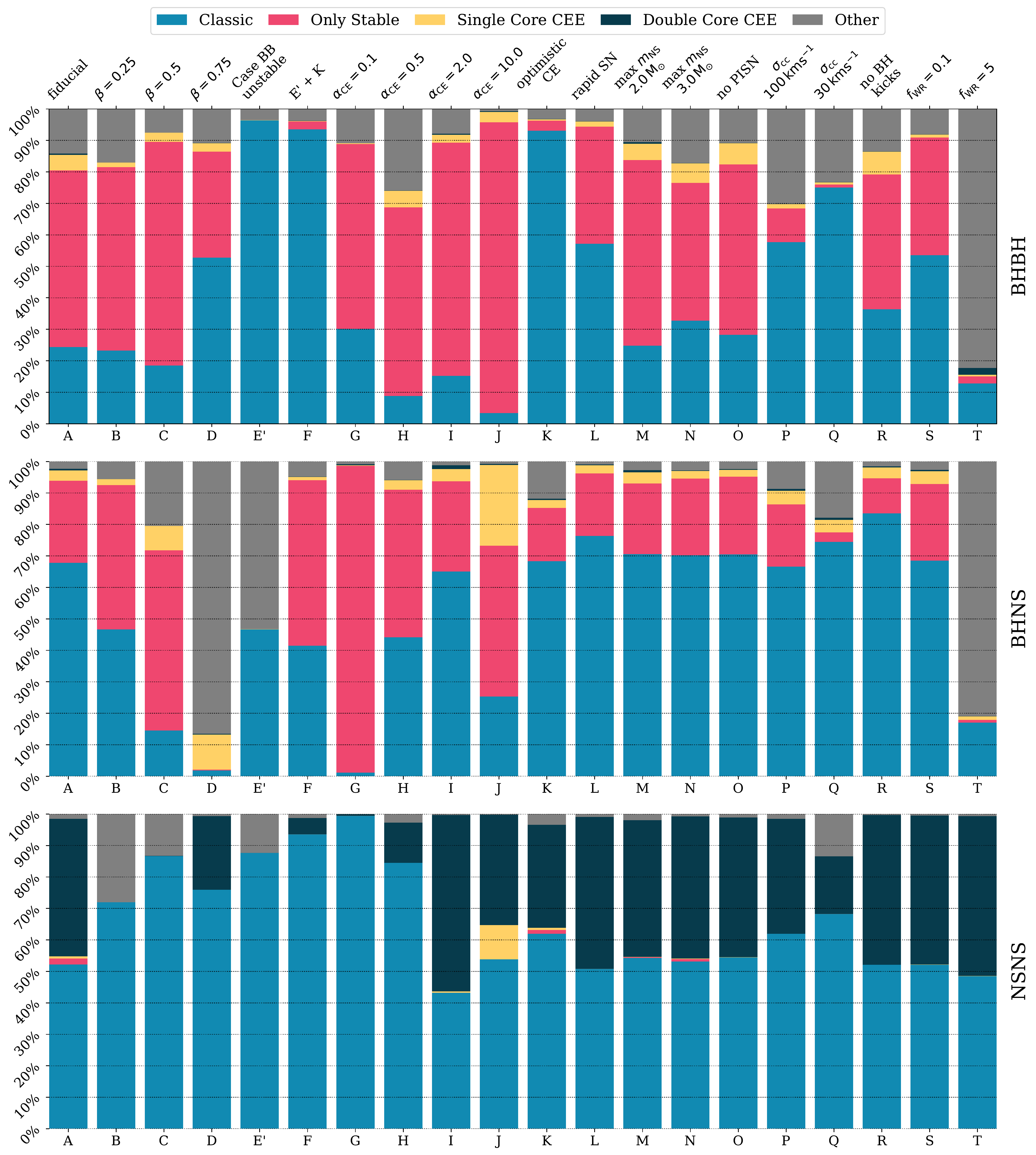}
    \caption{Fraction of each DCO type that is formed through different formation channels for all physics variations. Channels are described in detail in \citet{Broekgaarden+2021}. The classic, single core CEE and double core CEE channels all require at least one common-envelope event whilst only ``only stable'' consists of only stable mass transfer and ``other'' contains the remaining binaries which are mainly formed from case A ``classic'' binaries as well as ``lucky'' supernova kicks that shrink the binary. \href{https://github.com/TomWagg/detecting-DCOs-in-LISA/blob/main/paper/figures/figF1_formation_channels.pdf}{\faFileImage} \href{https://github.com/TomWagg/detecting-DCOs-in-LISA/blob/main/paper/figure_notebooks/formation_channels.ipynb}{\faBook}.}
    \label{fig:formation_channels}
\end{figure}

\begin{figure*}[p]
    \centering
    \includegraphics[width=\textwidth]{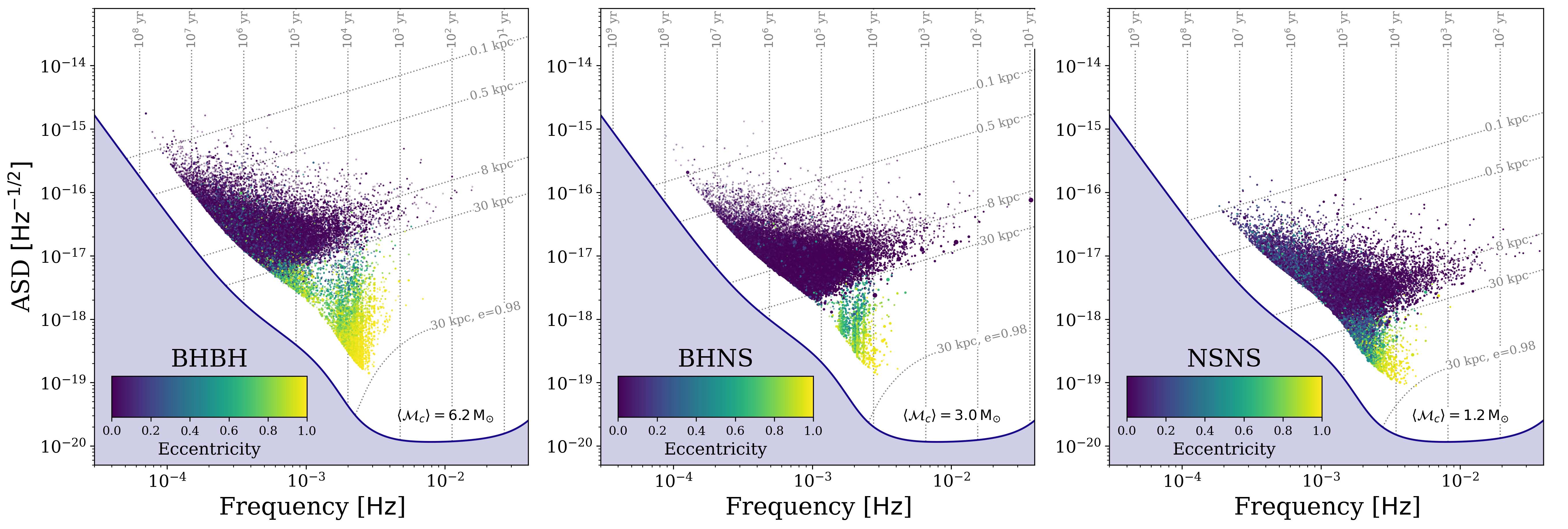}
    \caption{As the bottom panels of Fig.~\ref{fig:dcos_on_sc}, but without the density distributions and scatter points are coloured by their eccentricity. We show eccentric sources are located in an offshoot below the $30 \unit{kpc}$ around $2 \unit{mHz}$. \href{https://github.com/TomWagg/detecting-DCOs-in-LISA/blob/main/paper/figures/figF2_dcos_on_sc_eccentric_colours.png}{\faFileImage} \href{https://github.com/TomWagg/detecting-DCOs-in-LISA/blob/main/paper/figure_notebooks/sensitivity_curve.ipynb}{\faBook}.}
    \label{fig:dcos_on_sc_ecc_col}

    \includegraphics[width=0.8\textwidth]{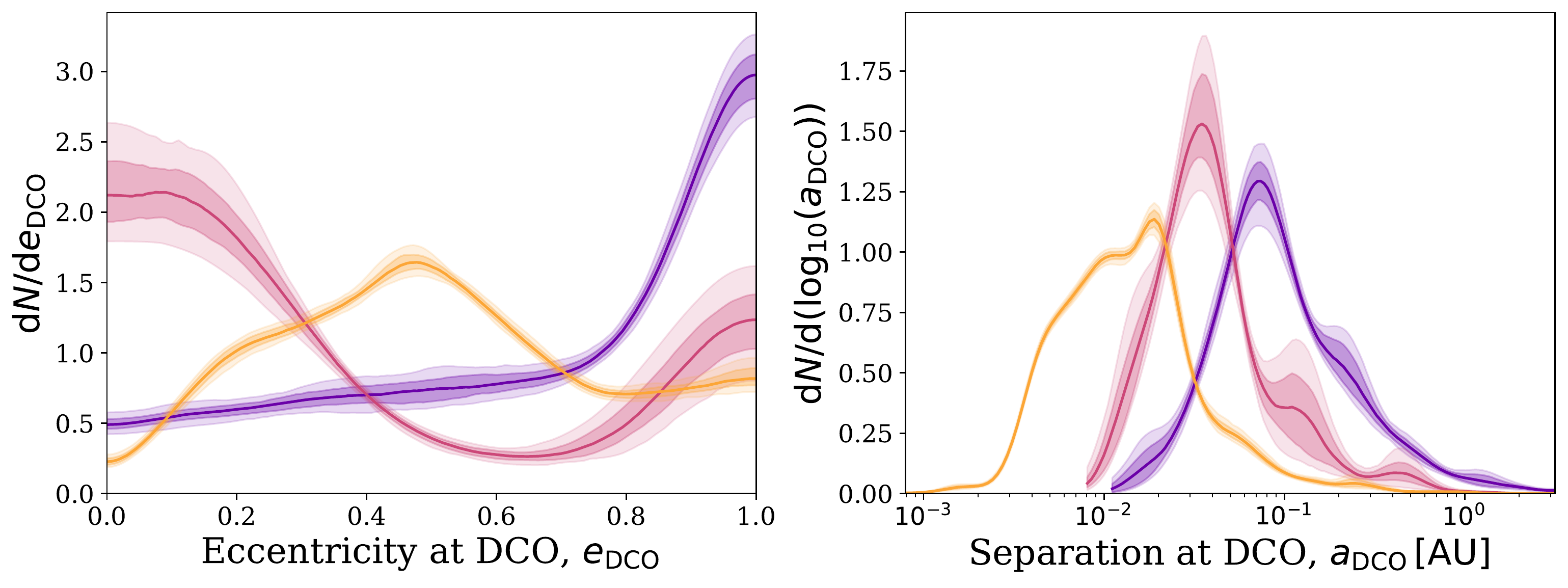}
    \caption{As Fig.~\ref{fig:fiducial_pdf_distributions}, but for the properties of the detectable systems at DCO formation. \href{https://github.com/TomWagg/detecting-DCOs-in-LISA/blob/main/paper/figures/figF3_dco_formation_distributions.pdf}{\faFileImage} \href{https://github.com/TomWagg/detecting-DCOs-in-LISA/blob/main/paper/figure_notebooks/fiducial.ipynb}{\faBook}.}
    \label{fig:dco_formation_properties}
\end{figure*}

\end{document}